\documentclass[twocolumn]{aastex631}
\usepackage{CLASSY_Preamble}

\begin{document}

\submitjournal{AASJournal ApJSS}
\shortauthors{Berg et al.}
\shorttitle{CLASSY~I}

\title{The COS Legacy Archive Spectroscopy SurveY (CLASSY) Treasury Atlas\footnote{
Based on observations made with the NASA/ESA Hubble Space Telescope,
obtained from the Data Archive at the Space Telescope Science Institute, which
is operated by the Association of Universities for Research in Astronomy, Inc.,
under NASA contract NAS 5-26555.}}

\author[0000-0002-4153-053X]{Danielle A. Berg}
\affiliation{Department of Astronomy, The University of Texas at Austin, 2515 Speedway, Stop C1400, Austin, TX 78712, USA}

\author[0000-0003-4372-2006]{Bethan L. James}
\affiliation{AURA for ESA, Space Telescope Science Institute, 3700 San Martin Drive, Baltimore, MD 21218, USA}

\author[0000-0003-0834-4150]{Teagan King}
\affiliation{Space Telescope Science Institute, 3700 San Martin Drive, Baltimore, MD 21218, USA}

\author[0000-0003-0557-3433]{Meaghan McDonald}
\affiliation{Space Telescope Science Institute, 3700 San Martin Drive, Baltimore, MD 21218, USA}

\author[0000-0002-2178-5471]{Zuyi Chen}
\affiliation{Steward Observatory, The University of Arizona, 933 N Cherry Ave, Tucson, AZ, 85721, USA}

\author[0000-0002-0302-2577]{John Chisholm}
\affiliation{Department of Astronomy, The University of Texas at Austin, 2515 Speedway, Stop C1400, Austin, TX 78712, USA}

\author[0000-0003-1127-7497]{Timothy Heckman}
\affiliation{Center for Astrophysical Sciences, Department of Physics \& Astronomy, Johns Hopkins University, Baltimore, MD 21218, USA}

\author[0000-0001-9189-7818]{Crystal L. Martin}
\affiliation{Department of Physics, University of California, Santa Barbara, Santa Barbara, CA 93106, USA}

\author[0000-0001-6106-5172]{Dan P. Stark}
\affiliation{Steward Observatory, The University of Arizona, 933 N Cherry Ave, Tucson, AZ, 85721, USA}

\collaboration{39}{the CLASSY Team:}
\author[0000-0003-4137-882X]{Alessandra Aloisi}
\affiliation{Space Telescope Science Institute, 3700 San Martin Drive, Baltimore, MD 21218, USA}
\author[0000-0001-5758-1000]{Ricardo O. Amor\'{i}n}
\affiliation{Instituto de Investigaci\'{o}n Multidisciplinar en Ciencia y Tecnolog\'{i}a, Universidad de La Serena, Raul Bitr\'{a}n 1305, La Serena 2204000, Chile}
\affiliation{Departamento de Astronom\'{i}a, Universidad de La Serena, Av. Juan Cisternas 1200 Norte, La Serena 1720236, Chile}
\author[0000-0002-2644-3518]{Karla Z. Arellano-C\'{o}rdova}
\affiliation{Department of Astronomy, The University of Texas at Austin, 2515 Speedway, Stop C1400, Austin, TX 78712, USA}
\author[0000-0003-1074-4807]{Matthew Bayliss}
\affiliation{Department of Physics, University of Cincinnati, Cincinnati, OH 45221, USA}
\author[0000-0002-3120-7173]{Rongmon Bordoloi}
\affiliation{Department of Physics, North Carolina State University, 421 Riddick Hall, Raleigh, NC 27695-8202, USA}
\author[0000-0003-4359-8797]{Jarle Brinchmann}
\affiliation{Instituto de Astrof\'{i]}sica e Ci\^{e}ncias do Espa\c{c}o, Universidade do Porto, CAUP, Rua das Estrelas, PT4150-762 Porto, Portugal}
\author[0000-0003-3458-2275]{St\'{e}phane Charlot}
\affiliation{Sorbonne Universit\'{e}, CNRS, UMR7095, Institut d'Astrophysique de Paris, F-75014, Paris, France}
\author[0000-0002-7636-0534]{Jacopo Chevallard}
\affiliation{Sorbonne Universit\'{e}, UPMC-CNRS, UMR7095, Institut d'Astrophysique de Paris, F-75014, Paris, France}
\author[0000-0003-3334-4267]{Ilyse Clark}
\affiliation{Department of Astronomy, The University of Texas at Austin, 2515 Speedway, Stop C1400, Austin, TX 78712, USA}
\author[0000-0001-9714-2758]{Dawn K. Erb}
\affiliation{Center for Gravitation, Cosmology and Astrophysics, Department of Physics, University of Wisconsin Milwaukee, 3135 N Maryland Ave., Milwaukee, WI 53211, USA}
\author[0000-0001-6865-2871]{Anna Feltre}
\affiliation{INAF - Osservatorio di Astrofisica e Scienza dello Spazio di Bologna, Via P. Gobetti 93/3, 40129 Bologna, Italy}
\author[0000-0001-8587-218X]{Matthew Hayes}
\affiliation{Stockholm University, Department of Astronomy and Oskar Klein Centre for Cosmoparticle Physics, AlbaNova University Centre, SE-10691, Stockholm, Sweden}
\author[0000-0002-6586-4446]{Alaina Henry}
\affiliation{Space Telescope Science Institute, 3700 San Martin Drive, Baltimore, MD 21218, USA}
\affiliation{Center for Astrophysical Sciences, Department of Physics \& Astronomy, Johns Hopkins University, Baltimore, MD 21218, USA}
\author[0000-0003-4857-8699]{Svea Hernandez}
\affiliation{AURA for ESA, Space Telescope Science Institute, 3700 San Martin Drive, Baltimore, MD 21218, USA}
\author[0000-0002-6790-5125]{Anne Jaskot}
\affiliation{Department of Astronomy, Williams College, USA}
\author[0000-0001-5860-3419]{Tucker Jones}
\affiliation{University of California, Davis}
\author[0000-0001-8152-3943]{Lisa J. Kewley}
\affiliation{Research School of Astronomy and Astrophysics, Australian National University, Cotter Road, Weston Creek, ACT 2611, Australia; ARC Centre of Excellence for All Sky Astrophysics in 3 Dimensions (ASTRO 3D), Canberra, ACT 2611, Australia}
\author[0000-0002-5320-2568]{Nimisha Kumari}
\affiliation{AURA for ESA, Space Telescope Science Institute, 3700 San Martin Drive, Baltimore, MD 21218, USA}
\author[0000-0003-2685-4488]{Claus Leitherer}
\affiliation{Space Telescope Science Institute, 3700 San Martin Drive, Baltimore, MD 21218, USA}
\author[0000-0003-1354-4296]{Mario Llerena}
\affiliation{Instituto de Investigaci\'{o}n Multidisciplinar en Ciencia y Tecnolog\'{i}a, Universidad de La Serena, Raul Bitr\'{a}n 1305, La Serena 2204000, Chile}
\author[0000-0003-0695-4414]{Michael Maseda}
\affiliation{Department of Astronomy, University of Wisconsin-Madison, 475 N. Charter St., Madison, WI 53706 USA}
\author[0000-0003-2589-762X]{Matilde Mingozzi}
\affiliation{Space Telescope Science Institute, 3700 San Martin Drive, Baltimore, MD 21218, USA}
\author[0000-0003-2804-0648]{Themiya Nanayakkara}
\affiliation{Swinburne University of Technology, Melbourne, Victoria, AU}
\author[0000-0002-1049-6658]{Masami Ouchi}
\affiliation{National Astronomical Observatory of Japan, 2-21-1 Osawa, Mitaka, Tokyo 181-8588, Japan}
\affiliation{Institute for Cosmic Ray Research, The University of Tokyo, Kashiwa-no-ha, Kashiwa 277-8582, Japan}
\affiliation{Kavli Institute for the Physics and Mathematics of the Universe (WPI), University of Tokyo, Kashiwa, Chiba 277-8583, Japan}
\author[0000-0003-0390-0656]{Adele Plat}
\affiliation{Steward Observatory, The University of Arizona, 933 N Cherry Ave, Tucson, AZ, 85721, USA}
\author[0000-0003-1435-3053]{Richard W. Pogge}
\affiliation{Department of Astronomy, The Ohio State University, 140 W 18th Avenue, Columbus, OH 43210, USA}
\affiliation{Center for Cosmology \& AstroParticle Physics, The Ohio State University, 191 W Woodruff Avenue, Columbus, OH 43210}
\author[0000-0002-5269-6527]{Swara Ravindranath}
\affiliation{Space Telescope Science Institute, 3700 San Martin Drive, Baltimore, MD 21218, USA}
\author[0000-0002-7627-6551]{Jane R. Rigby}
\affiliation{Observational Cosmology Lab, Code 665, NASA Goddard Space Flight Center, 8800 Greenbelt Rd, Greenbelt, MD 20771, USA}
\author[0000-0003-4792-9119]{Ryan Sanders}
\affiliation{University of California, Davis}
\author[0000-0002-9136-8876]{Claudia Scarlata}
\affiliation{Minnesota Institute for Astrophysics, University of Minnesota, 116 Church Street SE, Minneapolis, MN 55455, USA}
\author[0000-0002-9132-6561]{Peter Senchyna}
\affiliation{Carnegie Observatories, 813 Santa Barbara Street, Pasadena, CA 91101, USA}
\author[0000-0003-0605-8732]{Evan D. Skillman}
\affiliation{Minnesota Institute for Astrophysics, University of Minnesota, 116 Church Street SE, Minneapolis, MN 55455, USA}
\author[0000-0002-4834-7260]{Charles C. Steidel}
\affiliation{Cahill Center for Astronomy and Astrophysics, California Institute of Technology, MC249-17, Pasadena, CA 91125, USA}
\author[0000-0001-6369-1636]{Allison L. Strom}
\affiliation{Department of Astrophysical Sciences, 4 Ivy Lane, Princeton University, Princeton, NJ 08544, USA}
\author[0000-0001-6958-7856]{Yuma Sugahara}
\affiliation{Institute for Cosmic Ray Research, The University of Tokyo, Kashiwa-no-ha, Kashiwa 277-8582, Japan}
\affiliation{National Astronomical Observatory of Japan, 2-21-1 Osawa, Mitaka, Tokyo 181-8588, Japan}
\affiliation{Waseda Research Institute for Science and Engineering, Faculty of Science and Engineering, Waseda University, 3-4-1, Okubo, Shinjuku, Tokyo 169-8555, Japan}
\author[0000-0003-3903-6935]{Stephen M. Wilkins}
\affiliation{Astronomy Centre, University of Sussex, Falmer, Brighton BN1 9QH, UK}
\author[0000-0001-8289-3428]{Aida Wofford}
\affiliation{Instituto de Astronom\'{i}a, Universidad Nacional Aut\'{o}noma de M\'{e}xico, Unidad Acad\'{e}mica en Ensenada, Km 103 Carr. Tijuana-Ensenada, Ensenada 22860, M\'{e}xico}
\author[0000-0002-9217-7051]{Xinfeng Xu}
\affiliation{Center for Astrophysical Sciences, Department of Physics \& Astronomy, Johns Hopkins University, Baltimore, MD 21218, USA}

\correspondingauthor{Danielle A. Berg} 
\email{daberg@austin.utexas.edu}


\begin{abstract}
Far-ultraviolet (FUV;$\sim1200-2000$\AA) spectra are fundamental to our understanding of 
star-forming galaxies, providing a unique window on massive stellar populations, chemical 
evolution, feedback processes, and reionization. 
The launch of JWST will soon usher in a new era, pushing the UV spectroscopic 
frontier to higher redshifts than ever before, however, its success hinges on a 
comprehensive understanding of the massive star populations and gas conditions that power 
the observed UV spectral features. 
This requires a level of detail that is only possible with a combination of ample wavelength 
coverage, signal-to-noise, spectral-resolution, and sample diversity that 
has not yet been achieved by any FUV spectral database. 

We present the COS Legacy Spectroscopic SurveY (CLASSY) treasury and its first high 
level science product, the CLASSY atlas.
CLASSY builds on the HST archive to construct the first 
high-quality (S/N$_{1500\AA}\gtrsim5$/resel), 
high-resolution ($R\sim15,000$) FUV spectral database 
of 45 nearby ($0.002<z<0.182$) star-forming galaxies.
The CLASSY atlas, available to the public via the CLASSY website, is the result of optimally 
extracting and coadding 170 archival+new spectra from 312 orbits of HST observations.

The CLASSY sample covers a broad range of properties
including 
stellar mass ($6.2<$ log $M_\star$($M_\odot$) $<10.1$),
star formation rate ($-2.0<$ log SFR ($M_\odot$ yr$^{-1}$)  $<+1.6$),
direct gas-phase metallicity ($7.0<$ 12+log(O/H) $<8.8$),
ionization ($0.5<$ O$_{32} <38.0$),
reddening ($0.02<$ E(B-V) $<0.67$), and
nebular density ($10<n_e$(cm$^{-3}) <1120$).
CLASSY is biased to UV-bright star-forming galaxies, resulting in a sample
that is consistent with $z\sim0$ mass-metallicity relationship, but is offset 
to higher SFRs by roughly 2 dex, similar to $z\gtrsim2$ galaxies.
This unique set of properties makes the CLASSY atlas the benchmark 
training set for star-forming galaxies across cosmic time.
\end{abstract} 



\section{Introduction}\label{sec:1}


The interplay between gas and stars is one of the most fundamental, 
yet unsettled, drivers of galactic evolution. 
In the basic picture, gas is accreted onto galaxies from the cosmic web, 
settles into their gravitational wells, and is converted into stars. 
The massive stars in star-forming galaxies ionize the surrounding gas, 
producing nebular emission and driving outflows and radiative shocks. 
Such feedback drives chemical evolution and can modulate or limit accretion processes, 
thereby regulating the subsequent growth of galaxies. 
The far-ultraviolet (FUV), defined here as $\sim1200-2000$ \AA, is arguably 
the richest wavelength regime in diagnostic spectral features characterizing 
these processes and will provide an important window onto the first generation 
of galaxies in the forthcoming era of extremely large telescopes (ELTs) 
and the James Webb Space Telescope (JWST). 

\subsection{The Need for a Benchmark FUV Spectral Atlas}
While the full FUV spectrum is rich in diagnostic power, studies have thus far 
largely focused on specific and individual components of the galaxy evolution puzzle.    
As a result of differing requirements of various studies, the archive has been a piecemeal 
collection, missing significant pieces that reside in different parts of the FUV spectrum. 
This incompleteness has prevented us from understanding the interdependent nature of 
the physical processes in star-forming galaxies and their UV spectroscopic properties.
Recently, spectroscopic studies of $2<z<4$ star-forming galaxies have emerged 
as our most comprehensive rest-FUV spectral datasets \citep[e.g.,][]{rigby18a, steidel16}.
Of particular importance to developing a detailed understanding of UV emission lines 
has been the large sample VUDS \citep[e.g.,][]{lefevre15}, VANDELS \citep[e.g.,][]{mclure18}, 
and MUSE \citep[e.g.,][]{schmidt21} surveys.
However, the proper tool-set to interpret these rich datasets has been lacking.\looseness=-2

A scientifically useful FUV spectral atlas requires three components. 
First, it must have broad wavelength coverage to simultaneously probe the continuum, 
absorption, and emission features that characterize the stellar populations and gas of the 
galaxy.
Second, theory predicts that the velocities of galactic outflows scale with the circular 
velocities of their host galaxies \citep[e.g., ][]{murray05}. 
This means that very high spectral resolution ($R > 10,000$ or velocity resolution less than 
30~km~s$^{-1}$) is required to measure the impact of stellar feedback in galaxies with low stellar-mass \citep[see, e.g.,][]{mcgaugh01}.
Third, spectra with high signal-to-noise (S/N) in the continuum are needed to 
measure faint features. \looseness=-2

There are several existing compilations of FUV spectra of local star-forming galaxies. 
These collections include the 
the \citet{kinney93} atlas, which is perhaps the most impactful FUV atlas to date, 
of 143 star-forming and active galaxies observed with the 
International Ultraviolet Explorer (IUE),
the \citet{leitherer02} atlas of 19 galaxies observed with the 
0.9-m Hopkins Ultraviolet Telescope (HUT),
the \citet{grimes09} atlas of 16 galaxies with archival Far Ultraviolet Spectroscopic 
Explorer Spectrograph (FUSE) spectra, and
the \citet{leitherer11} atlas of 28 galaxies observed with the 
Goddard High Resolution Spectrograph (GHRS) and the Faint Object Spectrograph (FOS).
Unfortunately, each of these atlases is limited by the properties of the UV spectrograph 
used and so is insufficient for current science objectives.

\begin{figure*}
\begin{center}
    \includegraphics[width=1.0\textwidth]{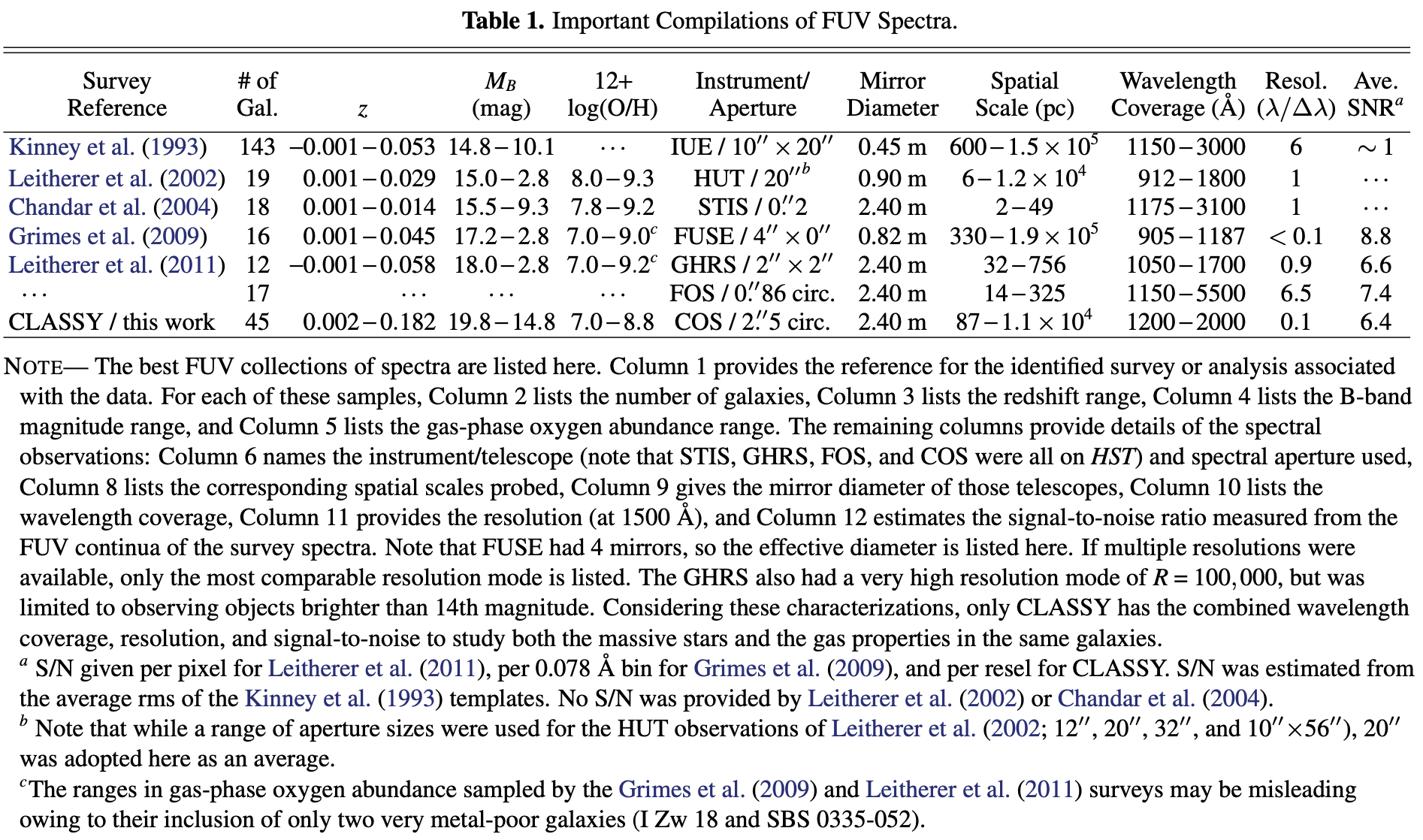} 
\end{center}
\caption*{
The best FUV collections of spectra are listed here.
Column 1 provides the reference for the identified survey or analysis associated with
the data.
For each of these samples, 
Column 2 lists the number of galaxies,
Column 3 lists the redshift range,
Column 4 lists the B-band magnitude range, and
Column 5 lists the gas-phase oxygen abundance range.
The remaining columns provide details of the spectral observations:
Column 6 names the instrument/telescope
(note that STIS, GHRS, FOS, and COS were all on \HST)
and spectral aperture used, 
Column 8 lists the corresponding spatial scales probed,
Column 9 gives the mirror diameter of those telescopes,
Column 10 lists the wavelength coverage,
Column 11 provides the resolution (at 1500 \AA), and
Column 12 estimates the signal-to-noise ratio measured from the FUV continua 
of the survey spectra.
Note that FUSE had 4 mirrors, so the effective diameter is listed here.
If multiple resolutions were available, only the most comparable resolution mode is listed.
The GHRS also had a very high resolution mode of $R=100,000$, but was limited to 
observing objects brighter than 14th magnitude.
Considering these characterizations, only CLASSY has the combined wavelength coverage,
resolution, and signal-to-noise to study both the massive stars and the gas properties
in the same galaxies. \newline
$^a$ S/N given per pixel for \citet{leitherer11}, per 0.078 \AA\ bin for
\citet{grimes09}, and per resel for CLASSY. 
S/N was estimated from the average rms of the \citet{kinney93} templates.
No S/N was provided by \citet{leitherer02} or \citet{chandar04}. \newline
$^b$ Note that while a range of aperture sizes were used for the HUT observations of
\citet[][12\arcsec, 20\arcsec, 32\arcsec, and 10\arcsec$\times$56\arcsec]{leitherer02},
20\arcsec\ was adopted here as an average. \newline
$^c$The ranges in gas-phase oxygen abundance sampled by the \citet{grimes09} and
\citet{leitherer11} surveys may be misleading owing to their inclusion of only
two very metal-poor galaxies (I~Zw~18 and SBS~0335-052).}
\label{tbl1}
\end{figure*}

The properties characterizing important compilations of FUV spectra are listed in Table~\ref{tbl1}
in the order of publication.
While the \citet{kinney93} atlas is the largest and still the most impactful,
the resolution ($\sim6$ \AA) and S/N ($\sim1$) of its spectra are too low to resolve 
the FUV spectra features and utilize their diagnostic power.  
The \citet{leitherer02} atlas using the HUT was ground-breaking for the FUV at the time
because it pushed blueward down to 912 \AA, but suffered from even lower spectral 
resolution ($\sim1$ \AA) than the \citet{kinney93} atlas.
In contrast, the \citet{grimes09} atlas was the first FUV atlas to achieve both
high S/N and spectral resolution, but observed only sub-\LYA\ ($<1215$ \AA) wavelengths. 
Finally, the \citet{leitherer11} atlas made progress in combining both large wavelength
coverage and moderately high spectral resolution (0.9 \AA), but only for a small subsample
of 12 galaxies with GHRS observations.
Sadly, the limited wavelength coverage, spectral resolution, and S/N of these past 
compilations are insufficient for the simultaneous study of the stellar, nebular, 
and outflow properties in star-forming galaxies needed to advance our understanding 
of galaxy evolution.


\subsection{The Cosmic Origins Spectrograph}
The installation of the Cosmic Origins Spectrograph (COS) on \HST\ in May 2009 during 
Servicing Mission 4 ushered in a new era of higher-sensitivity FUV spectroscopy. 
It is optimally designed for moderate-resolution UV spectroscopy 
of faint point sources with a resolving power of $R\sim15,000$
and a FUV spectroscopy channel covering approximately $900-2000$ \AA. 
Not only does COS have excellent FUV wavelength coverage and resolution,
as shown in Table~\ref{tbl1}, it also benefits from superior FUV sensitivity 
over all previous FUV spectrographs.
In fact, COS boasts more than 30 times the sensitivity of the Space Telescope
Imaging Spectrograph (STIS) for FUV observations of faint objects,
allowing its spectroscopic observations to unify the study of FUV stellar and 
gas-phase properties in a way that would not be possible with STIS\footnote{
Note, however, that STIS offers other capabilities over COS. 
Specifically, STIS has narrow slit settings available that offer higher spatial sampling 
than COS and optical gratings that provide redder wavelength coverage (1635--10137 \AA) 
than COS (913--3560 \AA).} 
or any other past FUV spectrograph. 

While COS has been used widely to observe many samples of star-forming galaxies and has
enabled key scientific discoveries, it lacks a high caliber atlas in the class
of those discussed above.
This is due in large part to the piecemeal nature of the current archive,
which, as a result, suffers from a variety of deficiencies. 
For one, archival COS data usually cover only a portion of the FUV owing to the 
diversity of science goals, the competitive nature of \HST, and the narrow
wavelength coverage of the individual medium-resolution gratings ($\sim300$\AA).
In addition, strong stellar features and interstellar absorption lines are more 
concentrated in the bluer end of the FUV, while
strong emission lines are clustered in the redder portion of the FUV.
Consequently, it is impossible to fully connect properties of the gas and stars 
within a single grating spectrum of a given galaxy.
Second, the S/N  in the continuum, and/or the resolution of archival COS spectra is 
often too low to accurately measure the ISM absorption or stellar features. 
For example, there are ample programs of low-resolution G140L grating spectra
that offer more efficient exposure times over the higher-resolution gratings, 
but data taken with G140L do not have sufficient spectral resolution to characterize 
the kinematics of the gas or disentangle multi-component emission and absorption features.
Third, the highest S/N FUV \HST\ spectra come from nearby galaxies, which can 
suffer the consequences of foreground Milky Way absorption lines contaminating 
those intrinsic to a galaxy. 

These problems are illustrated by the state of the Hubble Spectral Legacy Archive 
\citep[HSLA;][]{peeples17}, which {\it "provides the community with all raw and 
associated combined spectra for COS far-ultraviolet data publicly available in MAST"}. 
In the July, 2018 HSLA release, the number of nearby ($z<0.3$) galaxies with high-S/N 
($\gtrsim5$ per 100 km s$^{-1}$ resolution element) FUV COS spectroscopy totaled only 101. 
Of these, only 37 galaxies had observations with more than one grating setting, and poignantly, 
only two had full FUV $1200-2000$ \AA\ coverage.

\subsection{The CLASSY Treasury}\label{sec:1.2}
Here we present the COS Legacy Archive Spectroscopic SurveY (CLASSY) as the first 
high-S/N ($>$ 5), high-resolution ($R\sim15,000$) FUV
COS spectral atlas of star-forming galaxies.
CLASSY fills in the fundamental gaps in the 1200--2000 \AA\ FUV range 
that will achieve a holistic view of star-forming galaxies 
and enhance the legacy and utility of the COS archive.

CLASSY combines 135 orbits of new \HST\ data with 177 orbits of archival \HST\ data, 
for a total of 312 orbits, to complete the first atlas of 
high-quality rest-FUV COS spectra of 45 local star-forming galaxies. 
In order to achieve nearly-panchromatic FUV spectral coverage with the highest 
spectral resolution possible, CLASSY efficiently augments existing archival data 
with new \HST/COS observations, uniting the high-resolution G130M, G160M, and G185M gratings
for a sample of galaxies spanning broad parameter space.
As a result, CLASSY provides a well-controlled local FUV sample with the requisite sensitivity 
and spectral resolution to enable synergistic studies of stars and gas within the {\it same} galaxies.
The rest-FUV coverage of the CLASSY data provides a diversity of spectral features 
with unparalleled diagnostic power. 
With a wavelength range of roughly 1200--2000 \AA, the CLASSY spectra cover 
many emission and absorption features important for characterizing the ionizing stellar 
populations (e.g., metallicity, age), the properties of gas flows (velocities, 
column densities, ionization phases), and the physical conditions of the 
multiphase interstellar gas 
(e.g., metallicity, temperature, ionization) within the same galaxies.

The CLASSY atlas is the benchmark training set for star-forming galaxies across 
cosmic time, including the earliest galaxies that will be observed with the eminent James 
Webb Space Telescope (JWST) and the next generation of extremely large telescopes (ELTs).
Therefore, CLASSY will provide a number of state-of-the-art data products to the astronomical 
community via MAST and HSLA to ensure their enduring value and utility.
The primary data products are the high-resolution, high-S/N coadded multi-grating FUV spectral 
templates of the CLASSY sample of 45 star-forming galaxies presented here. 
Additional planned data products include stellar continuum fits, nebular abundance measurements, 
improved UV emission line diagnostics, feedback properties, radiative transfer models, and more.
With these high-level science products (HLSPs), the CLASSY Treasury will complete the vital 
picture that is needed to diagnose the suite of star-forming galaxy properties that will be 
offered to us in the next decade. 
Given the uncertain future of observed-frame UV capabilities, 
the CLASSY templates and models embody an indispensable toolset.

Here we present Paper~I of the CLASSY Treasury detailing 
the release of the first HLSP: the CLASSY coadded spectral atlas.
In Section~\ref{sec:2} we introduce the CLASSY atlas of FUV and optical spectra. 
We describe the sample selection in \S~\ref{sec:2.1},
the \HST/COS spectroscopic observations and coaddition process in \S~\ref{sec:2.2},
and the ancillary optical spectra in \S~\ref{sec:2.3}.
The details of the coadded spectra data release are described in Section~\ref{sec:3},
including the coadded spectra file format (\S~\ref{sec:3.1}),
digital access to the resulting data release (\S~\ref{sec:3.2})
for ease of utility, and
a comparison to previous FUV surveys in (\S~\ref{sec:3.3}).
In order to give a general overview of the CLASSY galaxies and their spectra,
we discuss optical measurements of their host galaxy properties in Section~\ref{sec:4}.
Specifically, we revisit the ancillary optical spectra used in \S~\ref{sec:4.1},
the emission line measurements (\S~\ref{sec:4.2}) and their reddening corrections 
(\S~\ref{sec:4.3}), and calculations of the electron densities (\S~\ref{sec:4.4}), 
temperatures (\S~\ref{sec:4.4}), oxygen abundances (\S~\ref{sec:4.5}), ionization parameters 
(\S~\ref{sec:4.6}), and stellar mass and star formation rates (\S~\ref{sec:4.7}).
The resulting global properties of the CLASSY sample are discussed in 
Section~\ref{sec:5}.
Finally, a demonstration of the subsequent scientific studies 
that can be performed with CLASSY is presented in Section~\ref{sec:6}.
A summary of Paper~I can be found in Section~\ref{sec:7}.

We refer the reader to the Appendix for additional CLASSY resources.
Specifically, 
Appendix~\ref{AppendixA} contains tables of the CLASSY FUV$+$optical 
spectral observations (Tables~\ref{tbl2} and \ref{tbl4}).
Appendix~\ref{AppendixB} contains tables of the CLASSY galaxy properties derived 
from optical spectra (Table~\ref{tbl5}) and photometry (Table~\ref{tbl6}).
Appendix~\ref{AppendixC} discusses the details of our spectral energy distribution fitting.  
Appendix~\ref{AppendixD} compiles useful notes on individual CLASSY galaxies. 
Finally, appendix~\ref{AppendixE} displays the CLASSY coadded spectrum for 
each galaxy in the sample. 
Note that, for consistency, the CLASSY objects are designated by an abbreviation of their
sexigesimal coordinates as J\texttt{HHMM}$\pm$\texttt{DDMM}.
A flat $\Lambda$CDM cosmology ($H_0$ = 70 km s$^{-1}$ Mpc$^{-1}$, $\Omega_m$ = 0.3) was 
assumed throughout this work.


\section{The CLASSY Atlas of UV+Optical Spectra}\label{sec:2}


The primary community data product of the CLASSY treasury is high-quality, 
coadded FUV plus optical spectra for the 45 star-forming galaxies in the CLASSY sample.
These CLASSY spectra will enable a number of important synergistic studies between
the stars and gas of galaxies that will result in additional CLASSY HLSPs (see \S~\ref{sec:6}) 
and an untapped wealth of potential community science investigations. 
Below we describe the CLASSY sample selection in \S~\ref{sec:2.1}, 
the \HST/COS FUV observations in \S~\ref{sec:2.2}, and
briefly describe the spectra coaddition process in \S~\ref{sec:2.3}.
We refer the reader to \citet[][hereafter, \PII]{james22} for a detailed
discussion of the data reduction and technical details of the coaddition process,
along with best practices for COS data of extended sources.


\subsection{Sample Selection}\label{sec:2.1}

\noindent{\it Overview:}
The CLASSY survey was motivated to build an atlas on the shoulders of the \HST/COS
archive to achieve comprehensive, high-resolution, high-S/N rest-FUV spectra of 
nearby ($0.002 < z < 0.182$) star-forming galaxies.  
Using careful sample selection to achieve these goals, the resulting CLASSY sample 
spans a wide range of stellar masses (log$M_\star\sim6-10 M_\odot$), 
star formation rates (log SFR $\sim -2 - +2$ M yr$^{−1}$),  
oxygen abundances (12+log(O/H)$\sim7-9$), 
electron densities ($n_e\sim10^1-10^3$ cm$^{−3}$), 
and reddening values (E(B--V)$\sim0.0-0.7$).

\begin{figure*}
\begin{center}
    \includegraphics[width=0.75\textwidth,trim=0mm 0mm 0mm 0mm,clip]{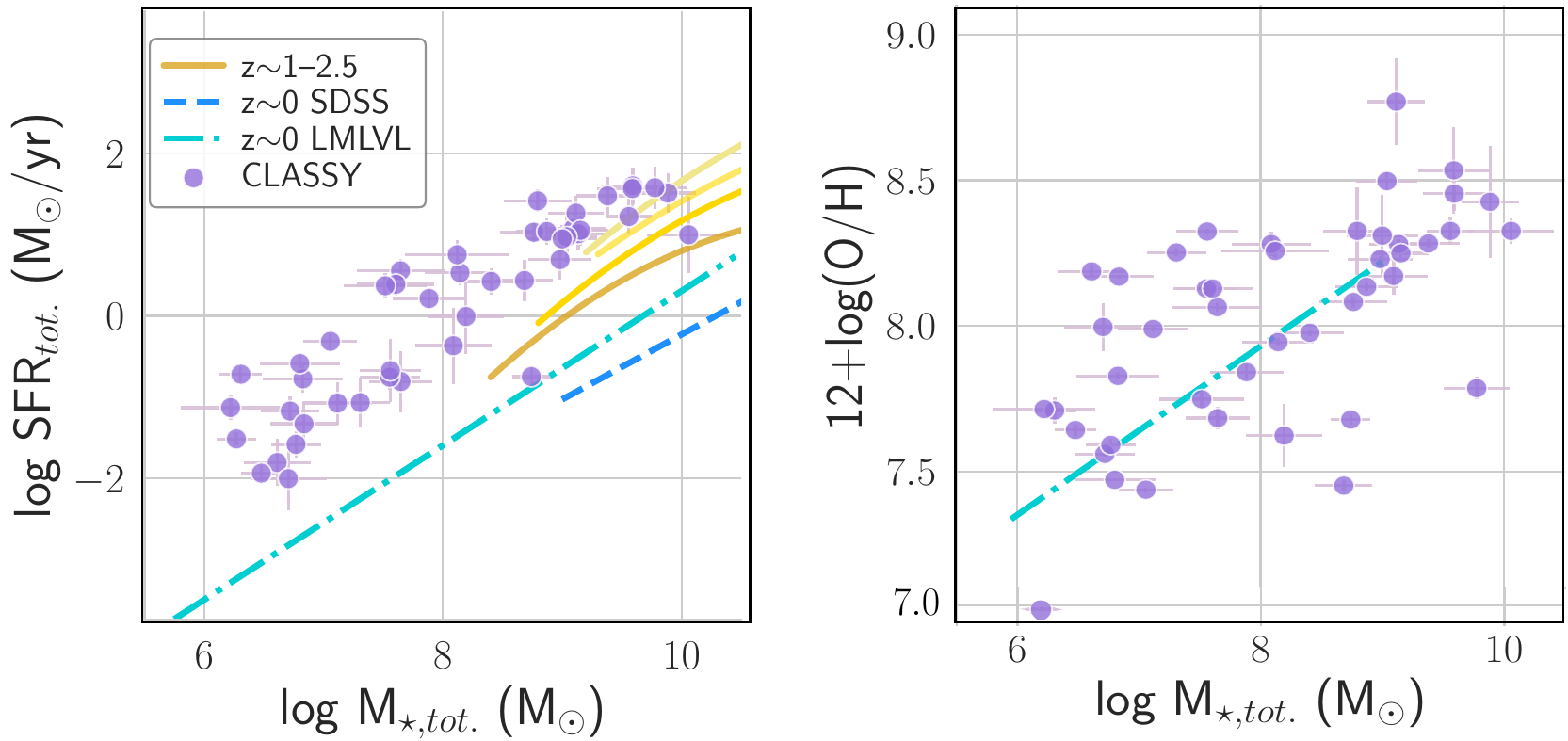} 
\end{center}
\vspace{-2ex}
\caption{
The CLASSY sample covers broad ranges in typical galaxy parameters, 
as measured from the optical spectra and UV$+$optical photometry 
(see Section~\ref{sec:4}).
Property trends are plotted for the total star formation rate versus total stellar mass (SFMR; {\it left})
and gas-phase oxygen abundance versus total stellar mass, or the mass-metallicity relationship (MZR; {\it right}).
For comparison, we plot the $z\sim0$ trends of 
\citet[][SDSS: blue dashed line]{chang15} and 
\citet[][LMLVL: green dot-dashed line]{berg12}, 
as well as the $z\sim1-2.5$ SFMR trends from \citet[][yellow lines]{whitaker14}, 
for the SFMR, and the \citet[][LMLVL; green dot-dashed line]{berg12} 
direct-metallicity trend for the MZR. 
Clearly the CLASSY MZR is typical of local low-mass star-forming galaxies,
but the sample is offset to higher SFRs that are more typical of galaxies
at cosmic noon. 
\label{fig1}}
\end{figure*}

\bigskip
In order to assemble an atlas with unprecedented S/N and spectral resolution
that efficiently utilizes the archive, we chose a sample of nearby, UV-bright galaxies, 
with existing high S/N spectra in at least one of the high-spectral-resolution gratings 
on {\it HST}/COS.
Note that we selected spectra from a single instrument to ensure a consistent aperture. 
We also highlight that COS was chosen over STIS due to its higher sensitivity in the FUV. 
Existing archival spectra were first examined using the HSLA, 
from which 101 nearby ($z<0.2$) galaxies were selected based on their 
high-S/N ($\gtrsim7$ per 100 km s$^{-1}$ resolution element) COS spectroscopy in at 
least one medium-resolution grating (G130M, G160M, or G185M). 
The following selection criteria were then implemented to ensure the efficient 
completion of high-quality, comprehensive rest-frame FUV spectra for a large, diverse 
sample of star-forming galaxies:
\setlist{nolistsep}
\begin{enumerate}[noitemsep,topsep=4pt,partopsep=4pt,leftmargin=15pt]
    \item Star-forming galaxies: 
    Any targets with secondary classifications of quasi-stellar object (QSO) or Seyfert 
    in the HSLA were removed. 
    Remaining targets were visually confirmed to not have obvious broad QSO emission 
    features. \looseness=-2
    \item Low redshift: 
    Redshifts from existing optical spectra of $z<0.2$ were required to capitalize on 
    the FUV sensitivity of COS, which is significantly higher than the NUV 
    detectors (COS and STIS), but falls off rapidly redward of 1950 \AA. \looseness=-2
    \item Compact: 
    Through a visual inspection of the existing {\it HST}/COS NUV acquisition images,
    we selected galaxies with single dominant emission regions and relatively compact 
    morphologies in the sense that the Gaussian distribution of their NUV light had a 
    full-width half-maximum (GFWHM) $<$2\farcs5.
    This ensured that the COS 2\farcs5 aperture (optimized for point sources) 
    will capture most of the FUV light from the galaxy. 
    Note, however, that this is a flawed criterion, as COS UV acquisition images are 
    vignetted and thus the optical sizes of the CLASSY sample are significantly
    larger for a few galaxies. \looseness=-2
    \item UV bright: 
    We required {\it Galaxy Evolution Explorer} \citep[{\it GALEX}; GR6;][]{bianchi14} 
    FUV magnitudes of $<20$ mag to allow for an efficient observing strategy,
    resulting in $< 3$ orbits per grating setting, on average. \looseness=-2
    \item Grating efficiency: 
    From the HSLA, we found three galaxies that had observations of sufficient quality in all 
    of the proposed gratings (J0337-0502, J0942+3547, and J0934+5514). 
    Additionally, 32 galaxies had observations of sufficient quality in two 
    of the proposed gratings, allowing their rest-frame FUV coverage to be completed with the 
    addition of a single grating. 
    Of these, 16 galaxies had existing G130M$+$G160M observations and so CLASSY completed 
    their FUV spectra with the addition of G185M and/or G225M observations, 
    while the remaining 16 had G160M$+$G185M observations and were completed with the addition of 
    G130M observations.
    Among the galaxies with only single-grating observations (G130M or G160M), 
    we preferentially chose the five with existing high-quality G160M observations 
    because new G130M observations are less time intensive due to the gratings' higher 
    sensitivity. \looseness=-2
    \item Broad parameter space coverage: 
    Measurements from existing optical spectra were used to select broad and efficient 
    coverage of galaxy properties to ensure the sample is suitable for comparison 
    to star forming galaxies near and far ($z<7$): star formation rates (SFRs), 
    specific SFRs (sSFRs), stellar masses, nebular metallicities, and ionization parameter. 
    The ranges of stellar mass, SFR, and metallicity are shown in Figure~\ref{fig1}.
    To better sample the higher nebular densities observed for $z>1$ galaxies 
    \citep[e.g.,][]{shirazi14,sanders16,kaasinen17,harshan20} and directly study 
    the effect of electron density on rest-FUV galaxy properties, we mined the
    Sloan Digital Sky Survey Data Release 12\footnote{\url{http://www.sdss.org/dr12/}}
    \citep[SDSS-III DR12;][]{eisenstein11,alam15} and existing literature in order to unearth 
    additional targets with [\ion{S}{2}] \W\W6717,6731 densities of $n_e > 400$ cm$^{-3}$.
    Although such targets are extremely rare at $z\sim0$, we were able to compile a sample of 
    10 galaxies in our so-called "high-density sample" that met criteria 1--4.
    Of these, five galaxies were already in our sample, while the other five had no previous 
    \HST/COS observations, and so required new G130M$+$G160M$+$G185M observation. \looseness=-2
\end{enumerate}

As a result of the search just described, the final CLASSY sample contains 
45 nearby ($z < 0.2$), UV-bright ($m_{FUV} <21$ AB arcsec$^{-2}$), relatively 
compact (GFWHM$_{NUV} < $2\farcs5), galaxies. 
Note, however, that COS Instrument Handbook defines compact objects as 
those having a GFWHM$<$0\farcs6; in this sense, the CLASSY sample includes 
a distribution of compact to extended morphologies. 
For comparison, we determined the optical size of each galaxy by measuring 
the average half light radius (r$_{50}$) from $u$ and $g-$band SDSS, DES, and PanSTARRS 
imaging (listed in Table~\ref{tbl6}); the optical sizes are 0\farcs22 larger 
on average than the UV.

The \HST/COS NUV acquisition images are shown in comparison to the optical 
imaging in Figure~\ref{fig2}, demonstrating that the peak of the optical 
and UV surface brightness profiles are aligned and that most of the stellar 
light is captured within the COS 2\farcs5 aperture.
While the optical imaging shown is extended due to the
mixture of nebular emission (some with EW(H$\alpha) > 10^3$ \AA) and 
light from older (relative to the UV) stellar populations, 
the similar 2\farcs5 COS aperture and 3\arcsec\ SDSS aperture will 
capture comparable nebular emission in their spectra.
This allows the FUV and optical spectra to be used in concordance to 
characterize the stellar and nebular properties of the CLASSY galaxies. 
Note, however, that four of the CLASSY galaxies 
(J0036-3333, J0337-0502, J0405-3648, and J934+5514)
have additional knots of star-formation clearly outside of the aperture. 

Additionally, the CLASSY sample covers a range of galaxy properties, listed in 
Tables~\ref{tbl5} and \ref{tbl6} in the Appendix (see Section~\ref{sec:4} for details 
of property calculations), including 
stellar masses of $6.22 < $ log $M_\star (M_\odot$) $< 10.06$, 
star formation rates of $-2.01 < $ log SFR ($M_\odot$ yr$^{-1}$) $< 1.60$, 
direct nebular metallicities of $6.98 < $12+log(O/H) $< 8.77$, 
electron densities of  $10 < n_e$ (cm$^{-3}$) $< 1120$, 
ionizations, characterized by $O_{32} = $ [\ion{O}{3}] \W5007/[\ion{O}{2}] \W3727, of
$0.54 < O_{32} < 38.0$, 
and reddening values of $0.001 <$ E(B$-$V) $< 0.673$.
This broad sampling of parameter space allows the CLASSY sample to uniquely serve as 
templates for interpreting galaxies across all redshifts.
However, it is important to note that CLASSY sample is biased, on average, 
to more extreme $O_{32}$ values (25th and 75th percentiles of 7.70 and 8.25),
low stellar masses (25th and 75th percentiles of 7.06 and 9.04),
higher SFRs (25th and 75th percentiles of -0.78 and 1.03),
and compact UV morphologies.

The range of the total stellar masses, SFRs, and oxygen abundances 
are shown in Figure~\ref{fig1}.
In comparison to other $z\sim0$ \citep[e.g., the Low-Mass Local Volume Legacy: LMLVL,][]{berg12}
and $z\sim2$ \citep[e.g.,][]{whitaker14} surveys, 
the CLASSY sample follows the expected mass-metallicity trend of local galaxies  
(left panel of Figure~\ref{fig1}), 
but is offset to higher SFRs that are more typical of galaxies at cosmic noon
\citep[e.g.,][right panel of Figure~\ref{fig1}]{madau14}. 


\begin{figure*}
  \centering
  \includegraphics[width=0.95\textwidth,trim=0mm 0mm 0mm 0mm,clip]{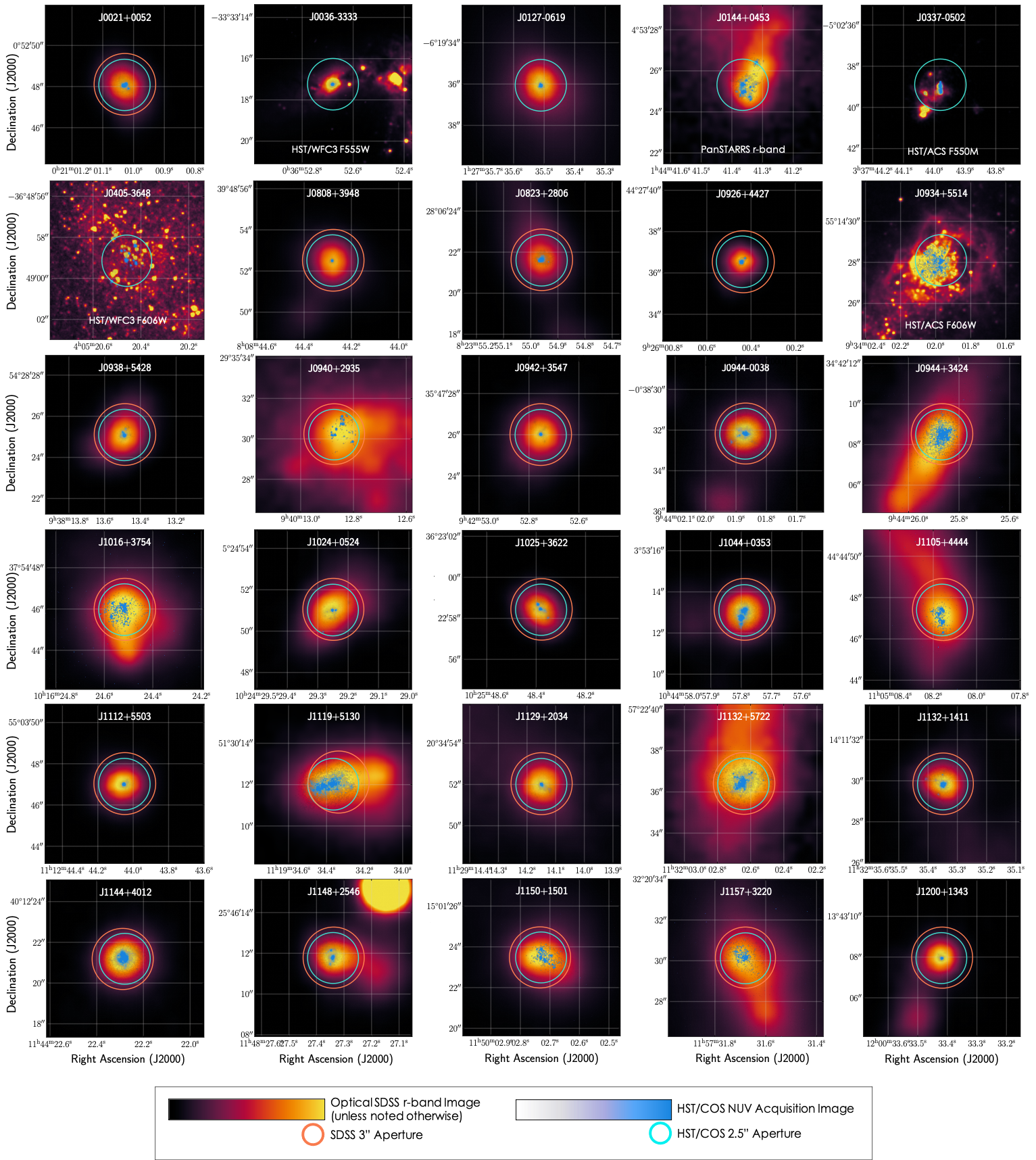}
  \caption{The {\it HST}/COS NUV acquisition images of the 45 galaxies in the CLASSY 
  sample are shown in blue, overlaid on top of optical SDSS {\it r}-band imaging 
  (unless otherwise noted; red-yellow color bar).
  The 2.5\arcsec\ COS aperture used for the UV CLASSY spectra is shown as a blue circle 
  and is very similar to the SDSS 3\arcsec\ aperture in orange.
  Galaxies missing SDSS optical spectra do not have a 3\arcsec\ orange circle.
  Relative to the {\it HST}/COS aperture, most of the NUV light is captured, but
  a significant fraction of the extended optical light, and subsequently nebular emission 
  from strong emitters (e.g., with EW(H$\alpha \gtrsim 10^3$ \AA), may still be missed. 
\emph{(Continued on next page.)}
 \label{fig2}}
\end{figure*}

\renewcommand{\thefigure}{2}
\begin{figure*}
\centering
  \includegraphics[width=0.95\textwidth,trim=0mm 0mm 0mm 0mm,clip]{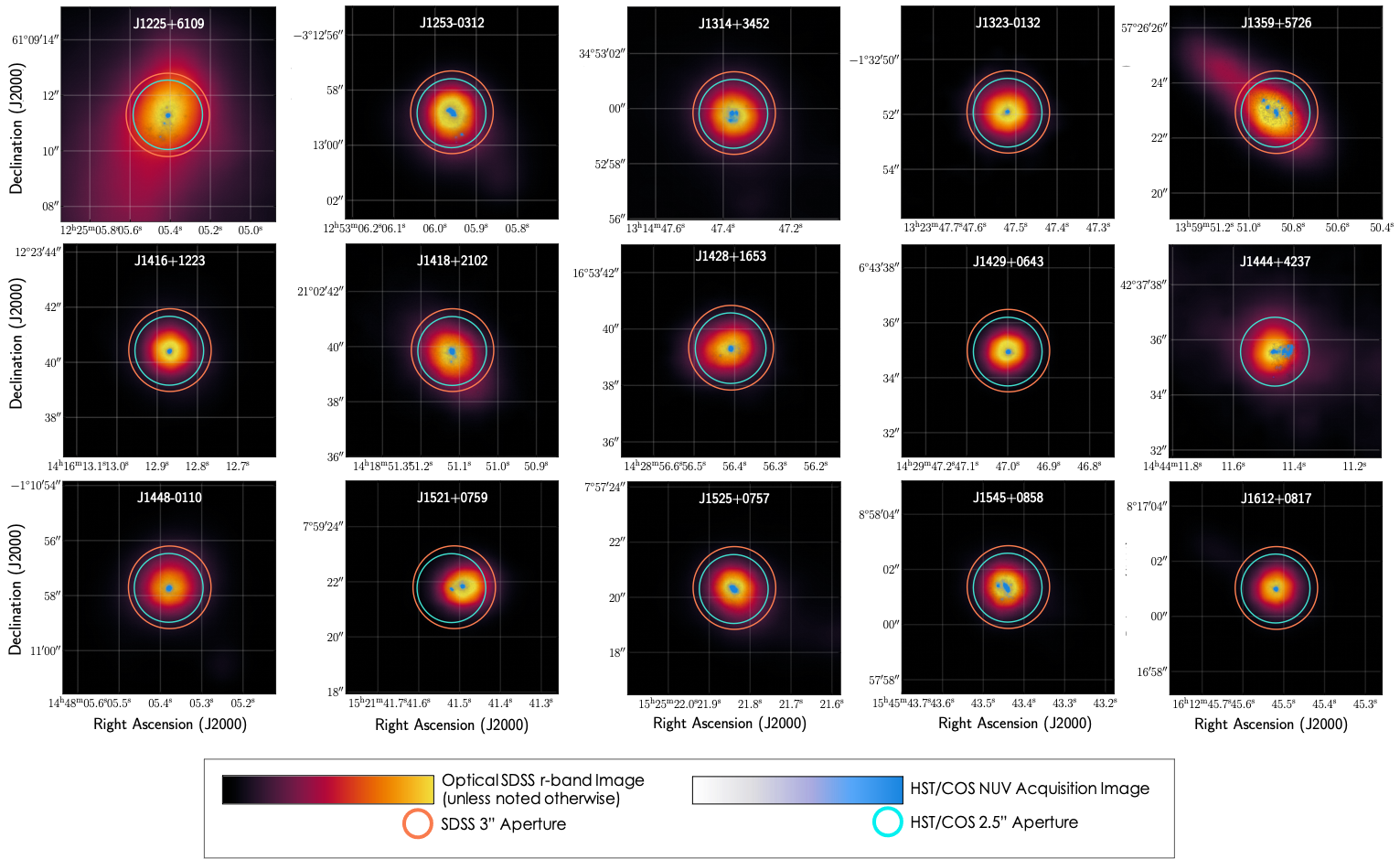}
  \caption{(\textit{continued}): 
  {\it HST}/COS NUV acquisition images of the 45 galaxies in the CLASSY sample.}
\end{figure*}


\subsection{\HST/COS FUV Spectroscopy}\label{sec:2.2}
\textit{Archival Data:}
The CLASSY Treasury program was designed to efficiently capitalize on the significant 
amount of existing FUV COS data in the HSLA.
As a result, the CLASSY spectra includes 96 archival datasets\footnote{
An \HST/COS dataset refers to a single \texttt{x1dsum} fits file, 
which is the combined 1-D extracted spectra for multiple exposures with the same grating, 
central wavelength, and aperture combining all FP-POS within a given visit.\looseness=-2} 
compiled from 177 HST orbits.
Of these archival datasets, 
3 galaxies have full G130M$+$G160M$+$G185M coverage (J0337-0502, J0934+5514, and J0942+3547),
17 galaxies have G130M (21 datasets) and G160M (22 datasets) observations,
16 galaxies have G160M (19 datasets) and G185M (16 datasets) observations, and 
5 galaxies have G160M (6 datasets) observations.
Additionally, six of the CLASSY galaxies have low resolution G140L spectra
(J1016+3754, J1044+0353, J1119+5130, J1323-0132, J1359+5726, and J1418+2102),
which were used for checking the relative flux calibration between medium-resolution 
gratings and filling in the continuum in the gaps between gratings and grating segments. 

\textit{New Data:}
In order to complete the FUV spectral coverage of the CLASSY sample,
the archival observations were supplemented with 76 new \HST/COS spectroscopic 
datasets of 42 galaxies, compiled over 135 orbits in Cycle 27 as part of PID 15840.
These data consist of 27 G130M datasets, 
5 G160M datasets, 29 G185M datasets, and 8 G225M datasets.
Note that the G225M datasets were necessary to ensure coverage of the 
\ion{C}{3}] \W\W1907,1909 emission lines for the 8 galaxies with $z>0.11$.

Details of all individual datasets are provided in Table~\ref{tbl2}.
For each CLASSY galaxy, the archival and CLASSY observations are denoted by their 
program ID (PID), followed by their respective acquisition image and FUV spectral 
observation details.\looseness=-2

\subsubsection{General Observing Procedure}

\noindent{\it Target Acquisition:} 
The entire CLASSY sample has precise and accurate target coordinates that 
enabled consistent pointings between archival and new data with a precision of $<0.1\arcsec$.
For the 37 targets with previous \HST/COS observations,
the archival data coordinates were used to acquire new data.
For the five CLASSY galaxies selected to target high [\ion{S}{2}] electron densities
($n_e > 300$ \cmcu),
no previous \HST/COS observations exist.
Instead, coordinates from the SDSS were used for four of these high-density galaxies 
(J0944+3442, J1200+1343, J1253-0312, J1323-0132) and coordinates for J0127-0619 
were adopted from the \HST\ Faint Object Spectrograph observations of \citet{thuan96}.
The overall excellent positional alignment of the SDSS and COS apertures is shown
in Figure~\ref{fig2}.

Target acquisitions were performed using the \texttt{ACQ/IMAGE} mode with the 
\texttt{PSA} aperture and \texttt{MIRRORA} when possible (87 datasets)
for the COS/NUV configuration. 
The resulting NUV images are useful for understanding the spatial structure of light entering
COS aperture.
However, \texttt{MIRRORB}, was employed out of caution for five galaxies 
(J1016+3754, J1105+4444, J1150+1501, J1157+3220, J1359+5726)
that are especially bright in the UV.
While the use of \texttt{MIRRORB} serves to protect the detector, it has the drawback 
of producing a double peaked image that impedes our interpretation of the spatial 
distribution of light\footnote{\url{https://hst-docs.stsci.edu/cosihb/chapter-8-target-acquisitions/8-4-acq-image-acquisition-mode}}.
The CLASSY NUV acquisition images are shown in Figure~\ref{fig2},
where targets observed with \texttt{MIRRORB} are easily recognizable by 
their more pixelated distribution of blue NUV light and their additional 
'ghost' peak along the spectral axis (e.g., J1016+3754; see, also, 
Table~\ref{tbl2} and \citetalias{james22}).
Considering the sample as a whole, the peaks of the UV light distributions of the CLASSY 
galaxies are clearly well aligned with the positioning of the COS 2\farcs5 aperture (blue circle).
As a result, the COS aperture captures the majority of the UV emission of the CLASSY galaxies.
In comparison to the GALEX fluxes of the CLASSY sample, we estimate that on average
more than 60\% of the total UV flux is captured by the COS aperture. 

\smallskip
\noindent{\it Spectral Observations:} 
The \HST/COS FUV observations used in the CLASSY survey were all taken in the 
\texttt{TIME-TAG} mode using the 2\farcs5 \texttt{PSA} aperture.
Note that no restrictions were placed on the position angle of the new CLASSY datasets;
the implications of this decision will be discussed in \citetalias{james22}.
The \texttt{FP-POS}=ALL setting was used for most of the datasets to take four spectral images 
offset from one another in the dispersion direction, increasing the cumulative S/N and 
mitigating the effects of fixed pattern noise. 
\texttt{TIME-TAG} mode was employed, allowing abnormal detection events to be filtered out,
eliminating cosmic rays in the process.
For the G130M grating with the \texttt{CENWAVE} set to 1291, only
\texttt{FP-POS} = 3 and 4 were available for the new observations
(see the COS2025 guidelines\footnote{\url{https://www.stsci.edu/hst/instrumentation/cos/proposing/cos2025-policies}} 
for further information). 
Each target was observed for the time predicted to reach a S/N $\gtrsim5$ per resolution element 
in the continuum for the G130M and G160M gratings.
The input continuum flux was measured at 1500 \AA\ from, in priority order,
existing COS spectra, GALEX fluxes, or an extrapolated model fit to existing optical data.
For the G185M and G225M gratings, observation times were determined to detect the \ion{C}{3}] \W\W1907,1909
emission lines with a strength of S/N $\gtrsim3$, based on observed C/O values for a given
metallicity \citep[see, e.g.,][]{berg19b}, the observed [\ion{O}{3}] \W5007 flux, and the 
theoretical emissivities. 
All raw data for the new and archival observations were retrieved from the {\it HST} archive and 
were reduced with the CALCOS pipeline (v3.3.10) using the standard 
\texttt{twozone} extraction technique.
In \citetalias{james22} we provide a detailed discussion on the optimized extraction techniques
employed for the extended sources within the CLASSY sample.


\subsubsection{Coaddition Process}\label{sec:2.2.1}
\noindent{\it Overview:}
Below we describe the main steps in the coaddition process, however,
the discussion of the other intermediate analyses and their impact on the reduced spectra
are discussed in \citetalias{james22}.

\bigskip
The primary HLSPs of the CLASSY treasury survey are high S/N multi-grating coadded spectra.
However, each \HST/COS grating has a different spectral resolution that must be accounted 
for when combining data from multiple gratings.
Additionally, while the CLASSY galaxies have relatively compact stellar clusters, 
their nebular emission generally fills the COS aperture (at minimum, with diffuse light;
see Figure~\ref{fig2}), and, consequently, experience varying degrees of vignetting 
and degradation in spectral resolution depending on the shape of their overall UV light 
profile entering the COS aperture.

To address the challenges associated with coadding multiple datasets that consist of
different grating configurations, exposure times, and resolutions, we developed a custom 
\texttt{python} reduction code (available as a \texttt{jupyter notebook}
on the CLASSY HLSP website).
The main steps followed in the code are depicted in Figure~\ref{fig:coadd} and include:
(1) joining the segments/stripes of individual grating datasets,
(2) coadding any multiples of individual grating datasets, including all cenwave 
configurations, 
(3) coadding datasets across gratings,
(4) binning the spectra, and
(5) correcting for Galactic contamination. 
Below we describe each step in more detail.

{\it 1. Joining grating segments and stripes:} 
The \HST/COS CLASSY spectra were acquired across multiple FUV detector segments 
and NUV detector stripes,
so the first step of the coaddition process unifies these components
for each dataset. 
For the G130M, G160M, and G140L gratings, 
the COS FUV detector has two segments (\texttt{FUVA}, \texttt{FUVB})\footnote{
Note, however, that the G140L grating with the 1105 cenwave configuration
only utilizes the \texttt{FUVA} segment.}, 
while the NUV detector used for the G185M and G225M gratings 
has three stripes (\texttt{FUVA}, \texttt{FUVB}, \texttt{FUVC}).
For each dataset, we extracted the \texttt{WAVELENGTH}, \texttt{FLUX}, \texttt{ERROR}, and 
\texttt{DQ\_WGT} (or data quality flag weights) columns for each segment or stripe
from the \texttt{x1dsum} files produced by the CALCOS pipeline and appended them.
Because the dispersion varies with wavelength,
we then used the \texttt{pysynphot} package in \texttt{python} \citep{lim13} to define a 
source spectrum and interpolate onto a new wavelength
grid defined by the largest step in the original wavelength array.
Note that there are wavelength gaps between each set of segments/stripes, as well as
gainsag holes resulting from detector damage due to extensive airglow emission exposure, 
so the full spectrum produced for each galaxy is contiguous rather than continuous.

{\it 2. Coadding single grating datasets:} 
If multiple datasets of a given grating exist for an individual target, 
such as observations acquired over multiple visits
or targets observed at multiple central wavelength configurations (\texttt{CENWAVE}s),
they must be combined in step two of Figure~\ref{fig:coadd} before proceeding.
Such collections of grating datasets exist for the G130M observations of one galaxy
(J0934+5514 / I Zw 18), the G160M observations of five galaxies, and the G185M 
observations of eight galaxies, or ten CLASSY galaxies in total.
These grating datasets were then coadded using a combined normalized data quality weight 
(using the \texttt{DQ\_WGT} array; to filter out or de-weight photons correlated with 
anomalies/bad data) 
and exposure-time-weighted 
calibration curve (to preserve the Poisson count statistics).
This weighting method was used for all instances where coadding was performed.

{\it 3. Coadding multiple grating datasets:} 
Accounting for the change in spectral dispersion between individual gratings, and 
datasets, is a particularly difficult task. 
Not only does dispersion change as a function of wavelength across the detector, 
it also changes as a function of lifetime position on the COS FUV detector
(the CLASSY data span all 4 lifetime positions thus far), 
and are degraded by extended CLASSY sources depending on their specific UV light profiles 
and aperture position angles.
To take this into account, at each stage of the co-adding process,
we performed a common resampling of the wavelengths to the highest dispersion 
of a given combination of gratings.
The different spectral dispersion of the coadds allow CLASSY to be used for many 
scientific studies that require a range of spectral resolutions.
Specifically, we produce four spectral coadds for each galaxy with different resolutions: 
\setlist{nolistsep}
\begin{enumerate}[noitemsep,topsep=4pt,partopsep=4pt,leftmargin=15pt]
    \item The Very High Resolution (VHR) Coadds: consisting of only G130M spectra, 
    which have a nominal point source resolution of 9.97 m\AA/pixel or 0.060 \AA/resel.
    This is the highest spectral resolution of the CLASSY gratings.
    \item The High Resolution (HR) Coadds: consisting of the CLASSY medium-resolution 
    FUV gratings, or G130M$+$G160M spectra, with a nominal point source resolution of 
    12.23 m\AA/pixel or 0.073 \AA/resel.
    \item The Moderate Resolution (MR) Coadds: consisting of the CLASSY FUV$+$NUV medium 
    resolution gratings, or G130M$+$G160M$+$G185M$+$G225M spectra, with a nominal point
    source resolution of 33 m\AA/pixel or 0.200 \AA/resel.
    \item The Low Resolution (LR) Coadds: consisting of the CLASSY medium and low 
    resolution gratings. Possible grating combinations include G130M$+$G160M$+$G140L and G130M$+$G160M$+$G185M$+$G225M$+$G140L, with a nominal point source resolution of 
    80.3 m\AA/pixel or 0.498 \AA/resel.
\end{enumerate}
Note that the flux calibration was performed for each dataset during the initial reduction by
calcos, however, the relative fluxing between gratings was also carefully considered during 
the coadding process. 
In short, the G160M spectrum was treated as the flux anchor of each spectrum and the continuum 
of all other datasets were fit and scaled to G160M prior to coadding.
See \citetalias{james22} for a detailed discussion of the flux calibration and its effects
on the final coadded CLASSY spectra.

{\it 4. Binning the spectra:} 
In order to gain signal-to-noise, the coadded spectra were also binned in the dispersion 
direction.
The COS medium-resolution gratings have a resolving power of $R\sim15,000$ for a perfect 
point source, which corresponds to six FUV detector pixels in the dispersion direction. 
All coadds were binned by a factor of six to reflect this
using the \texttt{SpectRes} \citep{carnall17} \texttt{python} function, 
which efficiently resamples spectra and their associated uncertainties onto an arbitrary 
wavelength grid while preserving flux.

{\it 5. Correcting for Galactic Contamination:} 
Finally, proper interpretation of UV spectral properties requires the removal of 
potential contamination from Geocoronal emission and Milky Way interstellar absorption.
It is important to note that this contamination affects the absorption spectra of the entire CLASSY sample, 
whether directly by deforming intrinsic features or 
indirectly by degrading adjacent regions of good continuum.
However, the high-ionization Milky Way absorption features (i.e., \ion{Si}{4} and \ion{C}{4})
are typically weak \citep{savage00} relative to the intrinsic profiles, and so can
often be ignored. 
To correct the CLASSY spectral templates, we create mask arrays assuming central 
velocities of $v_{cen}=0$ km s$^{-1}$ and conservative widths of $\Delta v = 500$ 
km s$^{-1}$ for the very-high, high, and moderate resolution coadds and 
$\Delta v = 1000$ km s$^{-1}$ for the low resolution coadds.

The final high-quality CLASSY coadded spectra are shown in the Appendix in Figure~\ref{fig12}. 
These coadds provide both high resolution spectra over the $\sim1200-1700$ \AA\ range of 
each object that is important to studies of stellar and ISM absorption features and 
contiguous FUV $\sim1200-2000$ \AA\ wavelength coverage at low resolution that is 
necessary to simultaneously interpret the stars and gas. 
However, due to the varying extended nature of the CLASSY sample, the observed spectral 
resolutions differ considerably amongst the four coadditions for different galaxies. 
Specifically, the extended source distribution causes the nominal point source line
spread function to broaden in complex, unpredictable ways
(see \citetalias{james22} for further discussion).
Therefore, to estimate observed spectral resolution of each galaxy,
we measure the Gaussian FWHM of the best (i.e., cleanest and highest S/N) Milky Way ISM 
absorption feature present.
Note that this measurement assumes unsaturated, unresolved lines. 
Table~\ref{tbl3} lists the nominal and measured spectral resolutions of the CLASSY coadded spectra;
41 galaxies had Milky Way ISM features that were well fit by a Gaussian, while the remaining
4 lacked good Milky Way features to fit.
While the choice of coadd dataset / spectral resolution is left to the user,
Table~\ref{tbl3} suggests that the moderate resolution coadds are the most appropriate spectra
for most of the CLASSY sample. 


\subsection{Optical Spectroscopy}\label{sec:2.3}
High-quality optical spectra have been compiled for the entire CLASSY sample
to ensure uniform determinations of galaxy properties. 
Initially, much of the sample was selected from the SDSS, and so archival SDSS spectra\footnote{Note that all CLASSY galaxies have 3\farcs0 aperture SDSS spectra; none were obtained with the 2\farcs0 BOSS spectrograph.}
exist for 38 of the CLASSY galaxies.
The exceptions are J0036-3333, J0127-0619, J0144+0453, J0337-0502, J0405-3648, J0934+5514, and J1444+4237.
Instead, we use 
VLT/VIMOS integral field unit (IFU) from \citet{james09} for J0127-0619,
MMT Blue Channel Spectrograph spectra from \citet{senchyna19a} for J0144+0453 and J1444+4237, 
Keck/KCWI IFU spectra from \citet{rickardsvaught21} for J0337-0502, 
and VLT/MUSE IFU spectra for the remaining four galaxies. 

To ensure robust direct $T_e$ measurements for J0808+3948, J0944+3442, and J1545+0858,
very high-S/N optical spectral were obtained using the Multi-Object Double Spectrographs 
\citep[MODS,]{pogge10} on the Large Binocular Telescope \citep[LBT,][]{hill10}. 
Simultaneous blue and red MODS spectra were obtained using the G400L (400 lines mm$^{-1}$, $R\approx1850$) 
and G670L (250 lines mm$^{-1}$, $R\approx2300$) gratings, respectively. 
Observations used the 1\arcsec$\times$60\arcsec\ long slit for 3$\times$900s exposures, 
or 45 min of total exposure per object. 
The slits were centered on the same coordinates as the COS apertures. 
Spectra were reduced, extracted, and analyzed using the beta version of the MODS reduction 
pipeline\footnote{\url{http://www.astronomy.ohio-state.edu/MODS/Software/modsIDL/}} which runs within 
the XIDL\footnote{http://www.ucolick.org/~xavier/IDL/} reduction package. 
One-dimensional spectra were corrected for atmospheric extinction and flux-calibrated based on 
observations of flux standard stars \citet{oke90}. 
Further details of the MODS reduction pipeline can be found in \citet{berg15}.

Additional optical spectra we gathered from archival VLT/MUSE IFU spectra and 
Keck/ESI spectra from \citet{sanders21}, and were prioritized over the SDSS spectra due to 
their superior S/N and spectral resolution.
For galaxies with IFU spectra, we extract a spectrum using a 2\farcs5 aperture to better match 
the COS aperture.
The details of the ancillary optical spectra adopted for the CLASSY sample
are listed in Table~\ref{tbl4}.

\renewcommand{\figurename}{Fig.}

\renewcommand{\thefigure}{3}
\begin{figure*}
\begin{center}
    \includegraphics[width=0.975\textwidth]{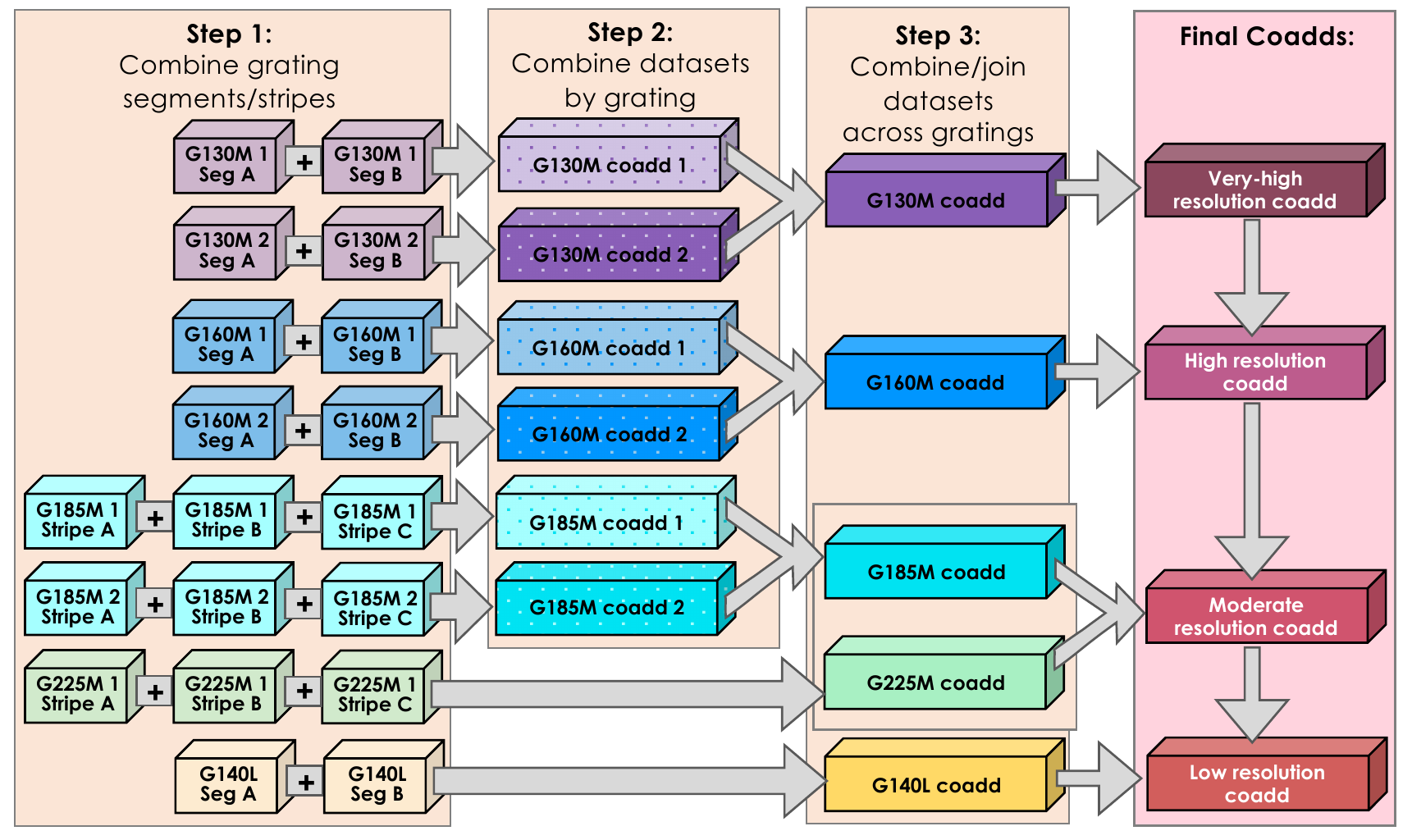} 
\end{center}
\caption{The main steps in the CLASSY coaddition process.
The CLASSY spectra are coadded from a combination of G130M, G160M, and one or more
of the G185M, G225M, and G140L \HST/COS gratings.
The first step highlights how the segments of the FUV datasets and the stripes of
the NUV datasets are resampled to a common wavelength grid.
Then, in step two, any sets of multiple grating datasets for a given galaxy are
coadded using the weighting method described in Section~\ref{sec:2.3}.
Finally, to provide the optimal combination of wavelength coverage and spectral 
resolution for different science cases, the gratings for a given galaxy are 
coadded into four spectra with different resolutions.}
\label{fig:coadd}
\end{figure*}


\section{Coadded Spectra Data Release}\label{sec:3}

\noindent{\it Overview:}
The result of the coaddition process described in Section~\ref{sec:2.3} is the first CLASSY HLSP:
a set of high-quality, moderately high-resolution, uniformly calibrated spectral templates with 
contiguous FUV coverage.
The CLASSY atlas spectra are stored as multi-extension fits (MEF) files containing
up to 20 total extensions of spectra,
with four possible spectral resolutions and five extensions per resolution set
corresponding to combinations of Galactic reddening and redshift corrections.
Any use of the CLASSY spectral atlas should cite this work (\citetalias{berg22}) and 
\citetalias{james22}.
Below we further describe the coadded spectra HLSPs and how to access them.
\bigskip

\subsection{Coadded Spectra File format}\label{sec:3.1}
The CLASSY coadded spectra HLSP consists of 45 MEF files, or one for each CLASSY galaxy.
For a given galaxy, its MEF file can have up to 21 extensions (0--20),
where the first extension (0) is the primary header containing general details relevant
to the spectrum and the remaining 20 extensions contain spectral data tables.
Information important to the spectra of a given galaxy are stored in the primary 
fits header keywords.
As shown in Figure~\ref{fig:coadd_struct}, there are two groups of keywords that 
describe either the target or the observations used to produce the coadded spectra.

The CLASSY coadded spectra are stored in many different forms to maximize
convenience for the user depending on their scientific needs.
Specifically, for each of the four potential coadd resolutions of a given galaxy, 
there are a set of five extensions corresponding to:
\setlist{nolistsep}
\begin{itemize}[noitemsep,topsep=4pt,partopsep=4pt,leftmargin=15pt]
    \item[] {\bf Ext.~1}: The observed-frame coadded spectrum.
    \item[] {\bf Ext.~2}: Ext.~1 $+$ $E(B-V)_{\rm{gal}}$ corrected: observed-frame coadd, 
    corrected for foreground Galactic extinction.
    \item[] {\bf Ext.~3}: Ext.~2 $+$ $z$ corrected: rest-frame coadd, corrected for foreground Galactic extinction.
    \item[] {\bf Ext.~4}: Ext.~2 $+$ binned: observed-frame coadd, corrected for foreground Galactic extinction, and binned by 6 native COS pixels.
    \item[] {\bf Ext.~5}: Ext.~3 $+$ binned: rest-frame coadd, corrected for foreground Galactic extinction, and binned by 6 native COS pixels.
\end{itemize}
Extensions 2 and 5 are corrected to the observed frame using the redshifts reported in 
Table~\ref{tbl2}, which were adopted from the SDSS, when available.
For galaxies not observed by the SDSS, redshifts were sourced from another optical spectrum
(see adopted references in Table~\ref{tbl4}).
Extensions 4 and 5 contain coadded and binned coadded spectra that were corrected for 
foreground Galactic extinction.
We used the \texttt{Python} \texttt{dustmaps} \citep{green18} interface to query the 
Bayestar 3D dust maps of \citet{green15} to determine the total foreground Galactic reddening along the 
line of sight of the coordinates of each CLASSY galaxy.
We note that the \citet{green15} map was adopted over more recent versions
due to its more optimal coverage of the CLASSY sample.
Specifically, the \citet{green15} map provides E(B--V) values for all but two
galaxies (J0405-3648 and J0036-3333) in the CLASSY sample,
whereas the most recent \citet{green19} map only covers 16 of the 45 galaxies.
Fortunately, for these 16 galaxies covered by both maps, the average difference in E(B-V)
is only 0.030 magnitudes, with 25th and 75th percentiles of 0.012 and 0.040 magnitudes, respectively.
For J0405-3648 and J0036-3333, the galactic E(B--V) values were taken from \citet{schlafly11}.
The foreground Galactic extinction correction was then applied using the \citet{cardelli89} reddening law.

As shown in Figure~\ref{fig:coadd_struct}, these five extensions form a set of
coadded spectra that are repeated for each resolution that exists for a given galaxy.
All 45 CLASSY galaxies have VHR and HR coadds, while 42 have MR coadds.
One of the three galaxies missing an MR coadd, J1112+5503, is missing G185M 
observations owing to an observational failure that was not made up as it was 
the final CLASSY observation.
The other two galaxies missing MR coadds, J1044+0353 and J1418+2102, already had significant 
\sfCTh\ \W\W1907,1909 detections in their archival G140L spectra that deemed follow-up
G185M spectra unnecessary. 
Including these three galaxies, six of the CLASSY galaxies have G140L spectra and subsequent LR coadds. 
In summary, 4 CLASSY galaxies have 20 spectral extensions (VHR, HR, MR, and LR coadds) and 
the remaining 41 have 15 extensions (VHR, HR, and MR coadds). 

Each extension in the CLASSY coadded spectra HLSPs is a simple data table with 5 columns.
The columns correspond to wavelength (\AA), flux density, error, S/N ratio, and a mask,
where the fluxes and errors are given in units of $10^{-15}$ erg \s\ \cmsq\ \AA$^{-1}$.
The spectral mask in the 5th column of each extension can be used to remove regions 
contaminated by Geocoronal emission from Ly$\alpha$ and \ion{O}{1} \W\W1301,1305,1306
and Milky Way interstellar absorption features.
For most of the CLASSY sample, the redshift is large enough to cleanly separate the
Geocoronal and Milky Way features from extragalactic features of interest.
However, for galaxies with $z\lesssim0.003$, and depending on the extent of the stellar 
wind and outflow features in a given galaxy, users should examine features for contamination.
Finally, each fits file has a table detailing the individual observations 
used in the coadds as the last extension. 
A detailed demonstration of the final CLASSY atlas spectra is shown for an
absorption-line galaxy, J0036-3333, and an extreme emission-line galaxy, J1323-0132, 
in Figures~\ref{fig5} and \ref{fig6}, respectively.
These moderate-resolution coadded spectra highlight the diversity of 
spectral features observed in the CLASSY sample.


\renewcommand{\thefigure}{4}
\begin{figure*}
\begin{center}
    \includegraphics[width=0.975\textwidth]{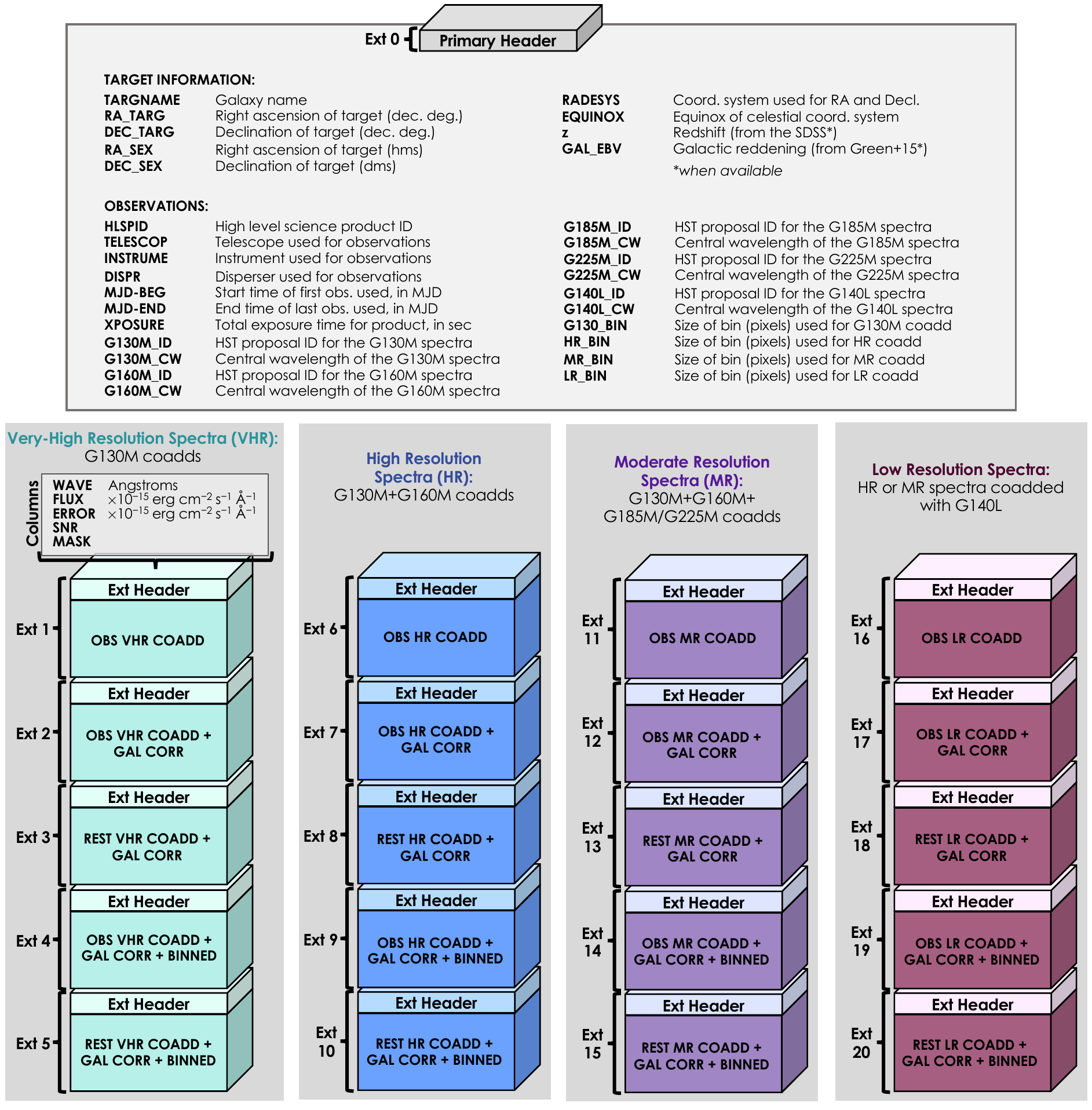} 
\end{center}
\vspace{-2ex}
\caption{
Example multi-extension fits (MEFs) file structure for the CLASSY coadd HLSPs.
The top shows the informative keywords included in the primary header (extension 0),
while the bottom shows the five extensions of spectra formats for each of the four 
resolution coadds.
The extensions can be accessed by extension number or name, as given in each block. 
Following the HLSP file naming convention, the CLASSY coadded spectra naming template is
\texttt{hlsp\_classy\_hst\_cos\_{\it<target>}\_scaled\_multi\_v1\_coadded.fits}.}
\label{fig:coadd_struct}
\end{figure*}


\renewcommand{\thefigure}{5}
\begin{figure*}
  \centering
  \includegraphics[width=0.95\textwidth,trim=0mm 20mm 0mm 0mm,clip]{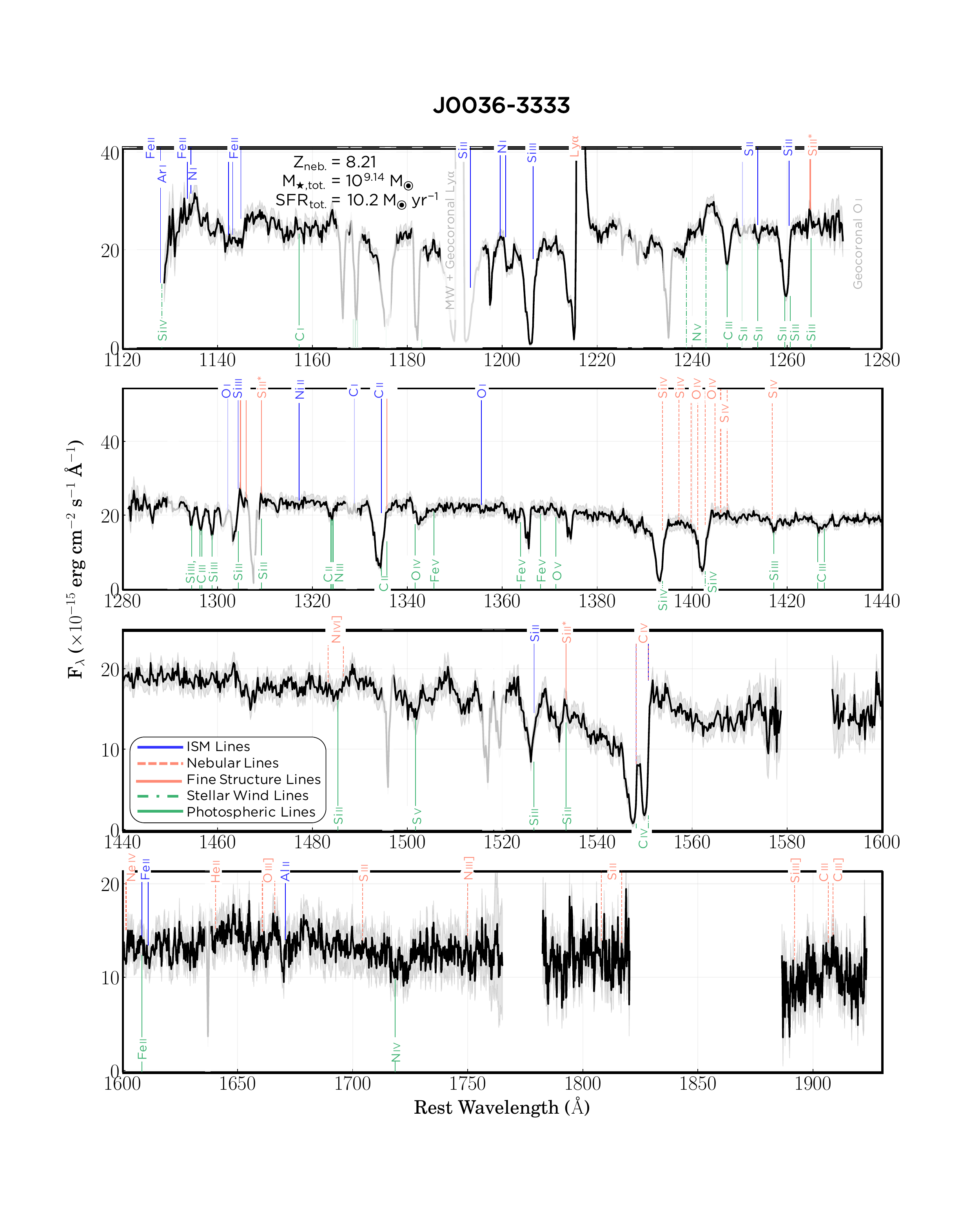}
  \caption{
  This moderate-resolution coadded CLASSY spectrum for J0036-3333 (black line) 
  is rich with stellar wind features, such as the \ion{Si}{4} and \ion{C}{4} features, 
  and interstellar medium absorption lines, including the strong \ion{Si}{2} and \ion{C}{2} features. 
  J0036-3333, however, contains no significant UV emission features besides \LYA. 
  Additionally, the flux$\pm$error spectrum, shown in gray, demonstrates the high S/N of CLASSY 
  and the ability to resolve absorption in J0036-3333 from contamination by Geocoronal and 
  Milky Way features (regions where the spectrum/black line isn't shown).}
  \label{fig5}
\end{figure*}

\renewcommand{\thefigure}{6}
\begin{figure*}
  \centering
  \includegraphics[width=0.95\textwidth,trim=0mm 0mm 0mm 0mm,clip]{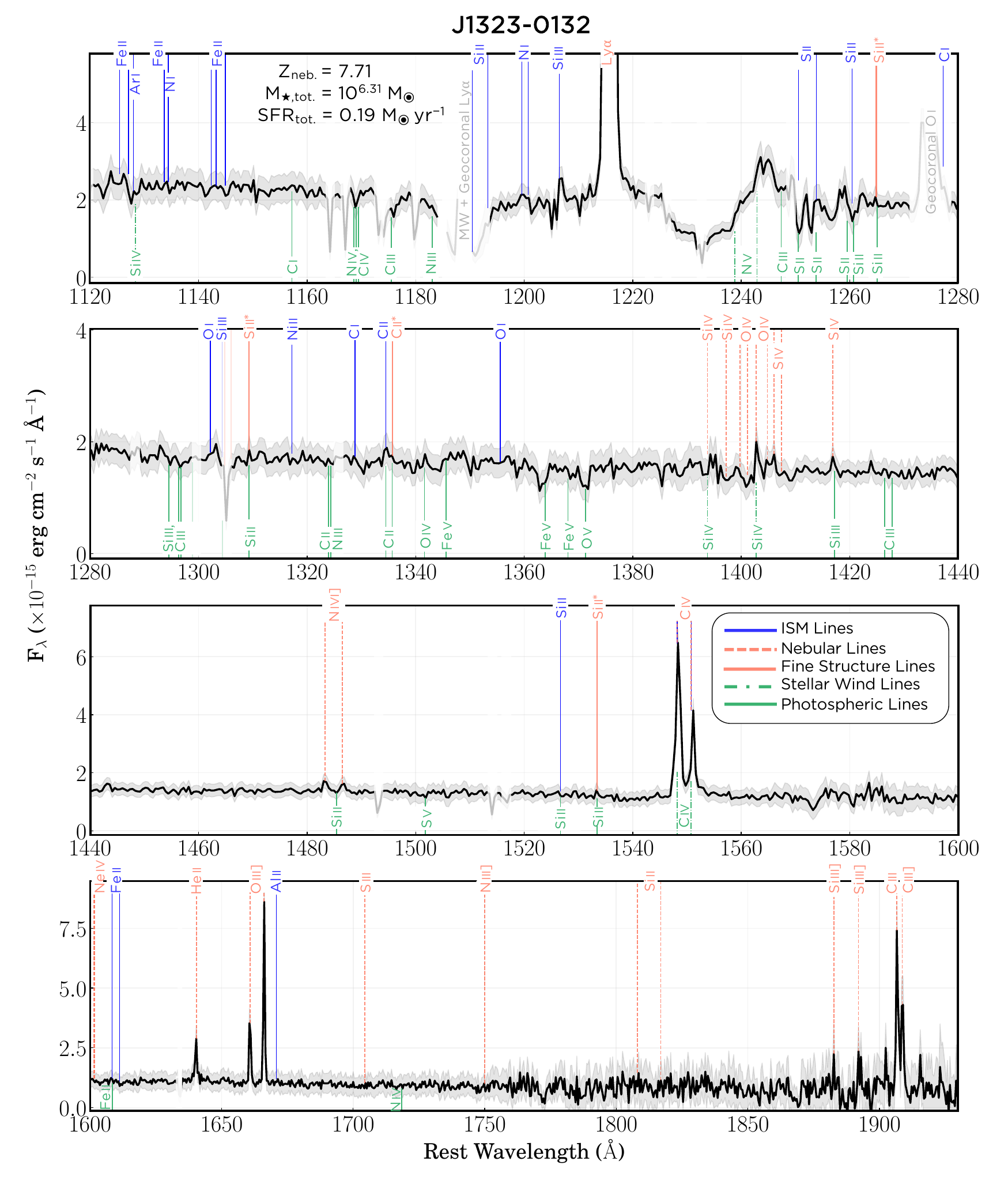}
  \caption{
  This moderate-resolution coadded CLASSY spectrum for J1323-0132 (black line) 
  is strikingly different than that of J0036-3333 in Figure~\ref{fig5}, highlighting some 
  of the emission lines that are observed at varying strengths for a subset of the CLASSY sample.  
  The spectrum has particularly strong emission features from   \LYA, \aCFo, \aHeTw, \sfOTh, 
  and \sfCTh, additional weak detections from the high-ionization \aOFo, \aSFo, and \sfNFo\ lines, 
  but minimal absorption features.
  Note the high S/N of this spectrum is shown by the gray flux$\pm$error spectrum, 
  while the spectrum/black line isn’t shown for regions contaminated by Geocoronal and Milky Way features.}
  \label{fig6}
\end{figure*}


\subsection{HLSP Digital Access}\label{sec:3.2}
All CLASSY Treasury HLSPs will be released to the astronomical community in digital
form as they are completed.
The publication of this paper marks the release of the first HLSP, the 
CLASSY FUV spectral atlas of star-forming galaxies, on two main platforms. 

First, all CLASSY HLSPs will be available via the CLASSY HLSP
home page: \url{https://archive.stsci.edu/hlsp/classy}, which also
serves as the main website for the CLASSY Treasury and relevant information.
Importantly, this website provides simple tarball downloads of individual CLASSY HLSPs,
as well as the entire CLASSY HLSP data collection.
In addition to full data access, the CLASSY HLSP home page provides 
machine-readable table downloads for the tables of properties in this paper
(Tables~\ref{tbl2}, \ref{tbl5}, and \ref{tbl6}), 
interactive spectra that allow quick-look examination of the coadded spectral features,
user-friendly \texttt{jupyter} notebook guides to accessing and using the 
CLASSY HLSPs, and important links such as to the NASA Astrophysics Database System (ADS) 
library of CLASSY publications.

The second platform for accessing the CLASSY HLPSs will be the Mikulski Archive 
for Space Telescopes (MAST) via the CLASSY MAST website: 
\url{https://mast.stsci.edu/search/ui/#/classy}
(expected early 2022).
The CLASSY MAST portal will provide a unique search-and-selection method to access 
the CLASSY HLSPs, where users can select a subsample of the data products
based on HLSP type, filename, or target properties such as coordinates
or stellar mass. 
This search tool will continue to be developed based on the specifics of each
HLSP release in order to allow more efficient and straightforward access to
data products that are of interest to the user. 
In this manner, appropriate subsets of the CLASSY HLSPs can easily be selected 
for a given scientific analysis, to match specific survey sample properties,
and more.


\subsection{Comparison to Previous FUV Spectroscopic Surveys}\label{sec:3.3}

The CLASSY atlas provides spectral templates with the powerful combination of
contiguous 1200--2000 \AA\ FUV wavelength coverage,
high resolution (\W/$\Delta$\W$=15,000$), and
high S/N (average S/N = 6.4 per resolution element).
These templates simultaneously probe the interactions between stellar winds,
galaxy-scale outflows, and nebular emission and provide a unique dataset that probe
spatial scales of 0.1--10.7 kpc 
across the COS aperture at the range of redshifts of the CLASSY sample.
As we discussed in Section~\ref{sec:1.2}, this combination of qualities of the
CLASSY survey is unique with respect to previous FUV spectra surveys. 

Interestingly, few CLASSY galaxies are in previous FUV spectral 
atlases, making it difficult to directly compare them.
Fortunately, IUE and CLASSY observations both exist for two galaxies: J0337-0502 and J0934+5514.
For a visual comparison, the HST and IUE spectra for J0337-0502 are plotted in 
Figure~\ref{fig7}, showing substantial, but expected differences. 
The smaller aperture of COS relative to the IUE (10\arcsec$\times$20\arcsec aperture) 
records lower UV continuum fluxes (however, this would not be expected
for most of the NUV-compact CLASSY sample)
and some features are present in only the IUE spectrum, which also contains the SE
cluster region.
For example, there appear to be weak emission features present just blueward of 
1400 \AA\ and at 1750 \AA, likely \ion{N}{3}], that are emitted by the SE region
but not the NW region.
Perhaps the most important difference, though, is the significant number of features
and details revealed by the higher resolution of the CLASSY spectra ($\sim0.1$ \AA\ 
resolution compared to the $\sim6$ \AA\ of the IUE spectrum).
Specifically, a number of significant ISM absorption lines and
nebular emission lines are recovered,
such as the \ion{O}{1} and \ion{Si}{2} lines shown in the left-most inset window
of Figure~\ref{fig7} and the \ion{C}{4}, \ion{He}{2}, \ion{O}{3}], and
\ion{C}{3}] lines shown in the other three inset windows. 

Figure~\ref{fig7} clearly demonstrates that the CLASSY survey provides a 
dramatic leap forward in FUV spectral atlases and is an invaluable tool for the 
interpretation and modeling of star-forming galaxies across redshifts.



\section{CLASSY Galaxy Properties}\label{sec:4}


\noindent{\it Overview:}
The CLASSY sample of star forming galaxies was carefully chosen to overcome
the common limitations of other surveys introduced by the biases of their samples.
Specifically, while CLASSY only targets relatively FUV-bright galaxies 
(GALEX $m_{\rm{FUV}}<20$ and $F_{1500}\gtrsim1\times10^{-15}$ erg \s\ \cmsq),
which also ensures the presence of young stellar populations, it probes a broad parameter
space in stellar mass, star formation rate, gas-phase abundance, ionization parameter,
and gas-phase electron density in a nearly uniform manner.

Below we describe the measurements of CLASSY nebular quantities and their resulting trends. 
All physical property calculations used the emission line ratios measured from the optical 
spectra listed in Table~\ref{tbl4} and the \texttt{PYNEB} package in \texttt{PYTHON} 
\citep{luridiana12,luridiana15} with the atomic data adopted in \citet{berg19b}.
\bigskip


\subsection{Contiguous FUV$+$ Optical Spectra}\label{sec:4.1}
A benefit of the CLASSY sample is the availability of ample ancillary data, 
including excellent optical spectra for the entire sample.
Optical spectra have long been the gold standard source for calculating important 
nebular properties, including the oxygen abundance, density, and reddening tabulated 
in Table~\ref{tbl1}.
As discussed in Section~\ref{sec:2.3}, the majority of the CLASSY sample was also observed 
as part of the SDSS, which provides full optical spectra using a similar aperture (3\arcsec) 
to that of the COS UV spectra.
Additionally, 8 of the CLASSY galaxies have IFU spectroscopy from 
either VLT/MUSE, VLT/VIMOS, or Keck/KCWI, while 9 of the CLASSY galaxies have
LBT/MODS or Keck/ESI longslit spectra (see Table~\ref{tbl4}).
We take advantage of these high-S/N spectra in order to calculate direct $T_e$ abundances 
when the SDSS spectra had insufficient signal-to-noise. 

There are obvious differences in the aperture sizes between the longslit optical spectra 
and the circular COS aperture, yet, these differences do not result in significant changes
in the nebular properties calculated.
A full analysis of the aperture effects on the nebular properties of CLASSY galaxies with 
multiple sources of optical spectra will be discussed in a forthcoming paper by 
\citet{arellano-cordova22}.
However, even when the UV and optical spectra have similar apertures, one significant 
difference is the vignetting of the COS aperture for the UV spectra.
Fortunately, for the current purpose of deriving galaxy properties, only relative flux ratios are needed,
and so the absolute flux calibration is irrelevant. 


\subsection{Emission Line Measurements}\label{sec:4.2}
Several of the galaxy properties of interest are best measured from the
photoionized nebular emission lines present in the adopted optical spectra of CLASSY galaxies.
Here in \citetalias{berg22} we focus on measuring the redshift ($z$), intrinsic 
dust reddening ($E(B-V)$), nebular electron density ($n_e$) and temperature ($T_e$), 
and gas-phase metallicity (12$+$log(O/H)) that characterize the CLASSY sample.
More detailed discussions of the optical and UV line measurements, their 
analysis, and their resulting HLSPs (see HLSPs 4, 6, 7, 8, and 9 in Section~\ref{sec:5})
will be presented in forthcoming papers by \citet{arellano-cordova22} and \citet{mingozzi22}.
Below we briefly describe the CLASSY optical emission line measurements.

For the purposes of calculating the nebular properties of the CLASSY sample,
we measured the strength of as many of the optical emission lines necessary to determine:
\setlist{nolistsep}
\begin{itemize}[noitemsep,topsep=4pt,partopsep=4pt,leftmargin=15pt]
    \item \ebv: H\D\ \W4101, H\G\ \W4340, H\B\ \W4861, H\A\ \W6563
    \item \Ne: \fSTw\ \W\W6717,6731
    \item \Te: \fOTh\ \W4363 $+$ \W\W4959,5007,
    \begin{itemize}[noitemsep,topsep=-4pt,partopsep=4pt,leftmargin=15pt]
            \item[] \fSTh\ \W6312 $+$ \W\W9069,9532,
            \item[] \fNTw\ \W5755 $+$ \W\W6548,6584, or
            \item[] \fOTw\ \W\W3727,3729 $+$ \W\W7320,7330
        \end{itemize}
    \item \LOH: required lines listed above.
\end{itemize}
For all of the optical spectra, regardless of telescope and instrument used,
we first modeled the stellar continuum using the 
\texttt{Starlight}\footnote{\url{www.starlight.ufsc.br}} spectral synthesis code
\citep{fernandes05} using stellar models from \citet{bruzual03}. 
This step is important for removing any absorption component present in the Balmer 
emission lines.
Emission lines were then fit in the continuum-subtracted spectrum with Gaussian 
profiles using \texttt{MPFIT} \citep{markwardt09}.
The fit parameters (i.e., velocity width, line center) of neighboring lines were 
tied together, 
allowing weak or blended features to be measured simultaneously. 
The uncertainty on the flux measurements is assumed to be the uncertainty on the
fit returned from the Monte Carlo minimization code.
The complete database of measured emission line fluxes will be released in a 
forthcoming HLSP via \citet{arellano-cordova22} and \citet{mingozzi22}.

We note that the dominant emission component is a narrow, nebular component for the
optical emission lines measured for the CLASSY sample.
However, broad emission can occur in the presence of stellar winds, outflows, and shocks, and is 
often clearly visible at the base of the strongest emission features. 
Because the broad components only contribute a small fraction to the total line flux,
we restrict the present analysis to the narrow components only and reserve the 
analysis of multi-component fits for the forthcoming work of \citet{mingozzi22}.


\renewcommand{\thefigure}{7}
\begin{figure*}
\begin{center}
    \includegraphics[width=1.0\textwidth]{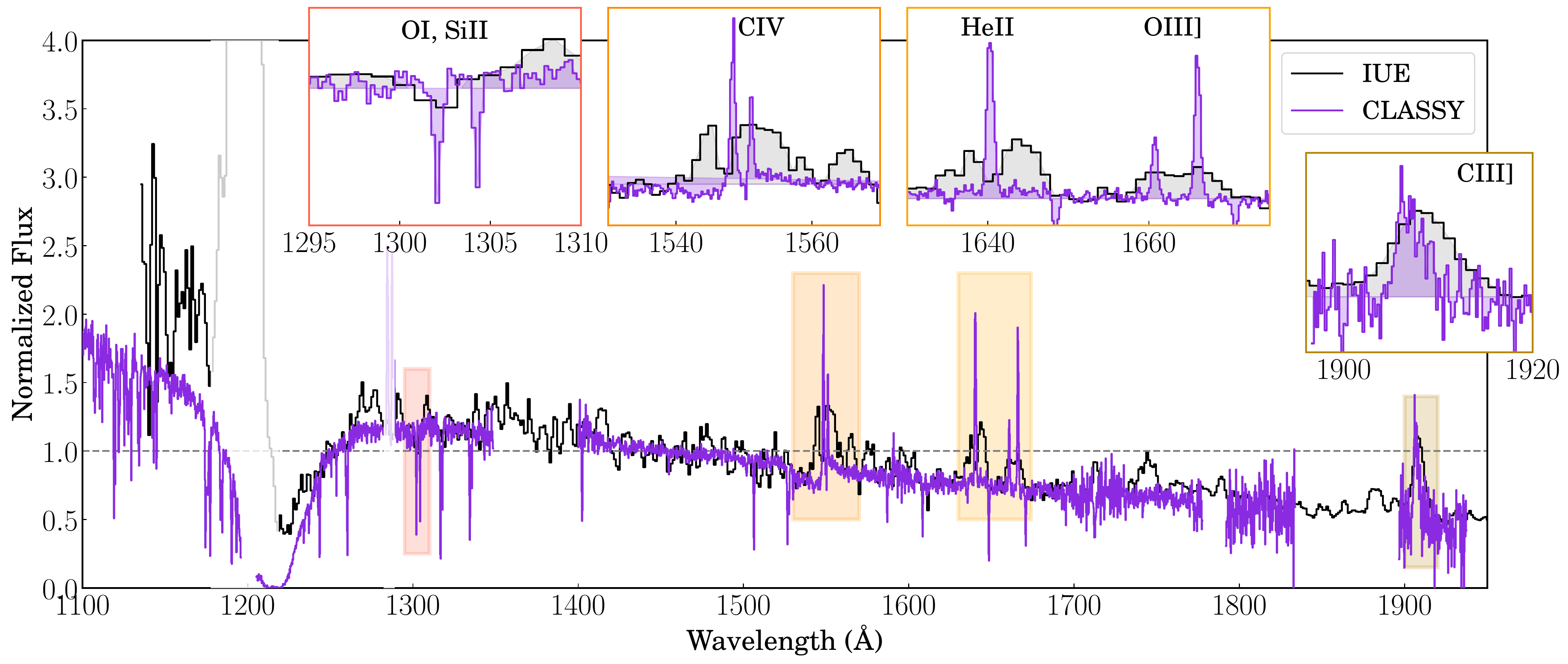}
\end{center}
\caption{
Comparison of rest-frame spectra for J0337-0502 to demonstrate
the improvement of CLASSY over previous FUV spectral atlases.
The IUE spectrum is relatively low resolution (\W/$\Delta$\W$\sim300$)
and consists of the integrated light within its large 10\arcsec$\times$20\arcsec\ aperture.
In contrast, the CLASSY coadded spectral templates are characterized by an unrivaled combination of
contiguous 1200--2000 \AA\ FUV wavelength coverage, high spectral resolution 
(\W/$\Delta$\W$=15,000$), and high S/N (average S/N = 6.4 per resel) for much more
compact regions that can be targeted with the 2\farcs5 COS aperture.
Both spectra are normalized at 1450 \AA.
Note that the breaks in the CLASSY spectrum are due to chip gaps and that the Geocoronal
\LYA\ and \ion{O}{1} emission have been whited-out.
While few features are apparent in the IUE J0337-0502 spectra, 
the CLASSY spectrum reveals a significant number of absorption and emission features
that enable simultaneous studies of the massive stellar population, ISM outflows,
and nebular emission.}
\label{fig7}
\end{figure*}


\subsection{Reddening Corrections}\label{sec:4.3}
Following the same method discussed in Section~\ref{sec:3.1} for the UV spectra,
flux measurements for the optical emission lines were first corrected 
for Galactic extinction using the \citet{cardelli89} reddening law with 
values queried from the \citet{green15} extinction map using the 
\texttt{PYTHON DUSTMAPS} interface \citep{green18}.
Subsequently, the emission lines were corrected for their corresponding 
intrinsic galaxy dust reddening.
The dust reddening values, $E(B-V)$, were determined using
the relative intensities of the strongest Balmer lines available
in a given spectrum (i.e., H$\alpha$/H$\beta$, H$\gamma$/H$\beta$, H$\delta$/H$\beta$) 
with an iterative application of the \citet{cardelli89} reddening law,
parameterized by an $A_v = 3.1\times E(B-V)$.
Note that \citet{wild11a} found that the nebular attenuation curve has a slope similar 
to the MW extinction curve, rather than that of the SMC or the \citet{calzetti00} curve, 
and so the \citet{cardelli89} extinction law is appropriate for correcting 
the CLASSY emission line fluxes.
The iterative approach allows for a careful consideration of deviations
due to electron temperature and density.
The initial reddening seed was determined from the Balmer line ratios using theoretical values from
\texttt{\sc{pyneb}}, assuming a starting temperature of $T_e = 10^4$ K and $n_e = 10^2$ cm$^{-3}$ for
Case B recombination.
In turn, this reddening value is then used to determine an initial estimate of 
the electron density and temperature (using the best measured auroral line 
available in a given spectrum).
This process is then repeated iteratively to modify the $E(B-V)$, $T_e$, and $n_e$
values until the electron temperature changes by less than 20 K between iterations. 
The final reddening estimate is an error-weighted average of the 
H$\alpha$/H$\beta$, H$\gamma$/H$\beta$, and H$\delta$/H$\beta$ reddening values
and was used to reddening-correct the optical emission line intensities.


\subsection{Electron Densities and Temperatures}\label{sec:4.4}
Detailed understanding of the nebular properties of star forming galaxies requires
knowledge of the electron temperature, $T_e$, and density $n_e$ structure in a galaxy
compiled from the values in each ionization zone.
Such a detailed examination of the physical properties across a nebula and comparison
of different $T_e$ and $n_e$ measurements will be the focus of a later work.
Here in \PI, we seek only to measure a characteristic $T_e$ and $n_e$ value for each 
CLASSY galaxy.
To do so, a standard 3-zone ionization model is adopted 
\citep[for a thorough discussion, see][]{berg21} and the high-ionization zone 
(characterized by O$^{++}$) is assumed to dominate in the young \ion{H}{2}
regions of the CLASSY galaxies.

The optical spectra compiled for the CLASSY sample are not uniform - the individual 
spectra vary significantly in their S/N, resolution, and wavelength coverage.
While it is desirable to calculate $T_e$ values in the best and most consistent manner 
for the entire sample, no single $T_e$-sensitive auroral line is uniformly detected 
for the CLASSY sample.
As such, direct $T_e$ values were determined using the highest S/N auroral line detection
for each galaxy.
In general, the [\ion{O}{3}] \W4363/\W5007 ratio is prioritized for metal-poor galaxies 
(12$+$log(O/H) $<8.0$) with blue wavelength coverage, while the [\ion{S}{3}] 
\W6312/\W\W9069,9532 ratio is prioritized in more metal-rich galaxies and/or spectra 
with red wavelength coverage, and the [\ion{N}{2}] \W5755/\W6584 ratio is useful for 
lower-ionization and more metal-rich galaxies.
The [\ion{O}{2}] \W\W3727,3729/\W\W7320,7330 ratio  was only used if no other $T_e$ 
measurements were possible, as the red [\ion{O}{2}] lines are sensitive to contamination 
from dielectric recombination at relatively low densities ($n_e > 300$ cm$^{-3}$).
Additionally, the red [\ion{S}{3}] lines are susceptible to contamination from atmospheric 
lines, and so were visually checked and corrected to the theoretical emissivity ratio 
when needed.

For simplicity, a single, uniform density is assumed and derived from the $n_e$-sensitive 
[\ion{S}{2}] \W6717/\W6731 ratio. 
The [\ion{S}{2}] ratios measured for \ion{H}{2} regions are typically consistent with the 
low-density limit ($n_e < 10^2$ cm$^{-3}$).
In this regime, fluctuations on the order of the ratio uncertainty have negligible impact 
on abundance calculations, and thus supports the assumption of a homogeneous density 
distribution expected to be $n_e\sim100$ cm$^{-3}$ throughout a nebula.

The uncertainties on the CLASSY $T_e$ and [\ion{S}{2}] $n_e$ measurements were determined 
using a Monte Carlo simulation:
An array of temperatures/densities were calculated for a 2-dimensional normal distribution 
of values generated with 500 values in each dimension, with centers and widths corresponding 
to the values and one sigma uncertainties, respectively, of the measured emission line flux 
ratio and electron temperature/density. 
The uncertainty was then taken as the standard deviation of the resulting distribution 
calculated.
The ionic temperatures used and calculated densities are reported in Table~\ref{tbl5}
in the Appendix.


\subsection{Metallicities}\label{sec:4.5}
Here we determine total oxygen abundances, 12+log(O/H), for the purpose
of characterizing the gas-phase metal content of the CLASSY galaxies.
The determination of a complete database of UV and optical ionic abundances and 
relative abundances will be reserved for forthcoming papers by \citet{arellano-cordova22}
and \citet{mingozzi22}.

The total oxygen abundance relative to hydrogen is calculated by summing the 
abundances of the individual ionic species ($X^i$) together relative to hydrogen as:
\begin{equation}
    \frac{N(X)}{N(H)} = \sum \frac{N(X^i)}{N(H^+)} = \sum \frac{I_{\lambda(i)}}{I_{H\beta}} \frac{j_{H\beta}}{j_{\lambda(i)}},
\end{equation}
where the emissivity coefficients, $j_{\lambda(i)}$, are determined assuming the characteristic 
$T_e$ and $n_e$ of the corresponding ionization zone. 
To determine the temperature of an ionization zone not directly measured, 
we adopt appropriate temperature relationships based on the $O_{32}$ ratio \citep[][]{yates20}.
For high-ionization ($O_{32} > 1.25$), high-temperature ($T_e > 2\times10^4$ K; low-metallicity) galaxies,
we adopt the theoretical $T_e-T_e$ relationships of \citet{garnett92}, while for
lower-ionization ($O_{32} < 1.25$), lower-temperature ($T_e < 2\times10^4$ K; higher-metallicity) galaxies,
we adopt the empirical $T_e-T_e$ relationships of \citet{berg20}.
To be precise, the total oxygen abundances (O/H) should be calculated from the sum of four 
ionization species that can be present and observed in \ion{H}{2} regions:
\begin{equation}
    \frac{\rm{O}}{\rm{H}} = \frac{\rm{O}^0}{\rm{H}^+} + \frac{\rm{O}^{+}}{\rm{H}^+} + \frac{\rm{O}^{+2}}{\rm{H}^+} + \frac{\rm{O}^{+3}}{\rm{H}^+}
\end{equation}
However, the ionization energy ranges of O$^+$ and O$^{+2}$ span the full range of a 
standard 3-zone ionization model \ion{H}{2} region and contributions from O$^0$ and 
O$^{+3}$ are typically negligible, even in very-high-ionization nebulae \citep{berg21}. 
Therefore, the generalized oxygen abundances for the CLASSY sample calculated here 
sum only the dominant ionic abundances, O$^+$/H$^+$ and O$^{+2}$/H$^+$, determined 
from either the [\ion{O}{2}] \W3727 or [\ion{O}{2}] \W\W7320,7330 lines and the 
[\ion{O}{3}] \W\W4959,5007 lines. 

The resulting oxygen abundances and their uncertainties are tabulated in Table~\ref{tbl5} in the Appendix.
Note that the uncertainties were simply propagated from the emission line measurements in
\S~\ref{sec:4.2}, accounting for errors in the continuum subtraction, flux calibration, 
and dereddening.
It may be appropriate to consider an additional systemic uncertainty of $+0.1-0.2$ dex 
to account for potential depletion onto dust grains (larger at higher metallicities) and
biases due to temperature inhomogeneities (larger at lower metallicities). 


\subsection{Ionization Parameters}\label{sec:4.6}
An important parameter for characterizing the physical nature of the ionized gas
in a galaxy is the ionization parameter, $q$, or the flux of ionizing photons 
(cm$^{-2}$ s$^{-1}$) per volume density of H, $n_H$ (cm$^{-3}$). 
Typically the dimensionless, volume-averaged ionization parameter, log$U$, is used 
and defined as $U = q/c$.
Using the photoionization model prescriptions from Table~3 of \citet{berg19a}, 
log$U$ values for the CLASSY sample were inferred from the relationship with the
observed light-weighted optical [\ion{O}{3}] \W5007/[\ion{O}{2}] \W3727 ratio.
Typical star-forming regions have been found to have average log$U$ values 
between $-3.5$ and $-2.9$ \citep[e.g.,][]{dopita00, moustakas10}. 
In comparison, for the CLASSY sample we measure ionization parameters of 
$-3.1 <$ log$U$ $< -1.6$.
The range of CLASSY ionization parameters extend to much high values than typical
star-forming regions, likely due to their enhanced star formation rates 
(see Figure~\ref{fig1}) and ensuing extreme nebular conditions. 


\subsection{Stellar Masses and Star Formation Rates}\label{sec:4.7}

Stellar mass (\Ms) and star formation rates (SFRs) were determined for the CLASSY sample
by constraining broad-band spectral energy distribution (SED) fitting via the 
BayEsian Analysis of GaLaxy sEds \citep[\texttt{BEAGLE, v0.24.0},][]{chevellard16}
code with UV$+$optical photometry.
Importantly, \texttt{BEAGLE} models both the stellar populations and the nebular emission.
Two sets of stellar mass and SFR parameters were computed assuming a constant star formation history (SFH):
(1) an aperture set focused on the star-forming regions within the COS 2\farcs5 aperture and 
(2) a galaxy set which characterizes entire host galaxies.
Details of the SED fitting process, model assumptions (e.g., a \citet{chabrier03} Galactic initial 
mass function was assumed), and tests are discussed in Appendix~\ref{AppendixC}.

The resulting stellar masses and SFRs are reported in Table~\ref{tbl6} in the Appendix
for both the aperture and total galaxy sets. 
We also measure the time since the onset of constant SFR, which we refer to here as the galaxy age. 
For host galaxies, we find ages of the constant SFH of our sample spanning $< 5$ Myr to 5 Gyr, 
with a 16\%-50\%-84\%\ range of $\log$ Age/yr = $8.18_{-0.75}^{+0.33}$.
According to these results, the youngest galaxy is J1323-0132,
(an extreme emission-line galaxy with EW H$\beta$ = 251 \AA) 
with an age of only $3.2_{-1.7}^{+19.5}$ Myr.
The other corresponding property ranges for CLASSY are 
$6.77 <$ log $M_\star/M_\odot < 9.15$, with a median of log $M_\star/M_\odot$ = 8.12,
$-2.01 <$ log $\rm{SFR}/(M_\odot$ yr$^{-1}) < 1.60$, with a median of 
log $\rm{SFR}/(M_\odot$ yr$^{-1}) = 0.39$, and
$-9.49 <$ log sSFR/Gyr$^{-1}$ $< -7.03$, with a median of 
log sSFR/Gyr$^{-1} = -8.05$.
Typical uncertainties for host galaxy measurements are 
$\sigma_{\log M_\star/M_\odot} = 0.27_{-0.04}^{+0.06}$ and 
$\sigma_{\log\rm{SFR}/(M_\odot \rm{yr}^{-1})} = 0.18_{-0.05}^{+0.13}$.
 
For aperture measurements, we find 
$\log$ Age/yr = $7.84_{-0.86}^{+0.55}$,
$4.91 <$ log $M_\star/M_\odot < 9.74$, with a median of log $M_\star/M_\odot = 7.39$,
$-3.04 <$ log $\rm{SFR}/(M_\odot$ yr$^{-1}) < 1.43$, with a median of 
log $\rm{SFR}/(M_\odot$ yr$^{-1}) = -0.05$, and
$-8.66 <$ log sSFR/Gyr$^{-1}$ $< -7.03$, with a median of 
log sSFR/Gyr$^{-1} = -7.78$.
The typical uncertainties are 
$\sigma_{\log M_\star/M_\odot} = 0.29_{-0.04}^{+0.05}$ and 
$\sigma_{\log\rm{SFR}/(M_\odot \rm{yr}^{-1})} = 0.17_{-0.06}^{+0.08}$,
similar to the values found for the host galaxy measurements.
The fact that aperture measurements lack UV photometry is partly mitigated by the large 
excesses in broad-band flux caused by strong nebular emission from the starburst,
and \texttt{BEAGLES}' ability to account for this emission in its models.
This excess is most evident in aperture flux, while in measuring total flux it is degraded 
by the relatively older stellar population outside of the aperture.
Note that we have adopted a simple constant SFH for simplicity and consistent 
comparison with high-redshift galaxies, however, this method can result in
large systematic uncertainties \citep[see, e.g.,][]{lower20}.
We will further assess the SED parameters relative to the optical spectra derived properties,
as well as explore the effects of different attenuation models, in a future paper.


\section{Global Properties of the CLASSY Galaxies}\label{sec:5}


\renewcommand{\thefigure}{8}
\begin{figure*}
\begin{center}
    \includegraphics[width=0.75\textwidth,trim=0mm 0mm 0mm 0mm,clip]{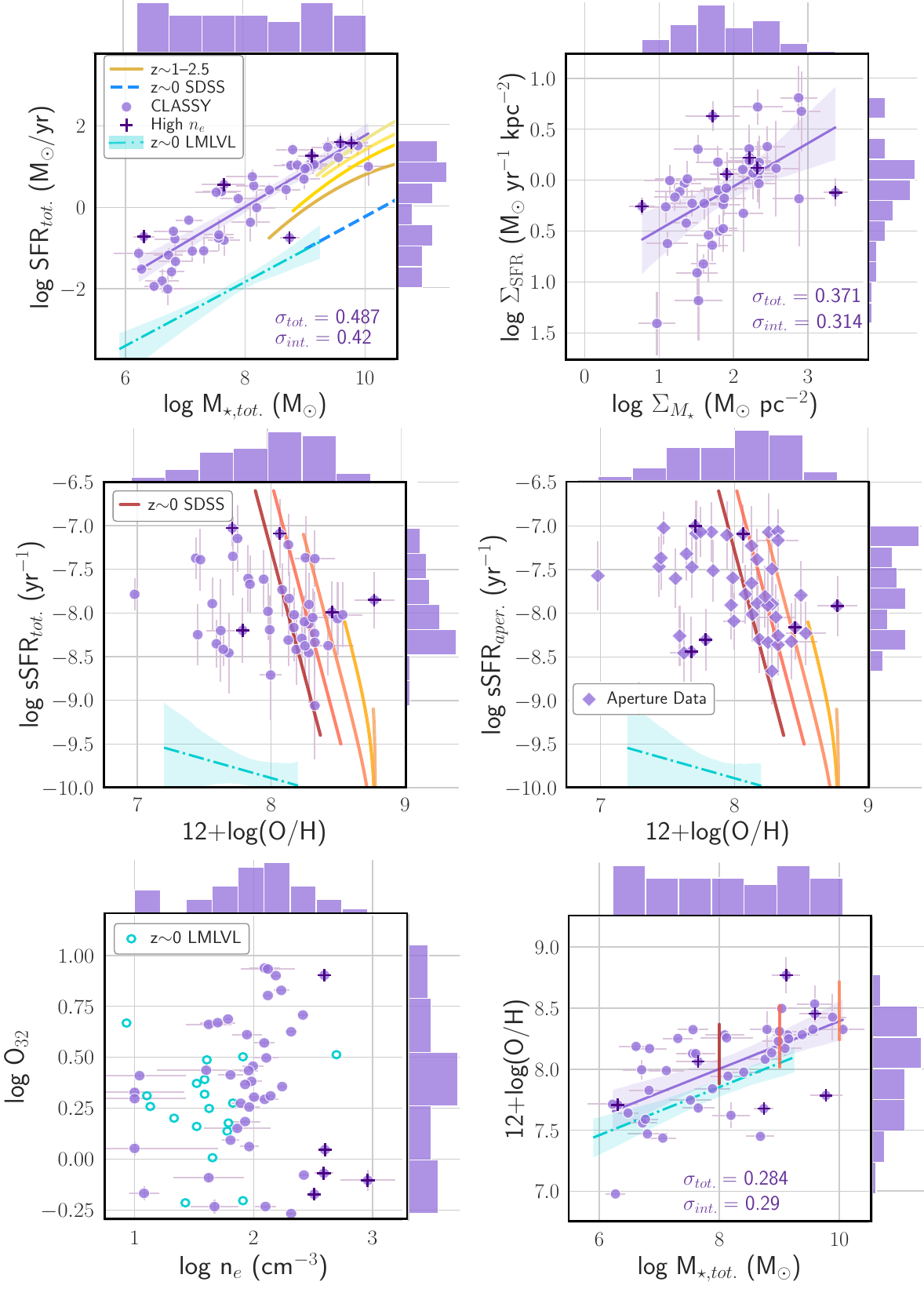} 
\end{center}
\vspace{-2ex}
\caption{
The CLASSY sample properties, as measured from the optical spectra 
and UV$+$optical photometry described in Section~\ref{sec:4},
span a broad range of parameter space.
{\it Top row:} Star formation rate is plotted versus
stellar mass for both the entire galaxy (left) and in terms of surface density (right).
The left plot shows a trend that extends to lower masses and
star formation rates, but lies above the previous galaxy sample studies 
at $z\sim0$ (SDSS: \cite{chang15}; blue dashed line and
LMLVL: \cite{berg12} and \cite{lee09}; cyan dot-dashed lines) and 
$z\sim1-2.5$ \citep[][yellow lines]{whitaker14}.
The total and intrinsic scatter of the observed trend are given as  
$\sigma_{tot.}$ and $\sigma_{int.}$, respectively.
{\it Middle row:} Specific star formation rate is plotted
against gas-phase oxygen abundance, using the total stellar
mass and SFR on the left and the aperture values on the right.
Here, the majority of the CLASSY sample aligns with or extends from the 
\citet[][]{curti20} relationships for $10^8 < M_\star/M_\odot < 10^{11}$ 
(orange to yellow lines, respectively) star-forming SDSS galaxies, 
but lie well above the dwarf galaxies of the LMLVL 
\citep[][]{berg12,lee09}.
{\it Bottom row:} Oxygen ionization ratio is plotted
versus density in the left-hand plot, showing the wide range in nebular 
conditions of the CLASSY sample.
In the bottom right-hand plot, gas-phase oxygen abundance is
plotted relative to total stellar mass, showing the well-known mass-metallicity relationship.
The CLASSY sample seems to follow the same trend as other normal star-forming galaxies
\citep[e.g.,][]{berg12,curti20}, but with larger scatter.
\label{fig8}}
\end{figure*}


By design, the CLASSY sample broadly probes many galaxy parameters.
Specifically, the CLASSY sample spans approximately four orders of magnitude
in both SFR and $M_\star$, roughly two  orders of magnitude
in both sSFR and gas-phase oxygen abundance, one order of magnitude in the O$_{32}$ 
ionization ratio, and two orders of magnitude in gas-phase electron density. 
This broad sampling of parameter space is demonstrated in Figure~\ref{fig8}
by plotting the properties calculated from the optical spectra 
and UV$+$optical photometry described in Section~\ref{sec:4}.

\subsection{The SFR--M${\star}$ Relationship}\label{sec:5.1.1}
The upper row of Figure~\ref{fig8} shows that the SFRs and stellar masses of 
the CLASSY sample form a tight, increasing trend when both the total galaxy properties
(left) and the total galaxy surface density properties (right) are considered. 
To parameterize the CLASSY stellar mass versus SFR relationship,
we used the \texttt{python} \texttt{LINMIX} Bayesian linear regression code.
\texttt{LINMIX} implements a linear mixture model algorithm \citep{kelly07} 
to fit data with uncertainties on two variables, and includes an explicit treatment 
of the intrinsic scatter. 
Using this analysis, we fit the linear relationship and determined the 1$\sigma$ dispersion. 
As a result, we find a relationship for the total galaxy properties of
\begin{equation}
    \mbox{log SFR}_{tot.} = (0.91\pm0.07)\times\mbox{log M}_{\star,tot.} 
    - (7.25\pm0.61),\nonumber
\end{equation}
with a total scatter of $\sigma_{tot.} = 0.49$ dex and an intrinsic scatter of
$\sigma_{int.} = 0.42\pm0.07$.
For the surface density properties, we find\looseness=-2
\begin{equation}
    \mbox{log} \Sigma_{\mbox{SFR}_{tot.}} = (0.44\pm0.13)\times\mbox{log} \Sigma_{\mbox{M}_{\star,tot.}} - (0.92\pm0.25),\nonumber
\end{equation}
with a total scatter of $\sigma_{tot.} = 0.37$ dex and an intrinsic scatter of
$\sigma_{int.} = 0.31\pm0.05$.
Due to the relatively small errors determined for the stellar mass and SFR measurements,
the scatter seems to be largely intrinsic, accounting for roughly 80\% of the total scatter.
However, these uncertainties do not incorporate contributions from the uncertainties in the 
stellar models, initial mass function, or attenuation curve assumptions, and so may be misleading.

The small dispersion in the CLASSY SFR--M${\star}$ relationship is unsurprising:
star-forming galaxies are known to follow the Star-forming Main Sequence 
\citep[MS; e.g.,][]{brinchmann04, noeske07, salim07}.
The MS has further been identified as a plane in the fundamental metallicity relationship
\citep[FMR;][]{Mannucci10}.
The tightness of the MS is thought to result from self-regulation that significantly
slows the conversion of gas to stars \citep[e.g.,][]{bouche10,dave12}.
Galaxies with elevated SFRs appear as shifted-up outliers from the star-forming MS.
It is, therefore, informative to compare CLASSY to the MS.

To this end we consider both nearby and intermediate redshift samples for comparison MSs.
At $z\sim0$ we consider two samples:
first, the \citet{chang15} sample, which derived stellar masses and
SFRs for approximately $10^6$ SDSS+WISE galaxies.
Second, and perhaps the best comparison sample, is the 19-galaxy low-mass 
subsample of the Local Volume Legacy \citep[LVL;][]{kennicutt08} analyzed to 
have direct abundances and improved mass measurements by \citet{berg12}. 
While most other surveys comprehensively cover massive, high-surface-brightness 
systems in flux-limited samples, the LVL survey provides a multi-wavelength 
inventory of a statistically robust, approximately volume-limited sample of 
star-forming galaxies, which provides an essential baseline of "typical" 
galaxies with a similar stellar mass range as the CLASSY sample. 
Plotted in the upper left panel of Figure~\ref{fig8}, 
the \citet{chang15} SFR--M$_\star$ trend (dotted red line) aligns well with and extends 
the LMLVL (dot-dashed blue line) tend to higher masses, forming the $z\sim0$ MS.
For the intermediate redshift ($0.5 < z < 2.5$) MS, we use the \citet{whitaker14} 
mass-complete sample of approximately $4\times10^4$ star-forming galaxies in the 
CANDELS fields (gold lines).

We note that it is important to compare properties derived using the same methodology.
The present work, \citet{chang15} sample, and \citet{whitaker14} sample all
use a \citet{chabrier03} IMF and SED fitting to determine masses and SFRs.
On the other hand, the low-mass LVL (LMLVL) sample adopted the \citet{salpeter55} 
IMF and used H$\alpha+$[\ion{N}{2}] fluxes to determine the SFRs \citep{lee09}.
We, therefore, converted the LMLVL SFRs to a \citet{chabrier03} IMF using the 
scaling from \citet{kennicutt12}.
Additionally, we also consider the H$\alpha$ SFRs determined for the CLASSY sample 
(see Section~\ref{sec:4.7}) for proper comparison to the LMLVL SFRs.
Interestingly, the CLASSY H$\alpha$ SFRs determined in Section~\ref{AppendixC} are
0.22 dex larger on average than the aperture SED-derived values, but with a large
standard deviation of 1.16 dex, and so would only serve to slightly increase the average 
offset of the CLASSY sample from the $z\sim0$ MS in Figure~\ref{fig8}.

The CLASSY SFR--M${\star}$ relationship in the upper left panel of Figure~\ref{fig8}
lies above the $z\sim0$ and $0.5 < z < 2.5$ MSs trends.
The CLASSY trend has a similar slope to the $z\sim0$ MS, but lies roughly 2 dex
higher in SFR, while the intermediate-$z$ MS relationships begin to approach the 
CLASSY relationship with increasing redshift.
This shows that CLASSY galaxies, selected to be compact and UV bright,
have significantly enhanced sSFRs relative to their star-forming MS
counterparts at $z\sim0$;
rather the CLASSY sSFRs are more comparable to a $z\sim2$ galaxy population.

The middle row of Figure~\ref{fig8} shows
the total specific SFRs (left) and the aperture sSFRs (right) versus the 
direct $T_e$-method gas-phase oxygen abundances of the CLASSY sample.
Here, the majority of the CLASSY sample aligns with or extends from the \citet{curti20} 
relationships for $10^8 < M_\star/M_\odot < 10^{11}$ (orange to yellow lines, respectively) 
star-forming SDSS galaxies, but lies $1-3$ dex above the LMLVL trend.
Given that both \citet{curti20}\footnote{\citet{curti20} employed a $T_e$-anchored 
method based on galaxy stacks, but did not measure $T_e$-abundances for each galaxy directly.} 
and \cite{berg12} used direct $T_e$ methods for their oxygen abundance determinations,
this offset in sSFR is likely due to the boosted SFR of the CLASSY sample
(see the SFR--M$_\star$ trend). 

\subsection{Ionization Parameter and Density Properties}\label{sec:5.1.2}
The bottom left hand panel of Figure~\ref{fig8} shows the CLASSY galaxies 
also probe a broad range of ionization parameter, as indicated by
the oxygen ionization ratio, O$_{32}$=\fOTh \W5007/\fOTw \W3727, and gas-phase
electron density.
Most \ion{H}{2} regions have measured densities near the low-density limit
($\sim100$ \cmcu); this can be seen by the LMLVL sample (blue circles), 
where all but one galaxy have $n_e < 100$ \cmcu. 
On the other hand, seven CLASSY galaxies have $n_e > 300$ \cmcu (purple plus symbols;
the high-density CLASSY sample), 
where additional effects such as dielectric recombination can become important.
Given the evidence that higher nebular densities may be more common at higher
redshifts \cite[e.g.,][median $n_e\sim250$ \cmcu\ at $z\sim2.3$]{sanders16},
this high-density sample was included in CLASSY to investigate the role of density. 

\subsection{The M$_{\star}$--Z Relationship}\label{sec:5.1.3}
In the bottom right hand panel of Figure~\ref{fig8}, 
gas-phase oxygen abundance is plotted relative to the total stellar mass, 
showing the well-known mass-metallicity relationship \citep[MZR; e.g.,][]{tremonti04}.
Interestingly, the CLASSY sample closely follows the \citet{berg12} direct-metallicity 
MZR  measured for the LMLVL, but with larger scatter.
To quantify the comparison, we employ \citet{kelly07} \texttt{\sc{linmix}} 
fitting method and find the CLASSY MZR to be\looseness=-2
\begin{equation}
    \mbox{12+log(O/H)} = (0.20\pm0.04)\times\mbox{log M}_{\star,tot.} + (6.40\pm0.35), \nonumber
\end{equation}
where $M_{\star}$ has units of $M_\odot$, and the total scatter is $\sigma_{tot.} = 0.28$ dex and consistent intrinsic scatter.
This suggests that the CLASSY sample, while biased to target UV-bright galaxies,
is consistent with typical star-forming galaxies evolving along the MZR.
The trends plotted in Figure~\ref{fig8} demonstrate the expansive galaxy
properties and nebular conditions of the CLASSY sample that allow it 
to uniquely serve as templates for interpreting galaxies with significant star
formation episodes across all redshifts.


\section{Potential Science With CLASSY}\label{sec:6}


\noindent{\it Overview:}
The main objective of CLASSY is to use FUV spectra to unify stellar and gas-phase physics, 
allowing a holistic understanding of massive stars as the drivers of the gaseous 
evolution of star-forming galaxies. 
Naturally, the pursuit of this goal will produce a number of additional HLSPs, beyond
the coadded spectra, that will be useful to the astronomical community.
The enduring value and utility of CLASSY will reside in these 
state-of-the-art HLSPs products. 
Below we summarize the main HLSPs the CLASSY team plans to produce and 
the scientific objectives they will enable studies of.

\bigskip
\noindent{\it Summary of Planned CLASSY HLSPs:}
\setlist{nolistsep}
\begin{enumerate}[noitemsep]
\item CLASSY spectral atlas 
\item Compiled ancillary data 
\item CLASSY stellar continuum fits
\item Database of emission and absorption feature properties
\item Database of \LYA\ emission profile fits
\end{enumerate}
Additionally, user-friendly CLASSY tutorials on interacting with and utilizing the data products
will be released in tandem with corresponding HLSPs.
These HLSPs will be used to diagnose a vast array of science objectives, include the 
(1) massive star astrophysics, 
(2) physical properties of outflows, 
(3) \LYA\ physics, 
(4) chemical evolution of galaxies, and 
(5) physics of reionization. 
Below we provide an early glimpse to the individual science cases that CLASSY will 
explore and motivate some of the upcoming science that can be done with the data.


\subsection{Understanding The Massive Star Properties}\label{sec:6.1}
Massive stars influence all facets of their host star-forming galaxies including
the shape of their UV spectra:
massive stars produce copious amounts of high energy photons that are reprocessed in 
the interstellar medium and power the nebular emission lines that trace the gaseous 
physical conditions \citep[e.g.,][]{stromgren39, seyfert43, tinsley80};
massive stars may have been important contributors to the ionizing photon budget 
that reionized the early universe 
\citep[e.g.,][]{ouchi09, robertson13, robertson15, finkelstein19};
massive star winds disrupt the gas cycle in galaxies and drive out multiphased 
material and more.
Despite the obvious importance of massive stars in understanding the evolution of galaxies 
and the early universe, their ionizing spectra are not well understood. 
Thus, uncertainties in the shape of the ionizing spectrum have a significant effect on the 
interpretation of UV spectra, including gas properties, stellar feedback, the production 
of H-ionizing photons, and the effects of dust. 
While the implementation of new physics in stellar population synthesis (SPS) models 
(i.e., rotation, binaries) continues to improve predictions of the extreme UV (EUV) 
radiation field \citep[e.g.,][]{levesque12, eldridge17, gotberg18}, 
the shape of the ionizing spectrum remains poorly constrained for the metal-poor 
(Z/Z$_\odot <0.2$) stellar populations that come to dominate at high-$z$. 
This shortcoming thwarts our understanding of current moderate redshift ($z\sim2-5$) 
studies and future spectra of high redshift galaxies expected from ELTs and the JWST.\looseness=-2

The CLASSY survey provides an important opportunity for progress:
with high-resolution, high signal-to-noise spectra covering the entire FUV
(although, not the ionizing continuum directly), 
CLASSY probes a large suite of emission and absorption features shaped by massive stars. 
These spectra include the \ion{N}{5} \W\W1238,1242, 
\ion{Si}{4} \W\W1393,1402, and \ion{C}{4} \W\W1548,1550 stellar wind features and the weak 
\ion{Si}{3} \W\W1290,1417, \ion{C}{3} \W\W1247,1426,1428, and \ion{S}{5} \W1502 
photospheric features, 
both of which are distinctly sensitive to either age or metallicity of the stellar 
population \citep[e.g.,][]{demello00, vidal-garcia17, chisholm19}, 
but are also broadly distributed in wavelength space (across 1240–1900 \AA).\looseness=-2

The power of these lines to characterize massive star populations is demonstrated 
in the top row of Figure~\ref{fig9} for J0036-3333.
The CLASSY spectrum is well-fit by a light-weighted combination of Starburst99 
\citep[SB99;][]{leitherer99,leitherer10} stellar population synthesis models 
following the method of \citet{chisholm19} (shown in turquoise). 
Such stellar fits can determine the light-weighted metallicity, age, and reddening 
of the the massive star population. 

Recent $z\sim0-3$ studies have presented the first demonstrations that the combination of 
UV stellar and nebular spectral features can be used to 
constrain the shape of the otherwise unseen 
ionizing spectra in the EUV \citep[e.g.,][]{steidel16, steidel18, olivier21}.
Building on this powerful tool, the CLASSY sample and its superior spectra will open an 
unprecedented window on the stellar populations that form in a range of environments, 
including the low metallicity systems that dominate at high redshift. 
The CLASSY spectra will enable measurements of the stellar photospheric and wind features 
that can be used as important direct constraints on the metallicity and age of the massive 
stars and the EUV radiation field they power. 
In turn, the CLASSY modeled intrinsic ionizing continua can provide the community 
with the key to understanding how massive stars influence feedback, 
the ionization and enrichment of nebular gas,
and the conditions leading to the escape of ionizing radiation.

\renewcommand{\thefigure}{9}
\begin{figure*}
\begin{center}
    \includegraphics[width=0.975\textwidth,trim=0mm 0mm 0mm 0mm,clip]{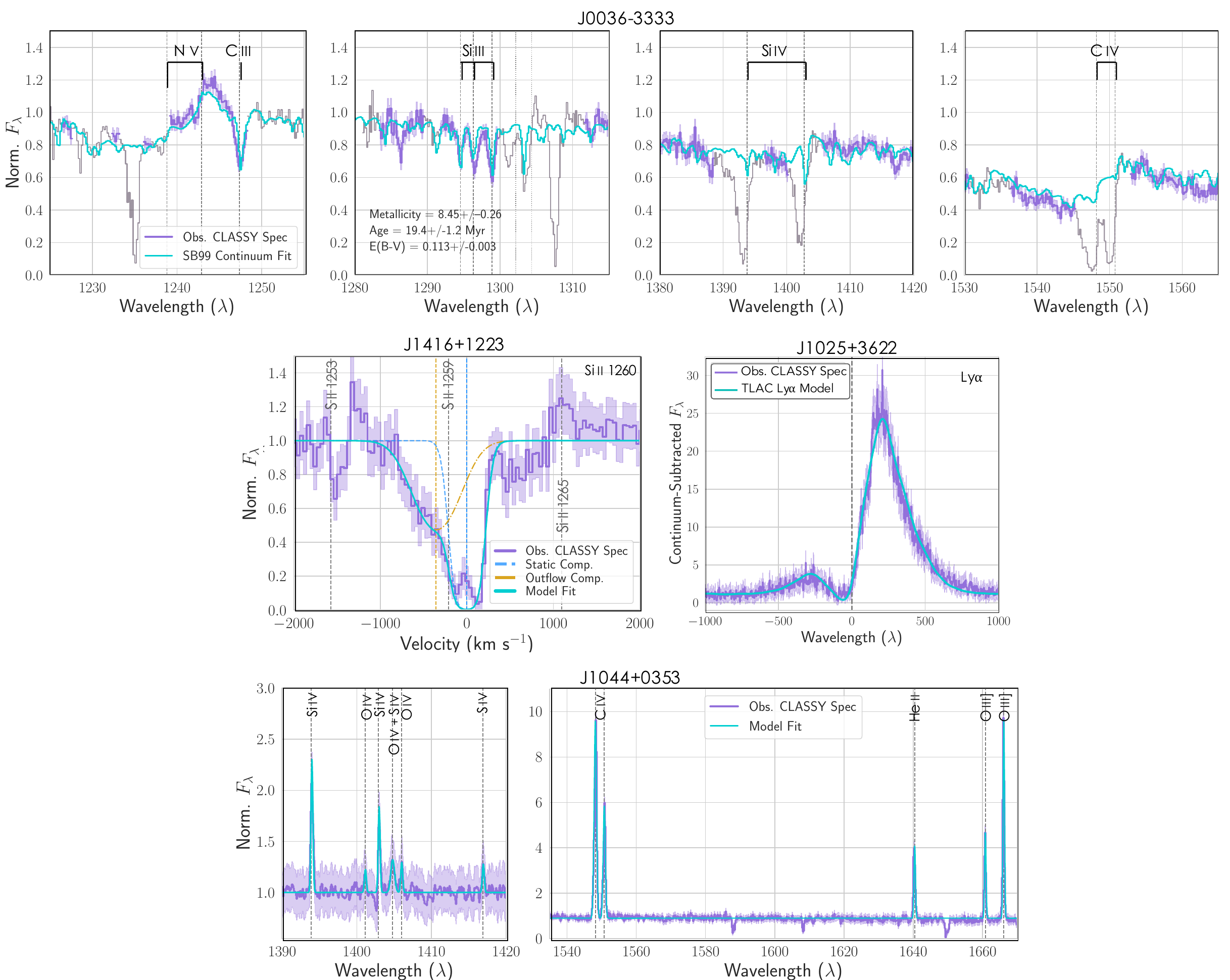} 
\end{center}
\vspace{-2ex}
\caption{
Demonstrations of science that can be derived from model fits to the CLASSY spectra.
{\it Top row:} 
The high-resolution coadded spectrum of J0036-3333 is well-fit by a
light-weighted combination of Starburst99 stellar population synthesis 
models (turquoise line). 
The regions of the continuum included in the fit are shown in purple, while masked regions 
are plotted in grey. 
The high S/N and spectral resolution of the CLASSY spectra allow both stellar wind
(\ion{N}{5} \W\W1238,1242, \ion{Si}{4} \W\W1393,1403, \ion{C}{4} \W\W1548,1550)
 and stellar photospheric (\ion{C}{3} \W1247, \ion{Si}{3} \W\W1295,1297,1299)
features to be fit.
{\it Middle left:}
The \ion{Si}{2} \W1260 interstellar absorption feature can be characterized in the 
high-resolution coadded spectrum of J1416+1223 using a double Gaussian fit.
This total fit (solid green line) includes a static component (dashed blue line) that 
is matched to the line spread function (LSF) and an outflowing component 
(dot-dashed gold line) that characterizes the kinematics of the low-ionization gas.
{\it Middle right:}
The Ly$\alpha$ profile in the very-high-resolution coadded spectrum of 
J1025+3622 is well-fit by a TLAC \citep[][]{gronke14} radiative transfer model. 
This model can be used to constrain the depth and kinematics of the neutral H gas.
{\it Bottom row:}
The high-resolution spectrum of J1044+0353 contains many high-ionization UV emission lines
that are well-fit by narrow, nebular Gaussian profiles.
These lines, including the \ion{Si}{4} \W\W1393,1403, \ion{O}{4} \W\W1401,1405,1407, 
\ion{S}{4} \W\W 1405,1406,1417, \ion{C}{4} \W\W1548,1550, \ion{He}{2} \W1640, and 
\ion{O}{3}] \W\W1661,1666 lines shown, can be used as nebular diagnostics.
\label{fig9}}
\end{figure*}


\subsection{The Physical Properties of Outflows}\label{sec:6.2}
The kinematics of the galaxy-scale outflows of gas driven by the massive stellar populations 
are encoded into the \LYA\ profiles and ISM resonant absorption lines observed in the FUV. 
These features currently provide the majority of the data on outflows at both low and high 
redshift.
The crucial question is how these outflows affect the evolution of galaxies. 
Answers require determination of the basic properties of the outflows 
(velocities, outflow rates, ionization state, etc.) and how these depend on the properties 
of the starburst and its host galaxy (star-formation rate, stellar mass, size, etc.).

The FUV spectral region covered by CLASSY contains a suite of resonance lines from atoms 
and ions spanning a large range in ionization potential. 
While several studies addressing the properties derived from the FUV exist 
\citep{heckman15,chisholm15,chisholm17}, CLASSY represents the opportunity for a major advance. 
Because CLASSY spans about four orders-of-magnitude in both stellar mass 
and SFR (see Figure~\ref{fig7}), 
it is the benchmark sample for investigations of how outflow and galaxy properties
scale together in the low redshift universe. 
Very few measurements of outflow velocities and mass outflow rates exist for galaxies 
with masses below $10^9 M_\odot$ 
\citep[e.g.,][]{martin99, martin05, heckman15, heckman16, bordoloi16}, 
where gravitational potential wells are the smallest and thus the galaxies are 
most impacted by supernovae feedback \citep{dekel86}.

Some absorption lines also have associated non-resonant fluorescence 
lines produced by radiative decay into a fine structure level just above the ground state.  
These lines are seen at both low and high redshifts 
\citep[e.g.,][]{shapley03,martin12,erb12,jones12,scarlata15,henry15,bordoloi16,finley17,wang20}, 
where the relative strengths of the fluorescent and absorption lines encode information
about the size/structure, geometry, and dust content of the outflow  
\citep{prochaska11,scarlata15,carr18,wang20,carr21}.
The CLASSY spectra will enable robust measurements of fluorescent emission that
can provide a powerful diagnostic of the structure and physical properties of outflows.\looseness=-2

The ability to model complex outflow signatures in the CLASSY galaxies is demonstrated
in the middle-left panel of Figure~\ref{fig9}.
The combination of CLASSY's high spectral resolution and S/N enables 
outflowing gas to be distinguished from static gas in the ISM and absorption
profiles to be corrected for the effects of in-filling by resonance scattering 
\citep{prochaska11, scarlata15}. 
For the \ion{Si}{2} \W1260.42 absorption feature in J1416+1223, a double Gaussian 
can be fit to the continuum-normalized spectrum to account for the static ISM 
component (dashed blue line) and characterize the outflow (dot-dashed gold line) 
kinematics of the low-ionization gas. 
Absorption line fits will be presented in the forthcoming CLASSY paper by \citet{xu22}. 
Such detailed absorption line analysis can then be used to analyze covering fractions,
ion column densities, and outflow rates. 
These measurements will constrain  stellar feedback on a wide-assortment of galaxy types, 
opening a new window onto the impact of stellar feedback on galaxy evolution with a specific 
emphasis on very low stellar mass galaxies.


\subsection{Ly\A\ Physics}\label{sec:6.3}
Outflows are predicted to promote the escape of Lyman-$\alpha$ (\LYA) 
photons before they are absorbed by interstellar dust.  
The shapes of the emergent \LYA\ line profiles therefore contain information about 
the physical properties of outflows. 
A comparison of outflow properties measured from \LYA\ 
emission lines and resonance absorption lines is a key CLASSY objective.

The CLASSY spectra cover \LYA\ for all targets, revealing a broad diversity
of \LYA\ profiles. 
Geocoronal emission from \LYA\ and \ion{O}{1} compromise a few \LYA\ profiles.  
Additionally, a significant fraction of the CLASSY spectra show broad \LYA\ absorption 
troughs underneath the \LYA\ emission that make continuum placement challenging, 
particularly in the lowest redshift targets where it blends with damped \ion{H}{1} 
absorption from the Milky Way. 
For the majority of the CLASSY sample, however, radiative transfer models can be used
to fit their complex \LYA\ profiles.
This technique is demonstrated for J1025+3622 in the middle-right panel of Figure~\ref{fig9},
where a Voigt profile was used to remove the underlying absorption and the \LYA\ emission
was modeled using the Monte-Carlo radiative transfer code, TLAC \citep[][]{gronke14}.
The details of such \LYA\ models offer the opportunity to constrain how the ISM and CGM 
modify the intrinsic \LYA\ emission through resonance scattering\citep[][]{gronke15}. 

Photons emitted by a central source in a simple shell model are scattered by an 
expanding shell of neutral hydrogen with variable dust opacity \citep{gronke15}. 
The fitted shell velocity and \ion{H}{1} column density have frequently been used to constrain 
galaxy properties \citep[e.g.,][]{kulas12, martin15, gronke17}. 
Yet direct comparisons to outflow properties derived from other spectral diagnostics 
remain limited to small samples \citep[][]{orlitova18, carr21}.
The combined \LYA, resonance absorption, and fluorescent emission profiles of the 
CLASSY sample will therefore allow discrepancies in outflow parameters to be addressed
via forward modeling of \LYA\ profiles with, e.g., 
the Semi-Analytical Line Transfer code \citep[SALT;][]{scarlata15} and 
the RAdiation SCattering in Astrophysical Simulations code \citep[RASCAS;][]{michel-dansac20}.

A key goal of CLASSY is to determine the strongest scaling relations between \LYA\ 
emission and galaxy properties.  
Many CLASSY spectra show double-peaked \LYA\ emission-line profiles, an indication of 
low \ion{H}{1} column density channels \citep[][]{verhamme17}. 
These profiles can be used to measure peak-to-peak velocity separations and, more 
generally, \LYA\ velocity offsets relative to galaxy redshifts.
Given that the peak-to-peak separation appears to be a promising indirect indicator of Lyman 
continuum leakage \citep[][]{verhamme15,izotov18,jaskot19}, the CLASSY peak separations along
with measurements of partial covering (via UV resonance lines), the age and metallicity 
of the massive star population, and the ionization structure inferred from UV and optical 
emission lines can be used to investigate this escape fraction indicator.
Examining the same galaxies from all these perspectives will provide unique insight 
into escape fractions, clarifying their relationship to radiative and mechanical feedback. 
A limitation imposed by the targets' proximity and the solid angle subtended by the COS PSA 
is that the measured equivalent widths and \LYA\ escape fractions will be aperture limited.

Looking back to the epoch of reionization (EoR), the increasing fraction of neutral 
hydrogen blocks \LYA\ transmission.  
The declining volume density of \LYA\ emitters at high-$z$ constrains the 
reionization history, and a better understanding of \LYA\ velocity offsets, specifically 
their dependence on galaxy properties, will reduce potential systematic errors on the 
inferred neutral fractions \citep[][]{mason18,mason19,naidu20}.
The CLASSY sample includes extreme emission-line galaxies, 
the closest local analogs of EoR galaxies, making them excellent environments to study 
the star-gas interplay that shaped cosmic reionization.


\subsection{UV Diagnostics of Chemical Evolution}\label{sec:6.4}
The CLASSY spectra contain strong UV emission-lines that characterize a plethora 
of gas properties, including temperature, density, and metal content 
\citep[both nebular and ISM, e.g.,][]{james14, byler18, byler20}, 
as well as reflecting the shape and hardness of the ionizing spectrum. 
The bottom panel of Figure~\ref{fig9} demonstrates Gaussian fits to several high-ionization
nebular emission lines in the continuum-subtracted spectrum of J1044+0353.
The full suite of emission-line measurements for the CLASSY sample will be released as
HLSPs as part of the forthcoming CLASSY papers by \citet{mingozzi22} and 
\citet{arellano-cordova22}.

Such detailed emission line analyses can be used to provide diagnostics of the 
chemical and physical properties of the CLASSY systems.
These include several temperature and density-sensitive line ratios.
When combined with optical line diagnostics afforded by the ground-based data, 
the CLASSY spectra will provide guidance on interpreting purely UV-based diagnostics 
of high-redshift galaxies. 
For example, the agreement of electron temperatures derived from the UV auroral \ion{O}{3}] 
\W\W1660,1666 emission lines with those derived from [\ion{O}{3}] in the optical can be tested. 

Recently, significant efforts have been invested into photoionization modeling of nebular 
UV emission lines to constrain the radiation fields, ionization sources, and metal content 
within star forming galaxies \citep[e.g.,][]{byler18, nakajima18, berg21, olivier21}. 
The CLASSY UV emission-line ratios can be calibrated against well-understood diagnostics 
from the existing optical spectra to provide the first stringent tests of proposed UV 
diagnostics, over a large range of physical environments 
(i.e., SFR, stellar mass, metallicity, gas density). 
The high-S/N detection of important combinations of UV emission lines, 
such as \ion{C}{4} \W\W1548,1550/\ion{He}{2} \W1640 versus \ion{O}{3}] 
\W\W1661,1666/\ion{He}{2} \W1640, are shown in Figure~\ref{fig9}.
With these lines CLASSY can assess the utility of proposed UV diagnostics
such as the UV versions of the canonical "BPT" diagram 
\citep[e.g.,][]{feltre16,nanayakkara19} to distinguish ionizing sources, \ion{C}{3}] 
\W\W1907,1909 equivalent widths to infer stellar age and nebular metallicity 
\citep[e.g.,][]{rigby15, jaskot16, senchyna17, ravindranath20}, and resonant emission lines, 
such as \LYA \W1215 and \ion{C}{4} \W\W1548,1550 as indicators of ionizing photon escape 
\citep[e.g.,][]{verhamme15, berg19b}. 

Uniquely, the CLASSY sample also includes a sub-set of high-density galaxies 
($n_e > 300$ \cmcu), with which a detailed study of the effects of density on 
FUV properties can be investigated. 
All together, CLASSY seeks to establish the essential UV diagnostic toolset needed 
to understand fundamental questions concerning the evolution, physical conditions, 
and ionization structure of star-forming galaxies across cosmic time.


\subsection{Reionization Physics}\label{sec:6.5}
At redshifts between $z=6-10$, ionizing photons escaped from galaxies 
to reionize the universe \citep{fan06}. 
Determining the sources of cosmic reionization is one of four key science goals of JWST. 
However, neither JWST nor ELTs will directly observe the Lyman Continuum (LyC) during the 
EoR owing to the increasing opacity of the intergalactic medium with redshift. 
The CLASSY survey will provide templates to understand the gaseous conditions in galaxies 
similar to high-redshift galaxies.
The high S/N and spectral resolution CLASSY observations will reveal complex geometric 
constraints and their relations to observables like the \LYA\ emission or the 
metal covering fraction.
Further, the stellar continua observations of the massive star populations can be 
combined with stellar population synthesis models (highlighted in Figure~\ref{fig9}
to predict the intrinsic number of ionizing photons produced by massive stars ($Q$) 
and the production efficiency of ionizing photons ($\xi_{\rm ion} = Q/L_{\rm UV}$). 
Extrapolating from the CLASSY stellar continuum fits, we will predict $\xi_{\rm ion}$ 
and investigate the correlation with UV emission lines 
\citep[e.g.,][]{jaskot16,schaerer18,ravindranath20}.
Finally, the comparisons between the FUV reddening and the reddening from the optical 
emission lines will inform on complex geometries and the origin of dust extinction. 
These results can then be compared to predictions from simulations to test how 
galaxies contribute to reionization \citep[e.g.,][]{fletcher19, finkelstein19}.

Theoretical arguments and small observational samples suggest that UV nebular emission 
and absorption features trace the escape fraction. 
CLASSY will indirectly infer escape 
fractions of a statistically significant sample using UV diagnostics accessible by ELTs 
and JWST: \LYA\ emission \citep{verhamme15, steidel18, izotov18}, the depth of 
low-ionization absorption lines \citep{heckman11, chisholm18a}, and the strength of 
high-ionization emission lines \citep{nakajima18}. 
The middle left panel of Figure~\ref{fig9} demonstrates that the CLASSY absorption lines 
estimates the porosity (or covering fraction) of the foreground neutral gas. 
In this example, the high signal-to-noise data suggests that the \ion{Si}{2} gas does not 
fully cover the background stellar continuum because the flux never reaches a value of zero. 
This implies that there are small regions of the stellar emission that propagate out of 
J1416+1223 without being absorbed by Si ions. 
Similarly, the TLAC \LYA\ models in the middle right of Figure~\ref{fig9} emphasizes 
how the neutral gas geometries can be estimated from the CLASSY profiles. 
These constraints will illustrate the neutral gas content in galaxies and provide a template 
to determine the escape of \LYA\ and ionizing photons at high redshift.


\section{Summary}\label{sec:7}


We have presented a new spectral atlas consisting of high-resolution, high-S/N 
contiguous FUV spectra of 45 local star-forming galaxies obtained with \HST/COS as 
part of the COS Legacy Archival Spectroscopic SurveY (CLASSY). 
As a result, the CLASSY atlas reaches far beyond previous FUV spectral libraries.
One of the considerable strengths of the CLASSY project is the careful selection of the
sample so that the value of the existing HST archival spectra was maximized. 
As such, the CLASSY atlas was constructed from 94 archival spectra from 177 orbits and 
76 new spectra from 135 orbits (170 spectra total from 312 orbits) of HST observations, 
or more than 600 total Cosmic Origins Spectrograph (COS) spectral images. 
Further, we uniformly reduced and coadded the combined archival and new observations
of the CLASSY atlas, 
producing full far-UV spectra that allow a large number and broad range of scientific 
analyses to follow.
The resulting CLASSY spectral atlas is the first of the CLASSY Treasury high level science 
products to be made publicly available to the astronomical community via the CLASSY HLSP 
website: \url{https://archive.stsci.edu/hlsp/classy}.

We have also compiled an ancillary set of high-quality optical spectra and photometry
and used them to derive updated galaxy properties for the CLASSY galaxies. 
As a result, we report properties for the CLASSY sample that span broad ranges in
stellar mass ($6.2<$ log $M_\star/M_\odot$ $<10.1$),
star formation rate ($-2.0<$ log SFR ($M_\odot$ yr$^{-1}$)$<+1.6$),
direct gas-phase metallicity ($7.0<$ 12+log(O/H) $<8.8$),
ionization ($0.5<$ O$_{32}<38.0$),
reddening ($0.01<$ E(B-V) $<0.12$), and
nebular density ($10<n_e (\rm{cm}^{-3})<1120$).
With these data, we showed that the CLASSY sample is consistent 
with the typical evolution of star-forming galaxies along the
mass-metallicity relationship in the Local Volume,
but seem to be currently experiencing a strong burst of star formation
that mimics the SFRs of moderate- to high-redshift galaxies.
Thus, the CLASSY spectral templates provide a unique dataset to study both local 
star-forming galaxies due to the specific spatial scales that they probe and 
distant galaxies due to their enhanced SFRs.

Looking to the future, the CLASSY Treasury will provide a suite of high level science
products to help the community prepare for future observatories and empirically connect 
the stellar and gas-phase properties of star-forming galaxies for the first time using 
a suite of UV absorption and emission line diagnostics, added to optical emission line 
measurements. 


\begin{acknowledgements}

The CLASSY team thanks the referee for thoughtful feedback that significantly improved 
both the paper and the HLSPs.
DAB is grateful for the support for this program, HST-GO-15840, that was provided by 
NASA through a grant from the Space Telescope Science Institute, 
which is operated by the Associations of Universities for Research in Astronomy, 
Incorporated, under NASA contract NAS5- 26555. 
BLJ thanks support from the European Space Agency (ESA).
The CLASSY collaboration extends special gratitude to the Lorentz Center for useful discussions 
during the "Characterizing Galaxies with Spectroscopy with a view for JWST" 2017 workshop that led 
to the formation of the CLASSY collaboration and survey.

Funding for SDSS-III has been provided by the Alfred P. Sloan Foundation, the Participating Institutions, the National Science Foundation, and the U.S. Department of Energy Office of Science. The SDSS-III web site is http://www.sdss3.org/.

SDSS-III is managed by the Astrophysical Research Consortium for the Participating Institutions of the SDSS-III Collaboration including the University of Arizona, the Brazilian Participation Group, Brookhaven National Laboratory, Carnegie Mellon University, University of Florida, the French Participation Group, the German Participation Group, Harvard University, the Instituto de Astrofisica de Canarias, the Michigan State/Notre Dame/JINA Participation Group, Johns Hopkins University, Lawrence Berkeley National Laboratory, Max Planck Institute for Astrophysics, Max Planck Institute for Extraterrestrial Physics, New Mexico State University, New York University, Ohio State University, Pennsylvania State University, University of Portsmouth, Princeton University, the Spanish Participation Group, University of Tokyo, University of Utah, Vanderbilt University, University of Virginia, University of Washington, and Yale University.

This work also uses the services of the ESO Science Archive Facility,
observations collected at the European Southern Observatory under 
ESO programmes 096.B-0690, 0103.B-0531, 0103.D-0705, and 0104.D-0503, and
observations obtained with the Large Binocular Telescope (LBT).
The LBT is an international collaboration among institutions in the
United States, Italy and Germany. LBT Corporation partners are: The
University of Arizona on behalf of the Arizona Board of Regents;
Istituto Nazionale di Astrofisica, Italy; LBT Beteiligungsgesellschaft,
Germany, representing the Max-Planck Society, The Leibniz Institute for
Astrophysics Potsdam, and Heidelberg University; The Ohio State
University, University of Notre Dame, University of
Minnesota, and University of Virginia.

This paper made use of the modsIDL spectral data reduction reduction pipeline
developed in part with funds provided by NSF Grant AST-1108693 and a generous
gift from OSU Astronomy alumnus David G. Price through the Price Fellowship in
Astronomical Instrumentation. 
This research has made use of the HSLA database, developed and maintained at STScI, Baltimore, USA.

\end{acknowledgements}

\facilities{HST (COS), LBT (MODS), APO (SDSS), KECK (ESI), VLT (MUSE, VIMOS)}
\software{
astropy (The Astropy Collaboration 2013, 2018)
BEAGLE (Chevallard \& Charlot 2006), 
CalCOS (STScI),
dustmaps (Green 2018),
jupyter (Kluyver 2016),
LINMIX (Kelly 2007) 
MPFIT (Markwardt 2009),
MODS reduction Pipeline,
Photutils (Bradley 2021),
PYNEB (Luridiana 2012; 2015),
python,
pysynphot (STScI Development Team),
RASCAS (Michel-Dansac 2020),
SALT (Scarlata \& Panagia 2015), 
STARLIGHT (Fernandes 2005), 
TLAC (Gronke \& Dijkstra 2014),
XIDL}


\newpage
\appendix


\section{CLASSY FUV+Optical Spectral Observations}\label{AppendixA}
\begin{figure*}
\begin{center}
    \includegraphics[width=0.95\textwidth]{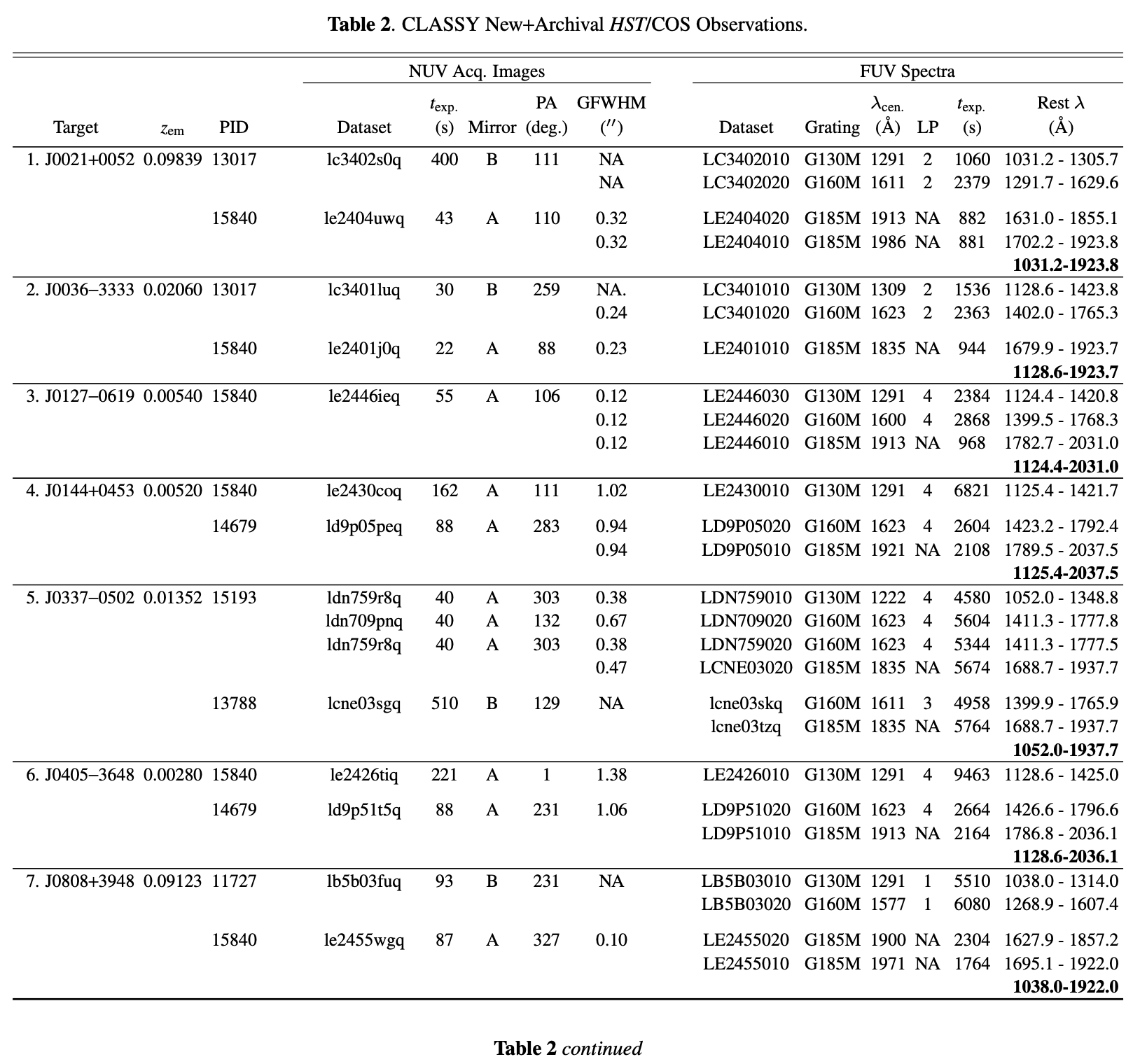}
\label{tbl2}
\end{center}
\vspace{-4ex}
\end{figure*}

Here we present the details of the CLASSY FUV \HST/COS spectroscopic observations in 
Tables~\ref{tbl2} and \ref{tbl3} and the optical spectroscopic observations in Table~\ref{tbl4}.

\clearpage

\begin{figure*}
    \includegraphics[width=1.0\textwidth]{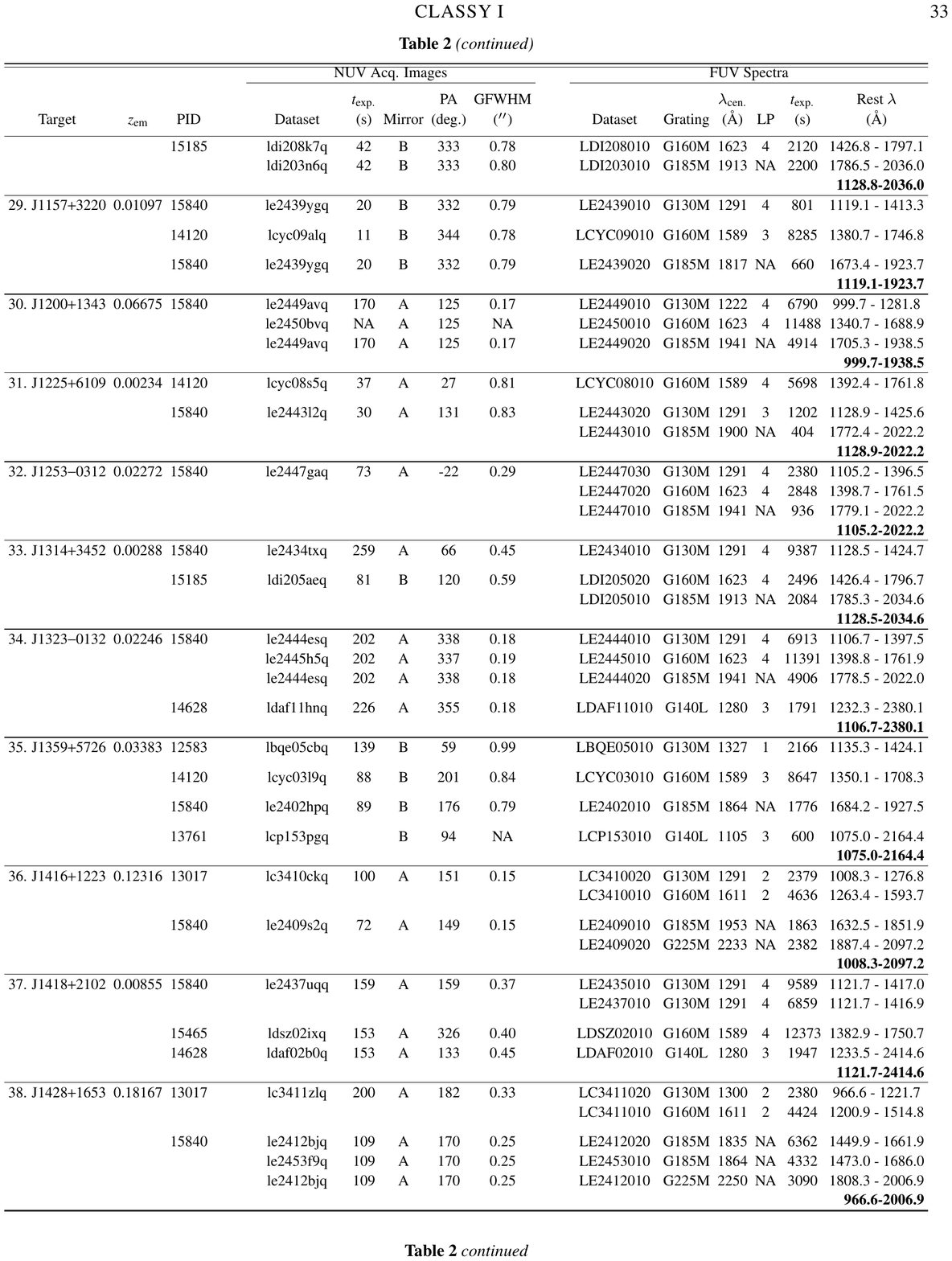} 
\end{figure*}

\begin{figure*}
    \includegraphics[width=1.0\textwidth]{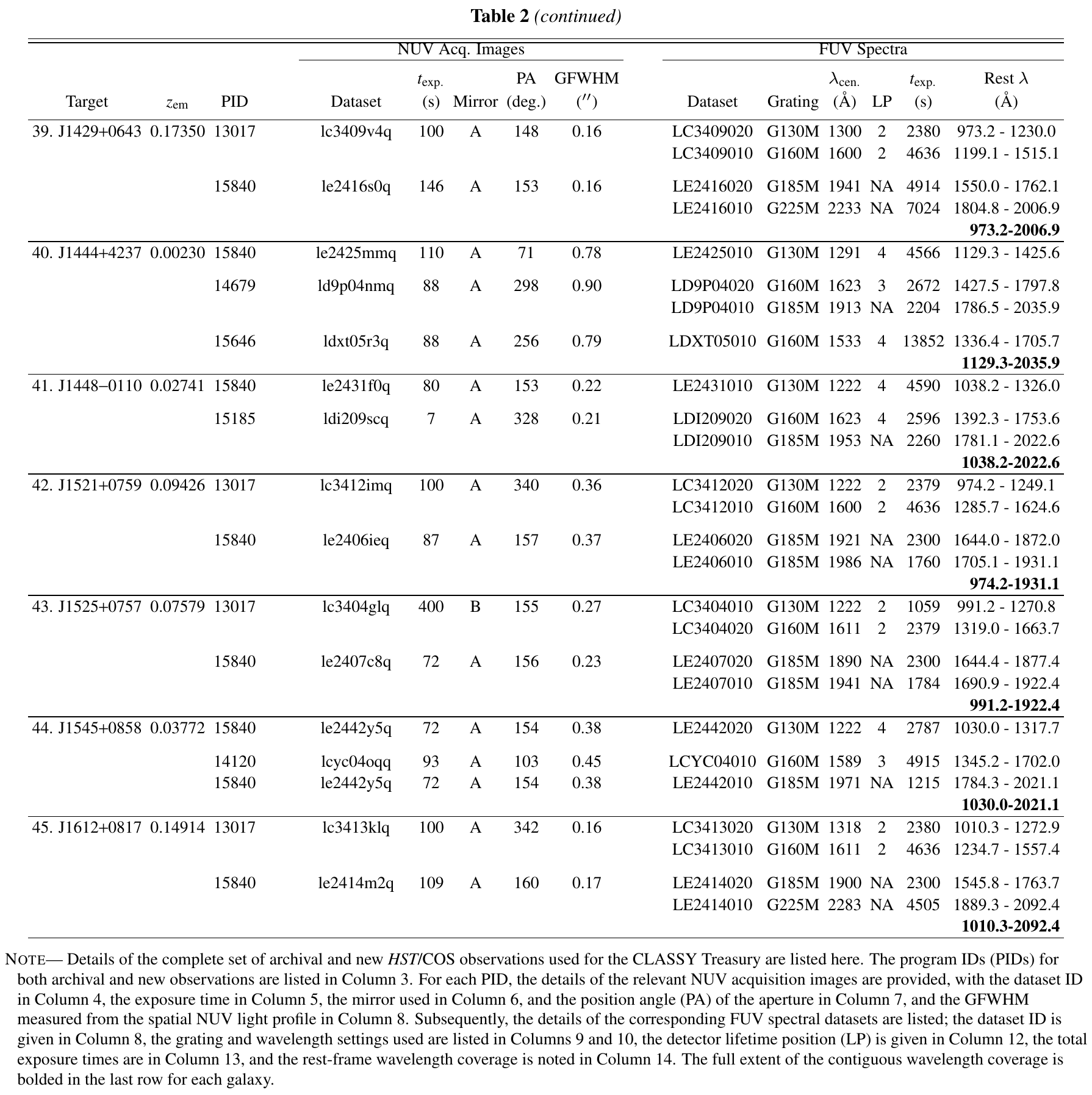} 
\end{figure*}

\clearpage

\renewcommand{\thetable}{3}
\begin{center}
\begin{deluxetable*}{rcccccc}
\centering
	\setlength{\tabcolsep}{3pt}
\caption{
Nominal and Measured Spectral Resolution of CLASSY Coadded Spectra.}
\tablehead{
\CH{}               & \CH{}         & \multicolumn{4}{c}{Nominal Resolution}    & \CH{Measured}   \\[-3ex] 
\CH{}               &\CH{Wavelength}& \multicolumn{4}{c}{(km s$^{-1}$/resel)}   & \CH{Resolution} \\[-1ex]  \cline{3-6} 
\\[-4ex] 
\CH{Target}         & \CH{(\AA)}    &\CH{VHR}&\CH{HR}&\CH{MR}&\CH{LR}   & \CH{(km s$^{-1}$/resel)} }
\startdata
1. J0021$+$0052     & 1260          & 14.3  & 17.4  & 47.6  & 119.6     & 42    \\
2. J0036$-$3333 	& 1260          & 14.3  & 17.4  & 47.6  & 119.6     & 128   \\
3. J0127$-$0619	    & 1334          & 13.5  & 16.4  & 45.0  & 112.0     & 74    \\
4. J0144$+$0453     & 1334          & 13.5  & 16.4  & 45.0  & 112.0     & 74    \\  
5. J0337$-$0502     & 1334          & 13.5  & 16.4  & 45.0  & 112.0     & 20    \\
\\[-1ex]
6. J0405$-$3648	    & 1334          & 13.5  & 16.4  & 45.0  & 112.0     & 65    \\
7. J0808$+$3948     & 1334          & 13.5  & 16.4  & 45.0  & 112.0     & 40    \\
8. J0823$+$2806     & 1334          & 13.5  & 16.4  & 45.0  & 112.0     & 67    \\
9. J0926$+$4427     & 1260          & 14.3  & 17.4  & 47.6  & 119.6     & 71    \\
10. J0934$+$5514    & 1334          & 13.5  & 16.4  & 45.0  & 112.0     & 105   \\
\\[-1ex]
11. J0938$+$5428    & 1334          & 13.5  & 16.4  & 45.0  & 112.0     & 67    \\
12. J0940$+$2935    & 1260          & 14.3  & 17.4  & 47.6  & 119.6     & 88    \\
13. J0942$+$3547    & N/A           &       &       &       &           & 66    \\
14. J0944$-$0038    & 1260          & 14.3  & 17.4  & 47.6  & 119.6     & 71    \\
15. J0944$+$3442    & N/A           &       &       &       &           &       \\
\\[-1ex]
16. J1016$+$3754    & 1260          & 14.3  & 17.4  & 47.6  & 119.6     & 71    \\
17. J1024$+$0524    & 1260          & 14.3  & 17.4  & 47.6  & 119.6     & 71    \\
18. J1025$+$3622    & 1260          & 14.3  & 17.4  & 47.6  & 119.6     & 52    \\
19. J1044$+$0353    & N/A           &       &       &       &           &       \\
20. J1105$+$4444    & 1548          & 11.6  & 14.2  & 38.8  & 96.5      & 52    \\
\\[-1ex]
21. J1112$+$5503    & 1260          & 14.3  & 17.4  & 47.6  & 119.6     & 61    \\
22. J1119$+$5130    & 1260          & 14.3  & 17.4  & 47.6  & 119.6     & 71    \\
23. J1129$+$2034    & 1334          & 13.5  & 16.4  & 45.0  & 112.0     & 33    \\
24. J1132$+$5722    & 1260          & 14.3  & 17.4  & 47.6  & 119.6     & 71    \\
25. J1132$+$1411    & 1260          & 14.3  & 17.4  & 47.6  & 119.6     & 71    \\
\\[-1ex]
26. J1144$+$4012    & N/A           &       &       &       &           &       \\
27. J1148$+$2546    & 1334          & 13.5  & 16.4  & 45.0  & 112.0     & 67    \\
28. J1150$+$1501    & 1334          & 13.5  & 16.4  & 45.0  & 112.0     & 85    \\
29. J1157$+$3220    & 1334          & 13.5  & 16.4  & 45.0  & 112.0     & 67    \\
30. J1200$+$1343    & 1334          & 13.5  & 16.4  & 45.0  & 112.0     & 83    \\
\\[-1ex]
31. J1225$+$6109    & 1548          & 11.6  & 14.2  & 38.8  & 96.5      & 58    \\
32. J1253$-$0312    & 1548          & 11.6  & 14.2  & 38.8  & 96.5      & 50    \\
33. J1314$+$3452    & 1260          & 14.3  & 17.4  & 47.6  & 119.6     & 45    \\
34. J1323$-$0132    & 1548          & 11.6  & 14.2  & 38.8  & 96.5      & 40    \\
35. J1359$+$5726    & 1548          & 11.6  & 14.2  & 38.8  & 96.5      & 42    \\	
\\[-1ex]
36. J1416$+$1223    & 1548          & 11.6  & 14.2  & 38.8  & 96.5      & 46    \\
37. J1418$+$2102    & 1548          & 11.6  & 14.2  & 38.8  & 96.5      & 46    \\	
38. J1428$+$1653    & 1548          & 11.6  & 14.2  & 38.8  & 96.5      & 65    \\	
39. J1429$+$0643    & 1548          & 11.6  & 14.2  & 38.8  & 96.5      & 65    \\ 	
40. J1444$+$4237    & 1548          & 11.6  & 14.2  & 38.8  & 96.5      & 32    \\ 	
\\[-1ex]
41. J1448$-$0110    & 1548          & 11.6  & 14.2  & 38.8  & 96.5      & 61    \\
42. J1521$+$0759    & 1548          & 11.6  & 14.2  & 38.8  & 96.5      & 65    \\
43. J1525$+$0757    & 1548          & 11.6  & 14.2  & 38.8  & 96.5      & 54    \\	
44. J1545$+$0858    & 1548          & 11.6  & 14.2  & 38.8  & 96.5      & 61    \\
45. J1612$+$0817    & 1548          & 11.6  & 14.2  & 38.8  & 96.5      & 50    	
\enddata
\tablecomments{
Comparison of the nominal (i.e., point source) spectral resolution of the CLASSY coadded spectra to the measured
resolution from Milky Way ISM absorption lines.
The best Milky Way ISM feature for each galaxy is listed in Column 2, followed by the
nominal resolution at that wavelength in km s$^{-1}$ per resolution element for each coadded 
dataset in Columns $3-6$.
Finally, Column 7 lists the Gaussian FWHM in km s$^{-1}$ measured from the HR, Galactic extinction-corrected, binned spectra (extension 9).}
\label{tbl3}
\end{deluxetable*} 
\end{center}

\renewcommand{\thetable}{4}
\begin{center}
\begin{deluxetable*}{rcccc}
\centering
	\setlength{\tabcolsep}{3pt}
\caption{Ancillary Optical Spectra for the CLASSY Galaxy Sample.}
\tablehead{
\CH{}               & \CH{Telescope/}   & \CH{}                 &\CH{Exposure} &\CH{} \\[-2ex] 
\CH{Target}         & \CH{Spectrograph} & \CH{Aperture}         &\CH{Time (s)} &\CH{Reference}	}
\startdata
1. J0021$+$0052     & VLT/MUSE      & 2.5\arcsec\ circ.             & 2449	        & ESO 0104.D-0503; PI Anderson  \\ 
2. J0036$-$3333 	& VLT/MUSE      & 2.5\arcsec\ circ.             & 2800          & ESO 096.B-0923; PI \"{O}stlin \\
3. J0127$-$0619	    & VLT/VIMOS     & 2.5\arcsec\ circ.             & 1608, 1488    & \citet{james09}               \\
4. J0144$+$0453     & MMT/BC	    & 1\arcsec$\times$180\arcsec\   & 6300	        & \citet{senchyna19a}           \\  
5. J0337$-$0502     & VLT/MUSE      & 2.5\arcsec\ circ.             & 5680          & ESO 096.B-0690; PI Hayes      \\
\\[-1ex]
6. J0405$-$3648	    & VLT/MUSE	    & 2.5\arcsec\ circ.             & 2000	        & ESO 0103.D-0705; PI Brinchmann \\
7. J0808$+$3948     & LBT/MODS	    & 1\arcsec$\times$60\arcsec\    & 2700	        & This work                     \\
8. J0823$+$2806     & APO/SDSS	    & 3\arcsec\ circ.               & 2700	        & \citet{eisenstein11}          \\
9. J0926$+$4427     & APO/SDSS	    & 3\arcsec\ circ.	            & 2700	        & \citet{eisenstein11}          \\
10. J0934$+$5514    & Keck/KCWI     & 2.5\arcsec\ circ.             & 1200          & \citet{rickardsvaught21}      \\
\\[-1ex]
11. J0938$+$5428    & APO/SDSS	    & 3\arcsec\ circ.	            & 2700	        & \citet{eisenstein11}          \\
12. J0940$+$2935    & APO/SDSS	    & 3\arcsec\ circ.	            & 2700	        & \citet{eisenstein11}          \\
13. J0942$+$3547    & Keck/ESI      & 1\arcsec$\times$60\arcsec\    & 2700	        & \citet{sanders21}             \\
14. J0944$-$0038    & Keck/ESI      & 1\arcsec$\times$60\arcsec\    & 2700	        & \citet{sanders21}             \\
15. J0944$+$3442    & LBT/MODS	    & 1\arcsec$\times$60\arcsec\	& 2700	        & This work                     \\
\\[-1ex]
16. J1016$+$3754    & APO/SDSS	    & 3\arcsec\ circ.	            & 2700	        & \citet{eisenstein11}          \\
17. J1024$+$0524    & Keck/ESI	    & 1\arcsec$\times$60\arcsec\    & 2700	        & \citet{sanders21}             \\
18. J1025$+$3622    & APO/SDSS	    & 3\arcsec\ circ.               & 2700	        & \citet{eisenstein11}          \\
19. J1044$+$0353    & VLT/MUSE	    & 2.5\arcsec\ circ.  	        & 4864          & ESO 0103.B-0531; PI Erb       \\
20. J1105$+$4444    & APO/SDSS	    & 3\arcsec\ circ.	            & 2700	        & \citet{eisenstein11}          \\
\\[-1ex]
21. J1112$+$5503    & APO/SDSS	    & 3\arcsec\ circ.	            & 2700	        & \citet{eisenstein11}          \\ 
22. J1119$+$5130    & MMT/BC	    & 1\arcsec$\times$180\arcsec\   & 4500	        & \citet{senchyna19a}           \\  
23. J1129$+$2034    & Keck/ESI	    & 1\arcsec$\times$60\arcsec\    & 2700	        & \citet{sanders21}             \\
24. J1132$+$5722    & MMT/BC	    & 1\arcsec$\times$180\arcsec\   & 5400	        & \citet{senchyna19a}           \\  
25. J1132$+$1411    & APO/SDSS	    & 3\arcsec\ circ.	            & 2700	        & \citet{eisenstein11}          \\
\\[-1ex]
26. J1144$+$4012    & APO/SDSS	    & 3\arcsec\ circ.	            & 2700	        & \citet{eisenstein11}          \\ 
27. J1148$+$2546    & Keck/ESI	    & 1\arcsec$\times$60\arcsec\    & 2700	        & \citet{sanders21}             \\
28. J1150$+$1501    & APO/SDSS	    & 3\arcsec\ circ.	            & 2700	        & \citet{eisenstein11}          \\ 
29. J1157$+$3220    & APO/SDSS 	    & 3\arcsec\ circ.               & 2700	        & \citet{eisenstein11}          \\ 
30. J1200$+$1343    & APO/SDSS	    & 3\arcsec\ circ.               & 2700	        & \citet{eisenstein11}          \\ 
\\[-1ex]
31. J1225$+$6109    & APO/SDSS	    & 3\arcsec\ circ.               & 2700      	& \citet{eisenstein11}          \\ 	
32. J1253$-$0312    & APO/SDSS	    & 3\arcsec\ circ.               & 2700          & \citet{eisenstein11}          \\ 	
33. J1314$+$3452    & APO/SDSS	    & 3\arcsec\ circ.               & 2700	        & \citet{eisenstein11}          \\ 	
34. J1323$-$0132    & APO/SDSS	    & 3\arcsec\ circ.               & 2700	        & \citet{eisenstein11}          \\    
35. J1359$+$5726    & APO/SDSS	    & 3\arcsec\ circ.               & 2700	        & \citet{eisenstein11}          \\  	
\\[-1ex]
36. J1416$+$1223    & APO/SDSS	    & 3\arcsec\ circ.               & 2700	        & \citet{eisenstein11}          \\ 
37. J1418$+$2102    & VLT/MUSE	    & 2.5\arcsec\ circ.             & 4864          & ESO 0103.B-0531; PI Erb       \\  	
38. J1428$+$1653    & APO/SDSS	    & 3\arcsec\ circ.               & 2700	        & \citet{eisenstein11}          \\  	
39. J1429$+$0643    & APO/SDSS	    & 3\arcsec\ circ.               & 2700	        & \citet{eisenstein11}          \\  	
40. J1444$+$4237    & MMT/BC	    & 1\arcsec$\times$180\arcsec\   & 6300	        & \citet{senchyna19a}           \\  	
\\[-1ex]
41. J1448$-$0110    & APO/SDSS	    & 3\arcsec\ circ.               & 2700	        & \citet{eisenstein11}          \\  
42. J1521$+$0759    & APO/SDSS	    & 3\arcsec\ circ.               & 2700	        & \citet{eisenstein11}          \\     
43. J1525$+$0757    & APO/SDSS	    & 3\arcsec\ circ.               & 2700	        & \citet{eisenstein11}          \\  	
44. J1545$+$0858    & LBT/MODS	    & 1\arcsec$\times$60\arcsec\    & 2700	        & This work                     \\ 
45. J1612$+$0817    & APO/SDSS	    & 3\arcsec\ circ.               & 2700	        & \citet{eisenstein11} 	
\enddata

\tablecomments{
Details of the best ancillary optical spectra available to accompany the CLASSY FUV spectra.
The telescope and spectrograph of the observations are listed in Column 2 and the cumulative 
exposure time is tabulated in Column 4.
The apertures listed in Column 3 are the ones used to extract the spectra used in this work to determine
the nebular properties of the CLASSY sample.
For IFU observations, the 1-dimensional spectra were extracted to aperture-match the FUV \HST/COS spectra.
References in Column 5 refer to the source of the observations.
}
\label{tbl4}
\end{deluxetable*} 
\end{center}


\section{CLASSY Sample Properties}\label{AppendixB}
Here we present properties of the CLASSY sample galaxies that were derived from the 
ancillary optical spectra in Table~\ref{tbl5} and \texttt{Beagle} SED fitting in
Table~\ref{tbl6}.
See Section~\ref{sec:4} for details of the calculations.
Additionally, we characterize the aperture bias of the CLASSY sample in Figure~\ref{fig10} by plotting the fraction of the optical flux measured
(from the $u$ and $g-$band SDSS, DES, and PanSTARRS imaging)
through a 2\farcs5 aperture relative to the total flux of the galaxy.
The points in Figure~\ref{fig11} are color-coded by their optical
half light radii, measured from the same imaging, showing that 
most of the CLASSY sample is relatively compact with 66\% of the 
galaxies having $r_{50,opt.} \lesssim 2.0\arcsec$.
As demonstrated in Figure~\ref{fig2}, the extent of the UV light is
much more compact than the optical light of the CLASSY sample.

\renewcommand{\thefigure}{10}
\begin{figure}
\centering
    \includegraphics[width=0.90\textwidth,trim=20mm 0mm 0mm 0mm,clip]{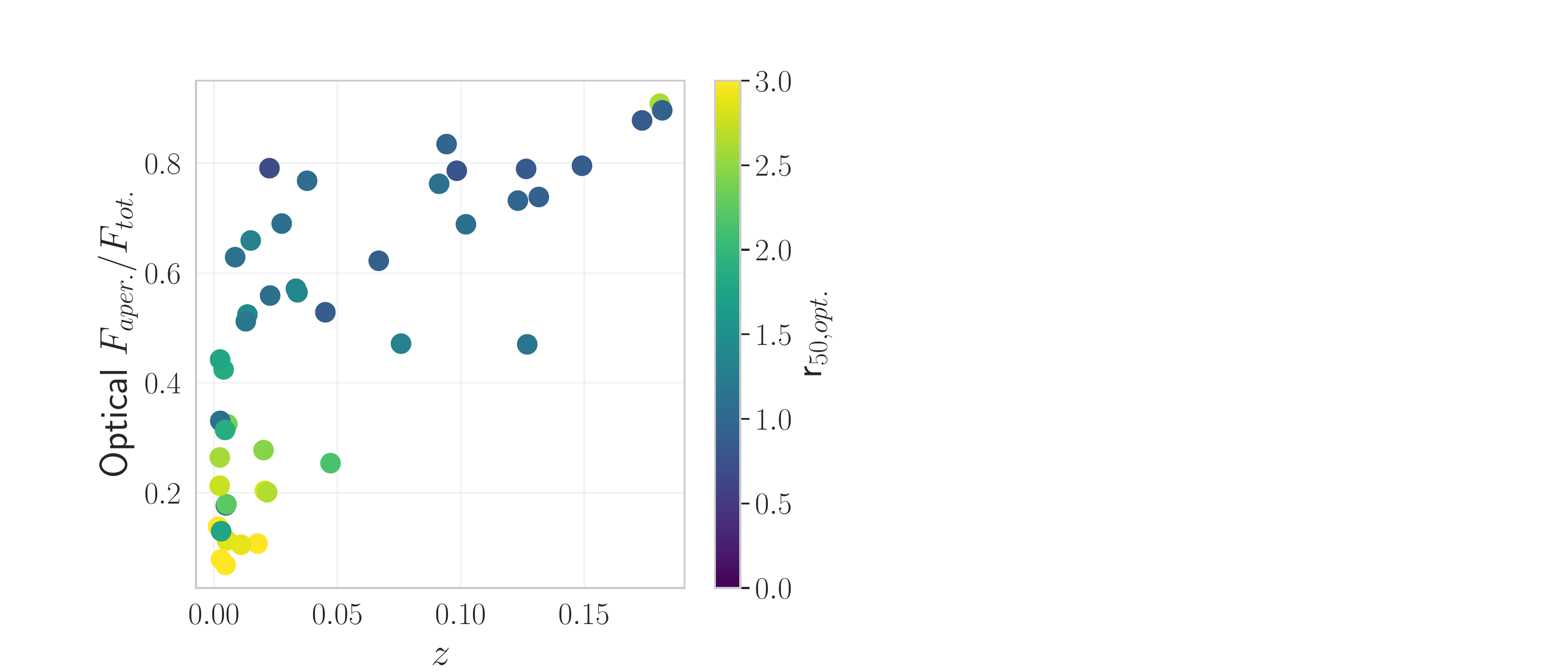} 
\caption{Aperture fraction of optical flux versus redshift for the
CLASSY sample.
The points are color-coded by the optical half light radii in arcseconds.
\label{fig10}}
\end{figure}

\renewcommand{\thetable}{5}
\begin{deluxetable*}{lcccccrlcc}
    \centering
	\setlength{\tabcolsep}{3pt}
    \tablewidth{0pt}
\caption{CLASSY Galaxy Spectroscopy-Derived Properties.}
\tablehead{
\CH{}           & \CH{Alt.} & \CH{R.A., Decl.}          & \CH{}         &\CH{}                  &\CH{}            &\CH{$n_e$} & \CH{}   &\CH{}    \\[-2ex]
\CH{Target}     & \CH{Name} & \CH{(J2000)}              & \CH{$z$}      &\CH{$F_{\lambda1500}$} &\CH{12$+$log(O/H)}   &\CH{(\cmcu)}& \CH{$O_{32}$}  & \CH{\ebv} }
\startdata
1. \ \ J0021$+$0052 &               & 00:21:01.03, $+$00:52:48.08 & 0.09839 & 3.94  & 8.17$\pm$0.07 | \TO & 120$\pm$40 & 2.0$\pm$0.1 & 0.131$\pm$0.006 \\
2. \ \ J0036$-$3333 & \scriptsize{Haro 11 knot}  
                                    & 00:36:52.68, $-$33:33:17.24 & 0.02060 & 16.6  & 8.21$\pm$0.17 | \TS & 10$\pm$10   & 1.1$\pm$0.1 & 0.298$\pm$0.012 \\
3. \ \ J0127$-$0619 & \scriptsize{Mrk 996}       
                                    & 01:27:35.51, $-$06:19:36.06 & 0.00540 & 4.04  & 7.68$\pm$0.02 | \TO & 400$\pm$20  & 1.1$\pm$0.1 & 0.476$\pm$0.006 \\
4. \ \ J0144$+$0453 & \scriptsize{UM133}         
                                    & 01:44:41.36, $+$04:53:25.32 & 0.00520 & 1.87  & 7.76$\pm$0.02 | \TO & 40$\pm$40   & 2.1$\pm$0.1 & 0.044$\pm$0.030 \\
5. \ \ J0337$-$0502 & \scriptsize{SBS0335-052 E} 
                                    & 03:37:44.06, $-$05:02:40.19 & 0.01352 & 7.99  & 7.46$\pm$0.04 | \TS & 140$\pm$40  & 6.2$\pm$0.2 & 0.053$\pm$0.006 \\
6. \ \ J0405$-$3648 &               & 04:05:20.46, $-$36:48:59.14 & 0.00280 & 0.96  & 7.04$\pm$0.05 | \TS & 10$\pm$10   & 0.6$\pm$0.1 & 0.106$\pm$0.005 \\
7. \ \ J0808$+$3948 &               & 08:08:44.28, $+$39:48:52.51 & 0.09123 & 3.42  & 8.77$\pm$0.12 | \TN & 910$\pm$160 & 0.8$\pm$0.1 & 0.241$\pm$0.07 \\
8. \ \ J0823$+$2806 & \scriptsize{LARS9}         
                                    & 08:23:54.96, $+$28:06:21.60 & 0.04722 & 3.85  & 8.28$\pm$0.01 | \TO & 140$\pm$10  & 2.0$\pm$0.1 & 0.209$\pm$0.004 \\
9. \ \ J0926$+$4427 & \scriptsize{LARS14}        
                                    & 09:26:00.44, $+$44:27:36.54 & 0.18067 & 1.14  & 8.08$\pm$0.02 | \TO & 130$\pm$30  & 3.1$\pm$0.1 & 0.104$\pm$0.008 \\
10. J0934$+$5514    & \scriptsize{I Zw 18 NW}      
                                    & 09:34:02.02, $+$55:14:28.10 & 0.00250 & 15.1  & 6.98$\pm$0.01 | \TO & 120$\pm$40  & 8.7$\pm$0.1 & 0.073$\pm$0.007 \\
11. J0938$+$5428    &               & 09:38:13.49, $+$54:28:25.09 & 0.10210 & 3.56  & 8.25$\pm$0.02 | \TO & 80$\pm$10   & 1.9$\pm$0.1 & 0.129$\pm$0.006 \\
12. J0940$+$2935    &               & 09:40:12.87, $+$29:35:30.21 & 0.00168 & 1.45  & 7.66$\pm$0.07 | \TO & 10$\pm$10   & 0.7$\pm$0.1 & 0.060$\pm$0.010 \\
13. J0942$+$3547    & \scriptsize{CG-274, SB 110}  
                                    & 09:42:52.78, $+$35:47:25.98 & 0.01486 & 3.80  & 8.13$\pm$0.03 | \TO & 20$\pm$10   & 2.6$\pm$0.1 & 0.055$\pm$0.011 \\
14. J0944$-$0038    & \scriptsize{CGCG007-025, SB 2}    
                                    & 09:44:01.87, $-$00:38:32.18 & 0.00478 & 1.40  & 7.83$\pm$0.01 | \TO & 100$\pm$20  & 2.9$\pm$0.1 & 0.160$\pm$0.010 \\
15. J0944$+$3442    &               & 09:44:25.87, $+$34:42:08.49 & 0.02005 & 0.69  & 7.62$\pm$0.11 | \TO & 70$\pm$60   & 1.4$\pm$0.1 & 0.162$\pm$0.013 \\
16. J1016$+$3754    & \scriptsize{1427-52996-221}     
                                    & 10:16:24.48, $+$37:54:46.08 & 0.00388 & 7.07  & 7.56$\pm$0.01 | \TO & 40$\pm$20   & 4.6$\pm$0.2 & 0.070$\pm$0.012 \\
17. J1024$+$0524    & \scriptsize{SB 36}             
                                    & 10:24:29.25, $+$05:24:51.02 & 0.03319 & 4.50  & 7.84$\pm$0.03 | \TO & 20$\pm$20   & 2.1$\pm$0.1 & 0.101$\pm$0.016 \\
18. J1025$+$3622    &               & 10:25:48.38, $+$36:22:58.42 & 0.12650 & 1.81  & 8.13$\pm$0.01 | \TO & 100$\pm$20  & 2.4$\pm$0.1 & 0.090$\pm$0.006 \\
19. J1044$+$0353    &               & 10:44:57.79, $+$03:53:13.10 & 0.01287 & 1.70  & 7.45$\pm$0.03 | \TS & 200$\pm$40  & 6.8$\pm$0.1 & 0.079$\pm$0.007 \\
20. J1105$+$4444    & \scriptsize{1363-53053-510}     
                                    & 11:05:08.16, $+$44:44:47.40 & 0.02154 & 4.68  & 8.23$\pm$0.01 | \TO & 100$\pm$10  & 2.0$\pm$0.1 & 0.167$\pm$0.005 \\
21. J1112$+$5503    &               & 11:12:44.05, $+$55:03:47.01 & 0.13164 & 1.91  & 8.45$\pm$0.06 | \TN & 390$\pm$80  & 0.9$\pm$0.1 & 0.225$\pm$0.016 \\
22. J1119$+$5130    &               & 11:19:34.36, $+$51:30:12.02 & 0.00446 & 2.63  & 7.57$\pm$0.04 | \TO & 10$\pm$10   & 2.0$\pm$0.1 & 0.095$\pm$0.008 \\
23. J1129$+$2034    & \scriptsize{SB 179}          
                                    & 11:29:14.15, $+$20:34:52.01 & 0.00470 & 1.87  & 8.28$\pm$0.04 | \TO & 90$\pm$20   & 1.8$\pm$0.1 & 0.227$\pm$0.011 \\
24. J1132$+$5722    & \scriptsize{SBSG1129+576}       
                                    & 11:32:02.64, $+$57:22:36.39 & 0.00504 & 2.57  & 7.58$\pm$0.08 | \TO & 120$\pm$30  & 0.8$\pm$0.1 & 0.095$\pm$0.008 \\
25. J1132$+$1411    & \scriptsize{SB 125}             
                                    & 11:32:35.35, $+$14:11:29.83 & 0.01764 & 1.75  & 8.25$\pm$0.01 | \TO & 90$\pm$20   & 2.7$\pm$0.1 & 0.127$\pm$0.008 \\
26. J1144$+$4012    &               & 11:44:22.28, $+$40:12:21.19 & 0.12695 & 1.20  & 8.43$\pm$0.20 | \TN & 130$\pm$30  & 0.6$\pm$0.1 & 0.223$\pm$0.010 \\
27. J1148$+$2546    & \scriptsize{SB 182}             
                                    & 11:48:27.34, $+$25:46:11.77 & 0.04512 & 2.07  & 7.94$\pm$0.01 | \TO & 120$\pm$30  & 3.7$\pm$0.1 & 0.096$\pm$0.021 \\
28. J1150$+$1501    & \scriptsize{SB 126, Mrk 0750}   
                                    & 11:50:02.73, $+$15:01:23.48 & 0.00245 & 12.6  & 8.14$\pm$0.01 | \TO & 90$\pm$10   & 2.3$\pm$0.1 & 0.039$\pm$0.004 \\
29. J1157$+$3220    & \scriptsize{1991-53446-584}     
                                    & 11:57:31.68, $+$32:20:30.12 & 0.01097 & 14.4  & 8.43$\pm$0.02 | \TO & 70$\pm$10   & 1.2$\pm$0.1 & 0.079$\pm$0.006 \\
30. J1200$+$1343    &               & 12:00:33.42, $+$13:43:07.95 & 0.06675 & 1.38  & 8.26$\pm$0.02 | \TO & 260$\pm$30  & 5.1$\pm$0.1 & 0.146$\pm$0.006 \\
31. J1225$+$6109    & \scriptsize{0955-52409-608}     
                                    & 12:25:05.41, $+$61:09:11.30 & 0.00234 & 9.50  & 7.97$\pm$0.01 | \TO & 50$\pm$20   & 4.7$\pm$0.1 & 0.110$\pm$0.005 \\
32. J1253$-$0312    & \scriptsize{SHOC391}            
                                    & 12:53:05.96, $-$03:12:58.84 & 0.02272 & 9.11  & 8.06$\pm$0.01 | \TO & 390$\pm$50  & 8.0$\pm$0.2 & 0.158$\pm$0.008 \\
33. J1314$+$3452    & \scriptsize{SB 153}             
                                    & 13:14:47.37, $+$34:52:59.81 & 0.00288 & 3.72  & 8.26$\pm$0.01 | \TO & 180$\pm$10  & 2.3$\pm$0.1 & 0.140$\pm$0.006 \\
34. J1323$-$0132    &               & 13:23:47.52, $-$01:32:51.94 & 0.02246 & 1.33  & 7.71$\pm$0.04 | \TO & 600$\pm$140 & 37.8$\pm$3.0 & 0.128$\pm$0.042 \\
35. J1359$+$5726    & \scriptsize{Ly 52, Mrk 1486}
                                    & 13:59:50.88, $+$57:26:22.92 & 0.03383 & 6.34  & 7.98$\pm$0.01 | \TO & 60$\pm$20   & 2.6$\pm$0.1 & 0.091$\pm$0.006 \\
36. J1416$+$1223    &               & 14:16:12.87, $+$12:23:40.42 & 0.12316 & 2.62  & 8.53$\pm$11.4 | \TN & 270$\pm$20  & 0.8$\pm$0.1 & 0.246$\pm$0.008 \\ 
37. J1418$+$2102    &               & 14:18:51.12, $+$21:02:39.84 & 0.00855 & 1.17  & 7.75$\pm$0.02 | \TS & 50$\pm$20   & 4.7$\pm$0.1 & 0.084$\pm$0.006 \\
38. J1428$+$1653    &               & 14:28:56.41, $+$16:53:39.32 & 0.18167 & 1.25  & 8.33$\pm$0.05 | \TO & 90$\pm$20   & 1.2$\pm$0.1 & 0.144$\pm$0.008 \\
39. J1429$+$0643    &               & 14:29:47.00, $+$06:43:34.95 & 0.17350 & 1.62  & 8.10$\pm$0.03 | \TO & 230$\pm$70  & 4.2$\pm$0.2 & 0.116$\pm$0.012 \\
40. J1444$+$4237    & \scriptsize{HS1442+4250}        
                                    & 14:44:11.46, $+$42:37:35.57 & 0.00230 & 2.08  & 7.64$\pm$0.02 | \TO & 90$\pm$40   & 4.1$\pm$0.1 & 0.081$\pm$0.053 \\
41. J1448$-$0110    & \scriptsize{SB 61}              
                                    & 14:48:05.38, $-$01:10:57.72 & 0.02741 & 4.08  & 8.13$\pm$0.01 | \TO & 150$\pm$20  & 8.0$\pm$0.1 & 0.148$\pm$0.005 \\
42. J1521$+$0759    &               & 15:21:41.52, $+$07:59:21.77 & 0.09426 & 3.52  & 8.31$\pm$0.14 | \TN & 90$\pm$30   & 1.5$\pm$0.1 & 0.153$\pm$0.008 \\
43. J1525$+$0757    &               & 15:25:21.84, $+$07:57:20.30 & 0.07579 & 3.52  & 8.33$\pm$0.04 | \To & 210$\pm$20  & 0.5$\pm$0.1 & 0.246$\pm$0.008 \\
44. J1545$+$0858  & \scriptsize{1725-54266-068}     
                                    & 15:45:43.44, $+$08:58:01.34 & 0.03772 & 4.37  & 7.75$\pm$0.03 | \TO & 130$\pm$20  & 8.6$\pm$0.3 & 0.110$\pm$0.036 \\
45. J1612$+$0817    &               & 16:12:45.52, $+$08:17:01.01 & 0.14914 & 2.70  & 8.18$\pm$0.19 | \TN & 340$\pm$60  & 0.7$\pm$0.1 & 0.290$\pm$0.008 
\enddata
\tablecomments{
    The CLASSY sample is composed of UV bright, nearby galaxies covering a range of metallicity, mass, SFR, and gas density.
    Columns 1$-$4 give the target name used in this work, alternative names used, coordinate R.A. and Decl., and redshift.
    Column 5 gives the continuum flux at 1500 \AA\, in units of $10^{-15}$ erg \s\ \cmsq\ \AA$^{-1}$, from the CLASSY coadded spectra. 
    Columns $6-9$ list list additional nebular properties derived from the optical spectra, namely 
    the direct $T_e$-method oxygen abundance (12+log(O/H)) and the ion temperature used,
    the [\ion{S}{2}] electron density, [\ion{O}{3}] \W5007/[\ion{O}{2}] \W3727 flux ratio, and the E(B$-$V) reddening value 
    determined from the Balmer decrement.}
\label{tbl5}
\end{deluxetable*} 

\renewcommand{\thetable}{6}
\begin{deluxetable*}{lcCCCRRRR}
    \centering
	\setlength{\tabcolsep}{3pt}
\caption{CLASSY Galaxy Photometry-Derived Properties.}
\tablehead{
\CH{}       & \CH{R.A., Decl.} & \CH{}    &\CH{$L_D$} & \CH{$r_{50}$}   &\CH{Aper. log \Ms} & \CH{Tot. log \Ms} &\CH{Aper. log SFR}   &\CH{Tot. log SFR}   \\[-2ex]
\CH{Target} & \CH{(J2000)}     & \CH{$z$} &\CH{(Mpc)} & \CH{($\arcsec$)}   &\CH{(\Mo)}         & \CH{(\Mo)}        & \CH{(\Mo\ \yr)} & \CH{(\Mo\ \yr)} }
\startdata
1. \ \ J0021+0052   	& 00:21:01.03, +00:52:48.08 	& 0.09839 & 452 & 0.78 & 8.88$\pm^{0.23}_{0.36}$ & 9.09$\pm^{0.18}_{0.38}$ &  0.99$\pm^{0.21}_{0.14}$ &  1.07$\pm^{0.14}_{0.11}$ \\
2. \ \ J0036-3333   	& 00:36:52.68, -33:33:17.24 	& 0.02060 & 89   & 2.85 & 8.77$\pm^{0.25}_{0.30}$ & 9.14$\pm^{0.26}_{0.23}$ &  0.10$\pm^{0.28}_{0.27}$ &  1.01$\pm^{0.19}_{0.21}$ \\
3. \ \ J0127-0619   	& 01:27:35.51, -06:19:36.06 	& 0.00540 & 23   & 2.37 & 7.38$\pm^{0.23}_{0.30}$ & 8.74$\pm^{0.18}_{0.15}$ & -1.06$\pm^{0.27}_{0.33}$ & -0.75$\pm^{0.15}_{0.13}$ \\
4. \ \ J0144+0453   	& 01:44:41.36, +04:53:25.32 	& 0.00520 & 23   & 2.85 & 5.61$\pm^{0.32}_{0.35}$ & 7.65$\pm^{0.24}_{0.29}$ & -1.86$\pm^{0.11}_{0.09}$ & -0.81$\pm^{0.29}_{0.46}$ \\
5. \ \ J0337-0502   	& 03:37:44.06, -05:02:40.19 	& 0.01352 & 58   & 1.43 & 6.89$\pm^{0.29}_{0.34}$ & 7.06$\pm^{0.24}_{0.21}$ & -0.57$\pm^{0.15}_{0.13}$ & -0.32$\pm^{0.07}_{0.11}$ \\
6. \ \ J0405-3648   	& 04:05:20.46, -36:48:59.14 	& 0.00280 & 11   & 3.56 & 5.34$\pm^{0.27}_{0.33}$ & 6.61$\pm^{0.28}_{0.28}$ & -2.96$\pm^{0.28}_{0.22}$ & -1.81$\pm^{0.31}_{0.27}$ \\
7. \ \ J0808+3948   	& 08:08:44.28, +39:48:52.51	& 0.09123 & 417 & 1.11 & 8.85$\pm^{0.32}_{0.23}$ & 9.12$\pm^{0.30}_{0.17}$ &  0.94$\pm^{0.20}_{0.22}$ &  1.26$\pm^{0.18}_{0.25}$ \\
8. \ \ J0823+2806   	& 08:23:54.96, +28:06:21.60 	& 0.04722 & 209 & 2.12 & 8.47$\pm^{0.28}_{0.27}$ & 9.38$\pm^{0.33}_{0.19}$ &  0.57$\pm^{0.26}_{0.23}$ &  1.48$\pm^{0.15}_{0.32}$ \\
9. \ \ J0926+4427   	& 09:26:00.44, +44:27:36.54 	& 0.18067 & 875 & 0.89 & 8.75$\pm^{0.31}_{0.28}$ & 8.76$\pm^{0.30}_{0.26}$ &  0.97$\pm^{0.18}_{0.15}$ &  1.03$\pm^{0.13}_{0.13}$ \\
10. J0934+5514      	& 09:34:20.02, +55:14:28.10 	& 0.00250 & 12   & 2.61 & 5.48$\pm^{0.37}_{0.39}$ & 6.27$\pm^{0.15}_{0.20}$ & -2.09$\pm^{0.11}_{0.08}$ & -1.52$\pm^{0.09}_{0.07}$ \\
11. J0938+5428      	& 09:38:13.49, +54:28:25.09 	& 0.10210 & 470 & 1.10 & 8.98$\pm^{0.23}_{0.46}$ & 9.15$\pm^{0.18}_{0.29}$ &  1.07$\pm^{0.25}_{0.26}$ &  1.05$\pm^{0.20}_{0.17}$ \\
12. J0940+2935      	& 09:40:12.87, +29:35:30.21 	& 0.00168 & 8     & 5.15 & 5.05$\pm^{0.36}_{0.21}$ & 6.71$\pm^{0.23}_{0.40}$ & -3.05$\pm^{0.19}_{0.12}$ & -2.01$\pm^{0.42}_{0.37}$ \\
13. J0942+3547      	& 09:42:52.78, +35:47:25.98 	& 0.01486 & 65   & 1.33 & 6.75$\pm^{0.43}_{0.40}$ & 7.56$\pm^{0.21}_{0.29}$ & -0.90$\pm^{0.14}_{0.11}$ & -0.76$\pm^{0.19}_{0.12}$ \\
14. J0944-0038      	& 09:44:01.87, -00:38:32.18 	& 0.00478 & 21   & 0.98 & 5.58$\pm^{0.38}_{0.21}$ & 6.83$\pm^{0.44}_{0.25}$ & -1.49$\pm^{0.21}_{0.15}$ & -0.78$\pm^{0.19}_{0.16}$ \\
15. J0944+3442      	& 09:44:25.87, +34:42:08.49 	& 0.02005 & 87   & 2.46 & 7.62$\pm^{0.31}_{0.28}$ & 8.19$\pm^{0.40}_{0.23}$ & -0.83$\pm^{0.27}_{0.43}$ & -0.01$\pm^{0.28}_{0.65}$ \\
16. J1016+3754      	& 10:16:24.48, +37:54:46.08 	& 0.00388 & 18   & 1.84 & 5.98$\pm^{0.28}_{0.36}$ & 6.72$\pm^{0.27}_{0.22}$ & -1.62$\pm^{0.13}_{0.10}$ & -1.17$\pm^{0.18}_{0.18}$ \\
17. J1024+0524      	& 10:24:29.25, +05:24:51.02 	& 0.03319 & 145 & 1.33 & 7.51$\pm^{0.37}_{0.28}$ & 7.89$\pm^{0.37}_{0.24}$ & -0.01$\pm^{0.19}_{0.12}$ &  0.21$\pm^{0.14}_{0.12}$ \\
18. J1025+3622      	& 10:25:48.38, +36:22:58.42 	& 0.12650 & 592 & 0.84 & 8.81$\pm^{0.27}_{0.25}$ & 8.87$\pm^{0.25}_{0.27}$ &  0.80$\pm^{0.21}_{0.16}$ &  1.04$\pm^{0.14}_{0.18}$ \\
19. J1044+0353      	& 10:44:57.79, +03:53:13.10 	& 0.01287 & 55   & 1.20 & 6.13$\pm^{0.23}_{0.09}$ & 6.80$\pm^{0.41}_{0.26}$ & -0.90$\pm^{0.07}_{0.11}$ & -0.59$\pm^{0.11}_{0.14}$ \\
20. J1105+4444      	& 11:05:08.16, +44:44:47.40 	& 0.02154 & 93   & 2.65 & 7.76$\pm^{0.35}_{0.25}$ & 8.98$\pm^{0.29}_{0.24}$ & -0.05$\pm^{0.22}_{0.17}$ &  0.69$\pm^{0.28}_{0.22}$ \\
21. J1112+5503      	& 11:12:44.05, +55:03:47.01 	& 0.13164 & 618 & 0.92 & 9.55$\pm^{0.26}_{0.28}$ & 9.59$\pm^{0.33}_{0.19}$ &  1.39$\pm^{0.22}_{0.30}$ &  1.60$\pm^{0.20}_{0.25}$ \\
22. J1119+5130      	& 11:19:34.36, +51:30:12.02 	& 0.00446 & 20   & 1.87 & 6.24$\pm^{0.25}_{0.28}$ & 6.77$\pm^{0.15}_{0.28}$ & -2.02$\pm^{0.26}_{0.17}$ & -1.58$\pm^{0.21}_{0.12}$ \\
23. J1129+2034      	& 11:29:14.15, +20:34:52.01 	& 0.00470 & 21   & 3.10 & 5.75$\pm^{0.36}_{0.40}$ & 8.09$\pm^{0.37}_{0.27}$ & -1.74$\pm^{0.15}_{0.12}$ & -0.37$\pm^{0.38}_{0.56}$ \\
24. J1132+5722      	& 11:32:02.64, +57:22:36.39 	& 0.00504 & 23   & 7.29 & 6.02$\pm^{0.18}_{0.35}$ & 7.31$\pm^{0.23}_{0.26}$ & -2.31$\pm^{0.28}_{0.19}$ & -1.07$\pm^{0.27}_{0.35}$ \\
25. J1132+1411      	& 11:32:35.35, +14:11:29.83 	& 0.01764 & 76   & 2.25 & 6.79$\pm^{0.41}_{0.28}$ & 8.68$\pm^{0.28}_{0.19}$ & -0.58$\pm^{0.13}_{0.10}$ &  0.44$\pm^{0.24}_{0.27}$ \\
26. J1144+4012      	& 11:44:22.28, +40:12:21.19 	& 0.12695 & 594 & 1.16 & 9.41$\pm^{0.29}_{0.22}$ & 9.89$\pm^{0.18}_{0.29}$ &  1.09$\pm^{0.29}_{0.41}$ &  1.51$\pm^{0.20}_{0.29}$ \\
27. J1148+2546      	& 11:48:27.34, +25:46:11.77 	& 0.04512 & 199 & 0.87 & 7.39$\pm^{0.46}_{0.23}$ & 8.14$\pm^{0.34}_{0.24}$ &  0.29$\pm^{0.18}_{0.15}$ &  0.53$\pm^{0.17}_{0.14}$ \\
28. J1150+1501      	& 11:50:02.73, +15:01:23.48 	& 0.00245 & 11   & 1.76 & 5.69$\pm^{0.38}_{0.30}$ & 6.84$\pm^{0.28}_{0.30}$ & -1.70$\pm^{0.11}_{0.09}$ & -1.33$\pm^{0.29}_{0.23}$ \\
29. J1157+3220      	& 11:57:31.68, +32:20:30.12 	& 0.01097 & 48   & 2.89 & 7.33$\pm^{0.41}_{0.31}$ & 9.04$\pm^{0.32}_{0.18}$ & -0.47$\pm^{0.19}_{0.13}$ &  0.97$\pm^{0.21}_{0.42}$ \\
30. J1200+1343      	& 12:00:33.42, +13:43:07.95 	& 0.06675 & 300 & 0.98 & 7.46$\pm^{0.59}_{0.21}$ & 8.12$\pm^{0.47}_{0.42}$ &  0.38$\pm^{0.21}_{0.15}$ &  0.75$\pm^{0.20}_{0.16}$ \\
31. J1225+6109      	& 12:25:05.41, +61:09:11.30 	& 0.00234 & 11   & 2.60 & 5.79$\pm^{0.35}_{0.28}$ & 7.12$\pm^{0.34}_{0.24}$ & -1.81$\pm^{0.15}_{0.13}$ & -1.08$\pm^{0.26}_{0.26}$ \\
32. J1253-0312      	& 12:53:05.96, -03:12:58.84 	& 0.02272 & 99   & 1.08 & 7.46$\pm^{0.33}_{0.25}$ & 7.65$\pm^{0.51}_{0.23}$ &  0.37$\pm^{0.15}_{0.18}$ &  0.56$\pm^{0.15}_{0.15}$ \\
33. J1314+3452     	& 13:14:47.37, +34:52:59.81 	& 0.00288 & 13   & 1.77 & 5.28$\pm^{0.25}_{0.17}$ & 7.56$\pm^{0.30}_{0.21}$ & -1.78$\pm^{0.14}_{0.14}$ & -0.67$\pm^{0.23}_{0.55}$ \\
34. J1323-0132      	& 13:23:47.52, -01:32:51.94 	& 0.02246 & 97   & 0.70 & 6.22$\pm^{0.12}_{0.09}$ & 6.31$\pm^{0.26}_{0.10}$ & -0.78$\pm^{0.10}_{0.09}$ & -0.72$\pm^{0.08}_{0.09}$ \\
35. J1359+5726     	& 13:59:50.88, +57:26:22.92 	& 0.03383 & 148 & 1.40 & 8.11$\pm^{0.36}_{0.30}$ & 8.41$\pm^{0.31}_{0.26}$ &  0.20$\pm^{0.25}_{0.18}$ &  0.42$\pm^{0.20}_{0.14}$ \\
36. J1416+1223      	& 14:16:12.87, +12:23:40.42 	& 0.12316 & 575 & 0.99 & 9.46$\pm^{0.23}_{0.30}$ & 9.59$\pm^{0.32}_{0.26}$ &  1.23$\pm^{0.24}_{0.26}$ &  1.57$\pm^{0.21}_{0.25}$ \\
37. J1418+2102      	& 14:18:51.12, +21:02:39.84 	& 0.00855 & 37   & 1.13 & 5.71$\pm^{0.48}_{0.21}$ & 6.22$\pm^{0.49}_{0.35}$ & -1.38$\pm^{0.13}_{0.15}$ & -1.13$\pm^{0.15}_{0.16}$ \\
38. J1428+1653      	& 14:28:56.41, +16:53:39.32 	& 0.18167 & 880 & 0.93 & 9.50$\pm^{0.19}_{0.30}$ & 9.56$\pm^{0.15}_{0.23}$ &  1.14$\pm^{0.27}_{0.23}$ &  1.22$\pm^{0.26}_{0.19}$ \\
39. J1429+0643      	& 14:29:47.00, +06:43:34.95 	& 0.17350 & 836 & 0.86 & 8.46$\pm^{0.37}_{0.21}$ & 8.80$\pm^{0.35}_{0.21}$ &  1.29$\pm^{0.11}_{0.12}$ &  1.42$\pm^{0.11}_{0.17}$ \\
40. J1444+4237      	& 14:44:11.46, +42:37:35.57 	& 0.00230 & 11   & 2.76 & 4.91$\pm^{0.33}_{0.23}$ & 6.48$\pm^{0.17}_{0.17}$ & -2.41$\pm^{0.11}_{0.11}$ & -1.94$\pm^{0.11}_{0.08}$ \\
41. J1448-0110      	& 14:48:05.38, -01:10:57.72 	& 0.02741 & 119 & 1.07 & 7.29$\pm^{0.54}_{0.30}$ & 7.61$\pm^{0.41}_{0.24}$ &  0.06$\pm^{0.20}_{0.13}$ &  0.39$\pm^{0.13}_{0.14}$ \\
42. J1521+0759      	& 15:21:41.52, +07:59:21.77 	& 0.09426 & 432 & 0.98 & 8.90$\pm^{0.35}_{0.26}$ & 9.00$\pm^{0.29}_{0.30}$ &  0.85$\pm^{0.21}_{0.21}$ &  0.95$\pm^{0.16}_{0.17}$ \\
43. J1525+0757      	& 15:25:21.84, +07:57:20.30 	& 0.07579 & 343 & 1.32 & 9.13$\pm^{0.31}_{0.24}$ & 10.06$\pm^{0.28}_{0.42}$&  0.87$\pm^{0.28}_{0.34}$ &  1.00$\pm^{0.69}_{0.24}$ \\
44. J1545+0858      	& 15:45:43.44, +08:58:01.34 	& 0.03772 & 166 & 1.08 & 7.39$\pm^{0.24}_{0.20}$ & 7.52$\pm^{0.43}_{0.26}$ &  0.32$\pm^{0.19}_{0.18}$ &  0.37$\pm^{0.13}_{0.17}$ \\
45. J1612+0817      	& 16:12:45.52, +08:17:01.01 	& 0.14914 & 708 & 0.88 & 9.74$\pm^{0.18}_{0.26}$ & 9.78$\pm^{0.28}_{0.26}$ &  1.43$\pm^{0.27}_{0.28}$ &  1.58$\pm^{0.28}_{0.24}$ 
\enddata
\tablecomments{
    CLASSY sample properties derived from UV$+$optical photometry,
    as described in Section~\ref{sec:4.6}. 
    Columns 1$-$3 give the target name used in this work, coordinate R.A. and Decl., and redshift.
    Column 4 lists the luminosity distance, corrected for the Hubble flow using the velocity
    field model of \citet{masters05}.
    Column 5 gives the half light radii in arcseconds, as derived from the optical imaging (see Figure~\ref{fig2}).
    Columns 6 and 7 list the stellar masses derived using \texttt{Beagle} SED fitting for both the 
    light within the COS aperture and the entire galaxy, respectively, of a given target.
    Similarly, columns 8 and 9 provide the SED-derived SFRs.}
\label{tbl6}
\end{deluxetable*}



\section{SED Fitting}\label{AppendixC}

Here we further describe the SED fitting method used to determine the stellar masses and star
formation rates presented in Section~\ref{sec:4.7}.
SED fitting was performed using the BayEsian Analysis of GaLaxy sEds \citep[\texttt{BEAGLE, v0.24.0},][]{chevellard16}.
\texttt{BEAGLE} is a new-generation SED tool that incorporates several novel approaches over past codes. 
First, the modular design of \texttt{BEAGLE} allows a physically consistent combination of different 
prescriptions for the production of starlight in galaxies and its transfer through the ISM 
(absorption and emission by gas, attenuation by dust) and the IGM (absorption by gas). 
Second, \texttt{BEAGLE} uses a flexible parameterization of several prescriptions to describe the 
star formation and chemical enrichment histories of galaxies, ranging from simple analytic 
functions to the predictions of sophisticated galaxy formation models, to interpret 
combinations of photometric and spectroscopic observation of galaxies.
As a result, \texttt{BEAGLE} fits can provide a physically consistent description of the contributions 
by stars, gas, and dust to the integrated emission from a galaxy.
Note that this updated SED fitting approach is particularly important for the subset
of CLASSY galaxies with very strong nebular continuum and line emission as
\texttt{BEAGLE} self-consistently accounts for these contributions in the synthetic model spectrum.
Using \texttt{BEAGLE}, two sets of stellar mass and SFR parameters were computed:
(1) an aperture set which provides insights for the star-forming regions specifically targeted 
within the COS 2\farcs5 aperture and 
(2) a host galaxy set which better characterizes the properties of the entire galaxies.

\subsection{Choice of Photometry}

The SDSS provides optical imaging in $ugriz$ for 42 galaxies in the CLASSY sample.
For the other three, J0036-3333 and J0405-3648 have $grizY$ images from the Dark Energy 
Survey \citep[DES,][]{des21}, and J0337-0502 is observed by the Pan-STARRS1 
\citep[PS1,][]{chambers16}.
For the UV, 43 galaxies (all but J0337-0502 and J1157+3220) have imaging in the NUV and 
FUV available from \citep[GALEX,][]{martin05}.
Additionally, WISE W1 and W2 near-infrared (NIR) data are also available for the entire CLASSY sample.
However, WISE photometry introduces large model uncertainties due to the contribution of
post-main-sequence stars to NIR light \citep[e.g.,][]{salim16} and does not significantly 
affect the CLASSY stellar mass ($\Delta_{\rm{log}\ M_\star} < 0.31$ dex) and 
SFR ($\Delta_{\rm{log SFR}} < 0.21$ dex) measurements, 
and so was not included in the CLASSY SED fitting.

\subsection{New Flux Measurements}

Due to the irregular morphologies and range of redshifts of the CLASSY sample,
the SDSS catalog photometry can have large uncertainties when a simple, 
symmetric model is assumed or only a portion (e.g., a single \ion{H}{2} region) 
of a host galaxy is measured.
We, therefore, remeasured the GALEX and SDSS fluxes. 
For the aperture stellar mass and SFR determinations,
$ugriz$ or $grizY$ photometry was measured for targeted star-forming regions in a 
3\arcsec{} aperture, well-matched to the size of COS aperture convolved with typical SDSS 
seeing \cite[see, e.g.,][]{senchyna19a}.
The sky background was estimated in a 1\arcsec{}-wide circular annulus located at 
1\arcmin{} using the median flux values after sigma-clipping (low 10$\sigma$, high 3$\sigma$). 
Corrections for Galactic foreground extinction were then applied, assuming an $R_V$ = 3.1 and
Milky Way extinction curve \citet{cardelli89}.
The GALEX UV imaging was not used for these aperture calculations due to its relatively 
poor resolution (4\farcs5 -- 6\farcs0 FWHM).

For the total stellar mass and SFR determinations of each galaxy,
total fluxes in the UV and optical were measured, but with a more sophisticated method 
to account for irregular morphologies and potential contamination from nearby objects.
First, segmentation maps were produced for each galaxy using the \texttt{Photutils} 
package \citep{bradley21}.
For the optical imaging, $g$-band segmentation was computed, which includes strong optical 
emission lines such as \fOTh\ and (or) H\B\ for the majority of the CLASSY sample.
This maximizes the ability to probe the often faint and extended nebular emission,
which are incorporated into the \texttt{BEAGLE} models.
To produce the $g$-band segmentation maps, a minimum of 8 connected pixels at 2$\sigma$ 
above the background after the image is smoothed by a 2D Gaussian kernel (FWHM = 10 pix) 
is required.
Similarly, NUV segmentation was computed for the UV photometry, but using a Gaussian 
kernel with FWHM = 3 pix.
Each map was visually inspected to ensure an adequate segmentation that captures the 
total galaxy light in both the optical and UV.
If a map failed the visual inspection, the segmentation parameters were adjusted and source
deblending was implemented in \texttt{Photutils} to exclude contamination.
Second, the $g$-band and NUV segmentations were used to measure optical photometry in 
$ugriz$ (or $grizY$) and NUV and FUV photometry, respectively.
In a similar manner as the aperture determinations, background subtraction and Galactic 
extinction corrections were applied.
In general, we found that the 3\arcsec\ aperture flux accounts for $\sim25\%$ of the total 
g-band flux at $z \sim 0.01$, and $\sim80\%$ at $z \sim 0.17$.

Finally, galaxy distances and absolute magnitudes were determined. 
Since nearly half of the sample (20/45) are very nearby galaxies with optical spectroscopic 
redshifts of $z<0.02$, they may have peculiar motions that will bias the luminosity 
distances derived from redshifts.
To uniformly estimate their distances, the heliocentric redshifts were first converted to 
the cosmic microwave background (CMB) frame using the CMB dipole measured by 
\cite{lineweaver96}.
The local velocity field modelled by \cite{masters05} was then adopted to compute the 
flow-corrected luminosity distances, with the amount of correction ranging from $< 1\%$  
to $20\%$ (median $6\%$).\looseness=-2

\subsection{SED Model Assumptions}

With the redshifts, corrected luminosity distances, and photometry for both the targeted 
regions and entire galaxies in hand, the \Ms\ and SFRs can be self-consistently constrained 
with the \texttt{BEAGLE} code.
Briefly, the \cite{gutkin16} model was adopted, which self-consistently incorporates the 
stellar emission from the latest version of the \cite{bruzual03} stellar population 
synthesis results and the nebular emission output by the photoionization code 
\texttt{Cloudy} \citep{ferland13}.
We assumed a \cite{chabrier03} Galactic disk initial mass function 
(mass range 0.1 -- 100 $M_\odot$), the two-component attenuation model 
from \cite{charlot00}, and a constant star formation history.
Free parameters are thus the stellar mass, age, metallicity, 
gas ionization parameter, log $U$, and effective attenuation $\hat{\tau}_V$. 
The interstellar (dust$+$gas-phase) metallicity was matched to the stellar 
metallicity of the constant star formation.
Of the three choices of dust-to-metal mass ratio provided by \texttt{BEAGLE}
($\xi_d=0.1, 0.3, 0.5$), we assume $\xi_d=0.3$ as the value closest to the solar value 
\citep[$\xi_d=0.36$][]{gutkin16}.

The \texttt{BEAGLE} SED fits for two CLASSY galaxies, local dwarf galaxy J0944-0383 and
$z\sim0.1$ emission-line galaxy J1416+1223, are shown in Figure~\ref{fig11}.
For comparison, the fit to the host galaxy UV$+$optical photometry is shown in the top panels
relative to the aperture optical photometry fit in the bottom panels.
Contamination to the broad-band flux from strong emission lines is clearly seen in the $gr$ bands 
for local systems (e.g., J0944-0038) or $r$ band for those at slightly high redshift 
(e.g., J1416+1223).
These models demonstrate the importance of properly accounting for strong nebular emission 
in our SED fitting with a self-consistent program like \texttt{BEAGLE}.
 
\renewcommand{\thefigure}{11}
\begin{figure*}
    \centering
    \includegraphics[width=1.0\textwidth]{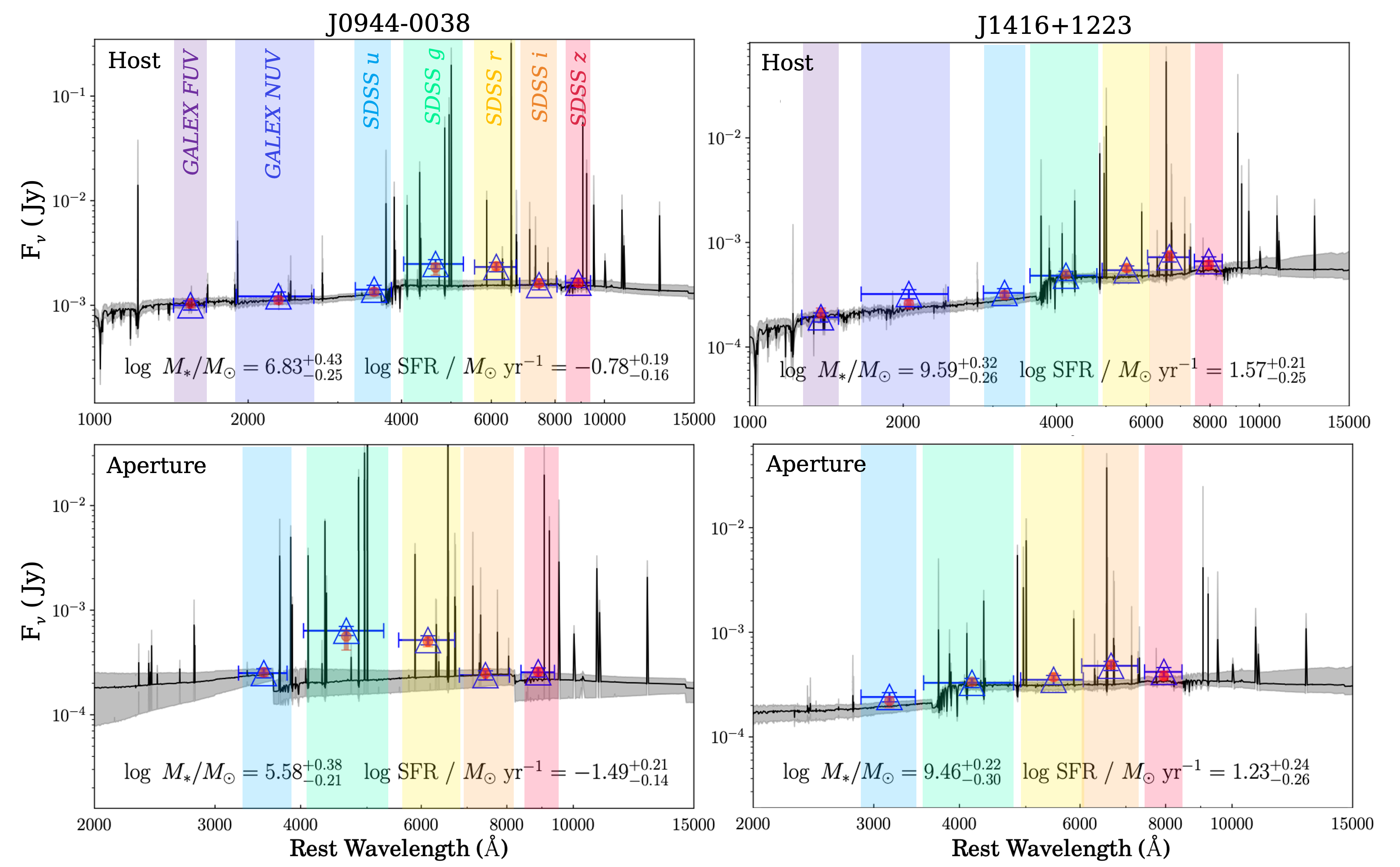}
    \caption{
    Example SED fits for two of the CLASSY galaxies, one local dwarf galaxy (J0944-0038, left) 
    and one emission-line galaxy at $z\approx0.1$ (J1416+1223, right).
    The fits to host galaxies are shown in the top row while the aperture fits are in the bottom.
    Blue triangles are the measured fluxes from broad-band photometry, and red circles give BEAGLE predictions.
    The median \texttt{BEAGLE} SED and 16-84 percentile of posterior distribution are shown in black line and gray-shaded area, respectively. 
    The significant excess to the broad-band flux in the $gr$ bands of J0944-0038 and $r$ band of J1416+1223
    is due to strong nebular emission lines, which are accounted for by \texttt{BEAGLE}.}
    \label{fig11}
\end{figure*}
From the output posterior distribution, the \Ms\ and SFRs were computed for both the 
targeted aperture regions and the entire galaxies.
The resulting stellar masses and SFRs are reported in Table~\ref{tbl6} in the Appendix
for both the aperture and total host galaxy sets.

\subsection{Choice of Star Formation History}
It is very difficult to recover accurate values of the SFRs through SED fitting for 
realistic galaxies, in large part due to their unconstrained SFHs.
Fortunately, previous SED fits using a constant SFH were shown to perform well for a subset of 
the CLASSY sample \citep{senchyna17, senchyna19a}, and allowed for a consistent comparison 
with the SED fitting conducted for high-redshift galaxies.
Following this method, we have assumed a constant SFH, which benefits from having the least free parameters,
yet provides satisfying fits to the broad-band photometry even when strong emission-lines 
are present. 
However, an important caveat is the potentially hidden old stellar populations in the CLASSY
galaxies, especially outside the burst of star formation targeted by the COS aperture. 
Our assumption of a simple constant SFH thus inevitably introduces additional systematic uncertainties 
to our estimates, often leading to a lower stellar mass compared to other more complicated SFHs.

\subsection{Exploration of SED Systematics}
To better understand the accuracy and precision of our SED-derived galaxy properties and
to provide consistency checks, we performed the following tests: 

{\it 1. Impact of assumed SFH:}
We investigated the impact of assuming just a simple constant SFH by comparing to more 
complicated models.
First, we allowed composite SFHs consisting of a constant SFH plus a single burst in the 
fit in order to allow the maximum flexibility for \texttt{BEAGLE} to properly account 
for any possible old stellar populations. 
In most cases, \texttt{BEAGLE} favored either a dominant constant SFH component or 
a dominant SSP component.
The resulting fitted masses are larger relative to the values assuming a constant SFH 
by $0.01^{+0.05}_{-0.04}$ dex.
Second, since a red component from an older stellar population would be primarily arise 
from stars outside the aperture, we also tested fitting the flux inside and outside 
aperture separately using two populations of constant star formation, one inside and 
one outside the aperture.
Using the four sources in our sample showing the most extreme extended components outside,
we found differences in both stellar masses and SFRs to be $\lesssim 1 \sigma$ and within the 
uncertainties. 
However, \citet{lower20} recently investigated the impact of SFH on SED fitting and found that 
assuming a constant SFH can underestimate the stellar mass by $0.48\pm0.61$ dex.
Unfortunately, other simple parametric models such as a delayed-$\tau$ SFH\footnote{
The common $\tau$-model, parameterized by $\dot{M}_\star \propto e^{-t/\tau}$,
has an exponentially-declining SFH with some characteristic e-folding time, $\tau$.
The delayed-$\tau$ model,
parameterized by $\dot{M}_\star \propto t\cdot e^{-t/\tau}$,
is an expansion of the $\tau$-model such that it
allows for linear growth at early times in addition to exponential decline at later times.}
also performed poorly in this work;
only complicated, non-parametric SFHs produce realistic values. 
Thus, for simplicity and for consistent comparison with high-redshift galaxies, 
a simple constant SFH was adopted for the CLASSY SED fitting, resulting in 
relatively large uncertainties.
	
{\it 2. Comparisons to other catalogs:}
We compared the stellar masses and SFRs to the values from the MPA-JHU catalog\footnote{
Data catalogs are available at \url{http://www.mpa-garching.mpg.de/SDSS/}. 
The Max Plank institute for Astrophysics/John Hopkins University(MPA/JHU) SDSS database 
was produced by a collaboration of researchers (currently or formerly) from the MPA and the JHU. 
The team is made up of Stephane Charlot (IAP), Guinevere Kauffmann and Simon White (MPA), 
Tim Heckman (JHU), Christy Tremonti (U. Wisconsin-Madison—formerly JHU), and Jarle Brinchmann 
(Universidade do Porto—formerly MPA).}.
On average, the CLASSY stellar masses are smaller by $-0.28$ dex, while the 16th and 84th 
percentile differences from the MPA-JHU values are $-0.60$ and $0.40$ dex, respectively.
For the SFRs, the CLASSY values are larger on average by 0.18 dex, while the 16th and 84th
percentile differences from the MPA-JHU values are $-0.09$ and $0.63$ dex, respectively.
While we find generally good agreement with the MPA-JHU values, there are some significant outliers
due to many differing assumptions in the photometry and SED modeling.
However, dust extinction is unlikely to be the cause given the strong agreement between the CLASSY aperture (host) 
values of effective optical depth ($\tau_{\rm V,eff}$) and those derived from optical spectra 
($\Delta \tau_{\rm V,eff} = -0.06_{-0.10}^{+0.14}$ ($0.09_{-0.22}^{+0.53}$)).

We also compared to the the GALEX-SDSS-WISE Legacy Catalog of \citet[][GSWLC]{salim16}.
The GSWLC contains SDSS galaxies more distant on average than the CLASSY sample 
at $0.01 < z < 0.30$ and $r_{\rm petro} < 18$. 
These distances ensure that the SDSS photometry are safe to use for SED fitting. 
A cross-match between the CLASSY sample and GSWLC yields 22 matches, of which 12 have 
GSWLC SED fits with $\chi_{\rm reduced}>30$.
For the remaining 10 galaxies, 4 have GSWLC SED fits with $\chi_{\rm reduced}>10$.
The large fraction of poor fits in GSWLC are likely due to its inability to account for 
large broad-band excess caused by strong nebular emission lines.
For the 6 galaxies with GSWLC $\chi_{\rm reduced}<10$, 
we find differences ($\Delta$ = CLASSY values $-$ GSWLC values) of
$\Delta M_{\star,host} = -0.27_{-0.39}^{+0.33}$ dex and
$\Delta SFR_{host} = 0.12_{-0.11}^{+0.13}$ dex.
On average, the CLASSY SFRs are larger by $\sim0.1$ dex compared to GSWLC1 SED SFRs, 
but smaller by $\sim0.2$ dex compared to GSWLC1 WISE SFRs. 
Extreme outliers and the sources of such discrepancies will be explored
in a future paper. 

{\it 3. Comparison to H$\alpha$ SFRs:}
We calculated SFRs in another way, using the reddening-corrected 
H$\alpha$ flux measured from the optical spectra, the luminosity distances 
listed in Table~\ref{tbl6}, and the SFR calibration from \citet{kennicutt12}.
For galaxies with optical spectra from the SDSS, we use the H$\alpha$ flux
values from the MPA-JHU catalog, where each spectrum has been tied back to the 
SDSS $r$-band \texttt{fibermag}.
Note that this H$\alpha$ SFR calibration assumes a \citet{kroupa03} IMF, 
which is consistent with the \citet{chabrier03} IMF used for the SED fitting \citep[e.g.,][]{chomiuk11}.
In comparison to the SED-derived SFRs, we find the H$\alpha$ SFRs to be 
larger on average, with a typical percent difference of $8$\%, and a few 
strong outliers extending from $-202$\% to $+48$\%.

{\it 4. Comparison to Nebular Properties:}
Finally, we compared the 12+log(O/H) and log$U$ values from the BEAGLE SED fits to those
derived from the optical spectra (see Sections~\ref{sec:4.5} and \ref{sec:4.6}).
Of our three comparisons (other catalogs, SFRs, and nebular properties), the latter shows 
the largest discrepancies, where the SED metallicities (ionization parameters) are 
systematically lower (higher) than the measured values, with median differences of
0.37 dex ($-0.36$ dex). 
However, the \texttt{BEAGLE} values have large uncertainties ($> 0.5$ dex) by design that give it
the flexibility necessary to test large parameter space and thus are still generally consistent
with the measured CLASSY values.
Note that the choice to not fix metallicity, ionization parameter, or dust extinction to spectroscopic 
values was deliberate in order to not introduce other systematic biases
(i.e., the burst of star formation within the spectroscopic aperture may have different ISM conditions than
the outer regions of a galaxy). 


\section{Notes on Individual Galaxies}\label{AppendixD}
For transparency and to improve usability of the CLASSY HLSPs,
we mention here any unique notes for individual galaxies.

\smallskip
\noindent{\it J0036-3333}: 
Also commonly known as the galaxy Haro~11, which contains multiple 
stellar populations. 
The CLASSY coadded spectrum only focuses on knot C \citep{vader93}.
Knot C is the most UV luminous source and main contributor to the 
Lyman alpha emission from Haro~11 \citep[][]{hayes07}.

\smallskip
\noindent{\it J0127-0619}:
The common name of J0127-0619 is Mrk 996.
This galaxy is known to have complex emission line kinematics, 
a large Wolf Rayet population, and a high electron density within its 
inner core region \citep[][]{james09,telles14}.

\smallskip
\noindent{\it J0144+0453}:
J0144+0453 is more commonly known as UM133 and has 
an extended tail to the North.

\smallskip
\noindent{\it J0337-0502}:
J0337-0502 is part of the well-known, metal-poor galaxy SBS~0335-052
that has two main star-forming clusters, E and W.
The CLASSY coadded spectrum only contains light from the E region.
It also combines data from COS FUV detector lifetime positions 3 and 4.
SBS 0335-052E consists of six young ($10<$ Myr) star clusters 
\citep[][]{adamo10}, four of which fall within the COS aperture 
\citep[][]{wofford21}.

\smallskip
\noindent{\it J0405-3648}: 
J0405-3648 is fairly extended, with many small ionized clumps seen in the 
optical imaging.
The CLASSY coadded spectrum combines data taken at position angles
that are offset by 50$^{\circ}$.

\smallskip
\noindent{\it J0808+3948}: 
The CLASSY coadded spectrum combines data taken at position angles
that are offset by 84$^{\circ}$.

\smallskip
\noindent{\it J0823+2806}: 
J0823+2806 is part of the Lyman Alpha Reference Sample 
\citep[LARS;][]{hayes14,ostlin14} and is identified as LARS9.
The CLASSY coadded spectrum combines data taken at position angles
that are offset by 34$^{\circ}$.
The COS aperture is dominated by emission from a single massive star cluster.

\smallskip
\noindent{\it J0926+4427}: 
J0926+4427 is part of the LARS and is identified as LARS14.
LARS14 is classified as a Lyman break analog \citep{heckman11}, 
and also a Green Pea galaxy \citep{henry15}.     
The CLASSY coadded spectrum combines data taken at position angles
that are offset by 82$^{\circ}$.

\smallskip
\noindent{\it J0934+5514}: 
This is the famous low-metallicity dwarf galaxy, I~Zw~18, which
is also known as Mrk 116 or UGCA 116.
I~Zw~18 has two main components that have been heavily studied:
the NW and SE components. 
The CLASSY coadded spectrum only contains light from the NW region.
It also combines data from COS FUV detector lifetime positions 1 and 4
and data taken at position angles that are offset by 67$^{\circ}$.

\smallskip
\noindent{\it J0938+5428}: 
J0938+5248, also known as SBS 0934+546, is part of the LARS
\citep[][]{hayes14,ostlin14} and is identified as LARS12.
The CLASSY coadded spectrum combines data taken at position angles
that are offset by 88$^{\circ}$.

\smallskip
\noindent{\it J0940+2935}: 
J0940+2935 has extended optical emission.
The CLASSY coadded spectrum combines data taken at position angles
that are offset by 18$^{\circ}$.

\smallskip
\noindent{\it J0942+3547}: 
This galaxy is also known as CG-274 and SB~110 in \cite{senchyna17}.
The CLASSY coadded spectrum combines data from COS FUV detector lifetime 
positions 3 and 4.
The CLASSY coadded spectrum combines data taken at position angles
that are offset by 29$^{\circ}$.

\smallskip
\noindent{\it J0944-0038}: 
J0944-0038, more commonly known as CGCG 007-025, is part of LMLVL \citep{berg12}.
It is also known as SB2 in \citet{senchyna17}.
The CLASSY coadded spectrum combines data from COS FUV detector lifetime 
positions 3 and 4.
The CLASSY coadded spectrum combines data taken at position angles
that are offset by 68$^{\circ}$.

\smallskip
\noindent{\it J0944+3424}: 
J0944+3424 has extended optical emission.
The CLASSY coadded spectrum combines data taken at position angles
that are offset by 76$^{\circ}$.

\smallskip
\noindent{\it J1016+3754}: 
The CLASSY coadded spectrum combines data from COS FUV detector lifetime 
positions 3 and 4 and combines data taken at position angles
that are offset by 84$^{\circ}$.

\smallskip
\noindent{\it J1024+0524}: 
This galaxy is also known as SB~36 in \citep{senchyna17}.
The CLASSY coadded spectrum combines data from COS FUV detector lifetime 
positions 3 and 4.

\smallskip
\noindent{\it J1025+3622}: 
The CLASSY coadded spectrum combines data taken at position angles
that are offset by 21$^{\circ}$.

\smallskip
\noindent{\it J1044+0353}: 
The CLASSY coadded spectrum combines data taken at position angles
that are offset by 23$^{\circ}$.
This galaxy is denoted as an extreme emission line galaxy in
\citet{berg21}.

\smallskip
\noindent{\it J1105+4444}: 
J1105+445, more commonly known as Mrk 162, has extended optical emission.
The CLASSY coadded spectrum combines data from COS FUV detector lifetime 
positions 3 and 4 and combines data taken at position angles
that are offset by 46$^{\circ}$.

\smallskip
\noindent{\it J1112+5503}: 
The COS G185M and G225M observations for J1112+5503 were executed with an 
error, but were not reacquired, and so no MR extensions are available
in the CLASSY coadds.

\smallskip
\noindent{\it J1119+5130}: 
J1119+5130, commonly known as Arp's Galaxy, has extended optical emission.
The CLASSY coadded spectrum combines data taken at position angles
that are offset by 88$^{\circ}$.

\smallskip
\noindent{\it J1129+2034}: 
This galaxy is identified in NED as IC 700, but is also 
known as SB~179 in \citep{senchyna17},
and is denoted as an extreme emission line galaxy.
The CLASSY coadded spectrum combines data from COS FUV detector lifetime 
positions 3 and 4.
The CLASSY coadded spectrum combines data taken at position angles
that are offset by 47$^{\circ}$.

\smallskip
\noindent{\it J1132+5722}: 
J1132+5722, also known as SBSG1129+576, has extended optical emission.
The CLASSY coadded spectrum combines data from COS FUV detector lifetime 
positions 3 and 4.

\smallskip
\noindent{\it J1144+4012}: 
The CLASSY coadded spectrum combines data taken at position angles
that are offset by 28$^{\circ}$.

\smallskip
\noindent{\it J1148+2546}: 
This galaxy is also known as SB~182 in \citep{senchyna17}.
The CLASSY coadded spectrum combines data from COS FUV detector lifetime 
positions 3 and 4 and data taken at position angles
that are offset by 27$^{\circ}$.

\smallskip
\noindent{\it J1150+1501}: 
J1150+1501 is more commonly known as Mrk 750.

\smallskip
\noindent{\it J1157+3220}: 
J1157+3220, also known as NGC 3991N, has extended optical emission.
The CLASSY coadded spectrum combines data from COS FUV detector lifetime 
positions 3 and 4.

\smallskip
\noindent{\it J1225+6109}: 
J1225+6109, also known as SBS 1222+614, has extended optical emission.
The CLASSY coadded spectrum combines data from COS FUV detector lifetime 
positions 3 and 4 and data taken at position angles
that are offset by 76$^{\circ}$.

\smallskip
\noindent{\it J1253-0312}: 
J1253-0312 is more commonly known as SHOC 391.

\smallskip
\noindent{\it J1314+3452}: 
This galaxy is also known as SB~153 in \citep{senchyna17}
and is denoted as an extreme emission line galaxy.
The CLASSY coadded spectrum combines data taken at position angles
that are offset by 54$^{\circ}$.

\smallskip
\noindent{\it J1323-0132}: 
J1323-0132, also known as UM 570, has extended optical emission.
The CLASSY coadded spectrum combines data from COS FUV detector lifetime 
positions 3 and 4.

\smallskip
\noindent{\it J1359+5726}: 
J1359+5726, also commonly known as Mrk 1486 and is part of the LARS
sample with ID Ly52, has extended optical emission.
The CLASSY coadded spectrum combines data from COS FUV detector lifetime 
positions 1 and 3 and data taken at position angles
that are offset by 82$^{\circ}$.

\smallskip
\noindent{\it J1418+2102}: 
The CLASSY coadded spectrum combines data from COS FUV detector lifetime 
positions 3 and 4 and data taken at position angles
that are offset by 26$^{\circ}$.
This galaxy is denoted as an extreme emission line galaxy in
\citet{berg21}.

\smallskip
\noindent{\it J1444+4237}: 
J1444+4237 is also known as HS1442+4250.
The CLASSY coadded spectrum combines data from COS FUV detector lifetime 
positions 3 and 4 and data taken at position angles
that are offset by 47$^{\circ}$.
Note that an additional dataset, LDXT06010, was not used in the CLASSY
coadd because there was a data quality note that the Take Data Flag was not on throughout observation
and so the COS light path was blocked during the exposure.

\smallskip
\noindent{\it J1448-0110}: 
J1448-0110 is also known as SHOC 486.

\smallskip
\noindent{\it J1545+0858}: 
The CLASSY coadded spectrum combines data from COS FUV detector lifetime 
positions 3 and 4 and data taken at position angles
that are offset by 51$^{\circ}$.


\section{CLASSY Coadded Spectra}\label{AppendixE}
In Figure~\ref{fig12} we display one of the moderate- or low-resolution coadded spectra 
for each of the 45 galaxies in the CLASSY sample.  
The rest-frame, Galactic extinction corrected, binned, moderate-resolution 
spectrum is plotted in grey when available.
For galaxies with archival G140L spectra, the low-resolution coadd is shown.
In order to visually detect emission and absorption features more easily,
the spectrum has also been convolved with a 1D Gaussian 2-element kernel and overplotted in black. 
The most prominent absorption and emission features are labeled in the top panel of each spectrum,
with nebular (dashed pink lines) and fine structure (solid pink lines) emission features marked
above the spectra, while ISM (solid blue lines), stellar wind (dot-dashed green lines), and
photospheric (solid green lines) are marked below the spectra.
Contamination from Geocoronal and \ion{O}{1} emission and Milky Way ISM absorption are whited-out. 
The bottom panel plots the S/N ratio as a function of wavelength, where each vertical
shaded bar shows the average S/N within its wavelength coverage. 
Since the plotted spectrum is binned by the 6 pixels of a resolution element, 
these S/N values are per resel.

\renewcommand{\thefigure}{12}
\begin{figure*}
\begin{center}
    \includegraphics[width=0.90\textwidth,trim=10mm 10mm 10mm 0mm,clip]{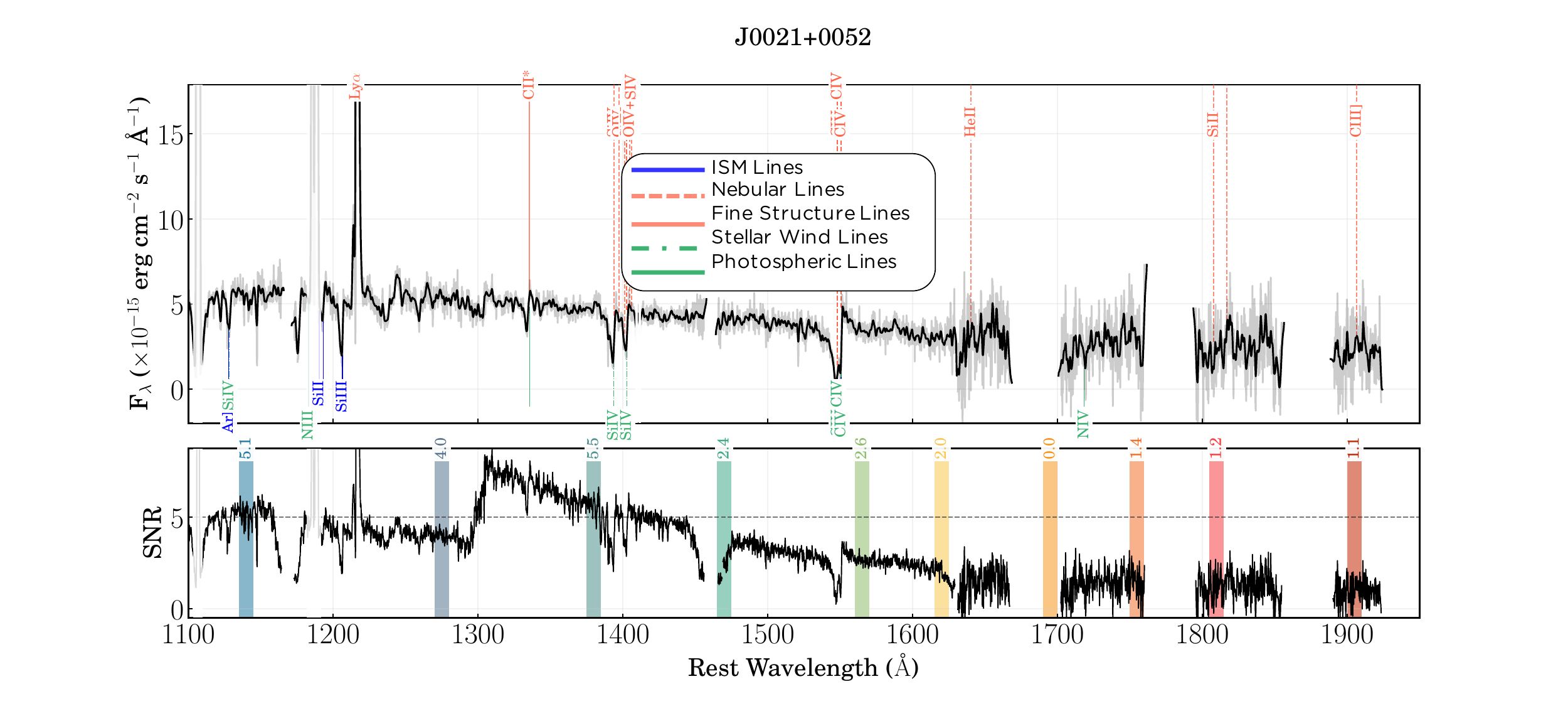} \\
    \includegraphics[width=0.95\textwidth,trim=10mm 15mm 10mm 0mm,clip]{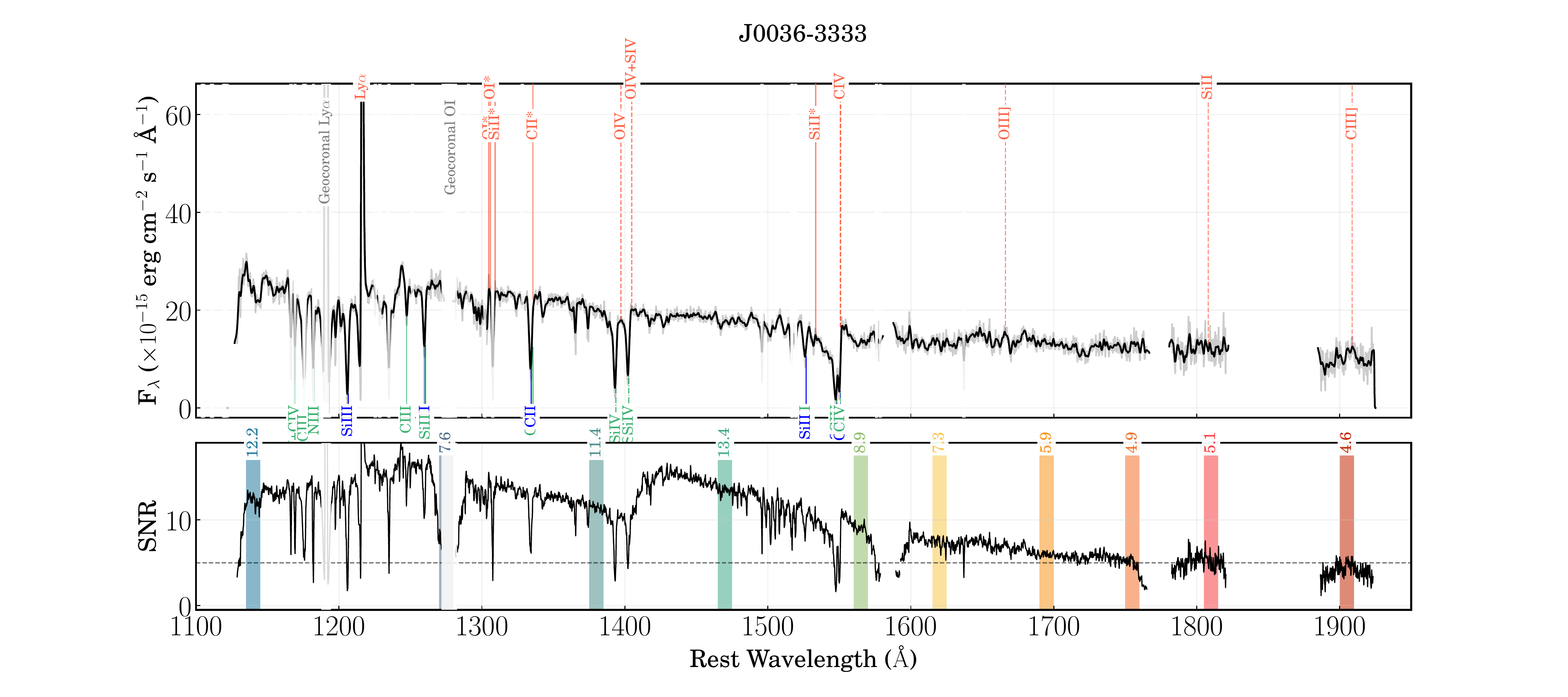} \\
    \includegraphics[width=0.95\textwidth,trim=10mm 0mm 10mm 0mm,clip]{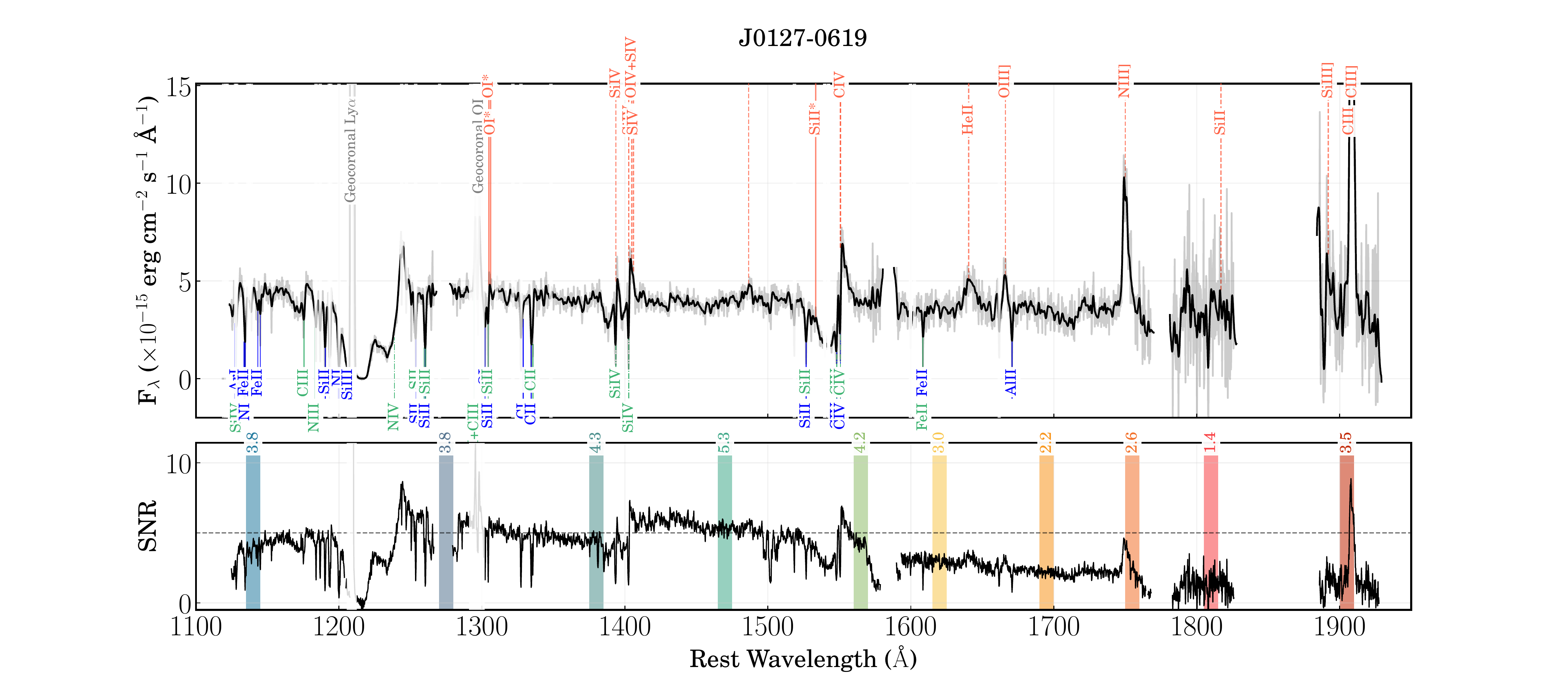} 
\end{center}
\vspace{-4ex}
\caption{Coadded spectra for all 45 galaxies in the CLASSY sample.
The moderate resolution coadd is shown for most galaxies, highlighting
the full FUV spectral range covered. 
For galaxies with G140L spectra, the low resolution coadd is shown, 
highlighting the full FUV spectral range covered in the CLASSY atlas. 
The binned, Galactic extinction-corrected, rest-frame spectrum is shown in gray
and smoothed version that has been convolved with a 1D Gaussian 2-element kernel 
is overplotted in black to make spectral features more apparent.
Prominent absorption and emission features are labeled in the top panel.
The bottom panel shows the S/N ratio as a function of wavelength,
where the colored vertical bands note the S/N value at their respective locations. 
\label{fig12}}
\end{figure*}

\begin{figure*}
\begin{center}
    \includegraphics[width=1.0\textwidth,trim=10mm 15mm 10mm 1mm,clip]{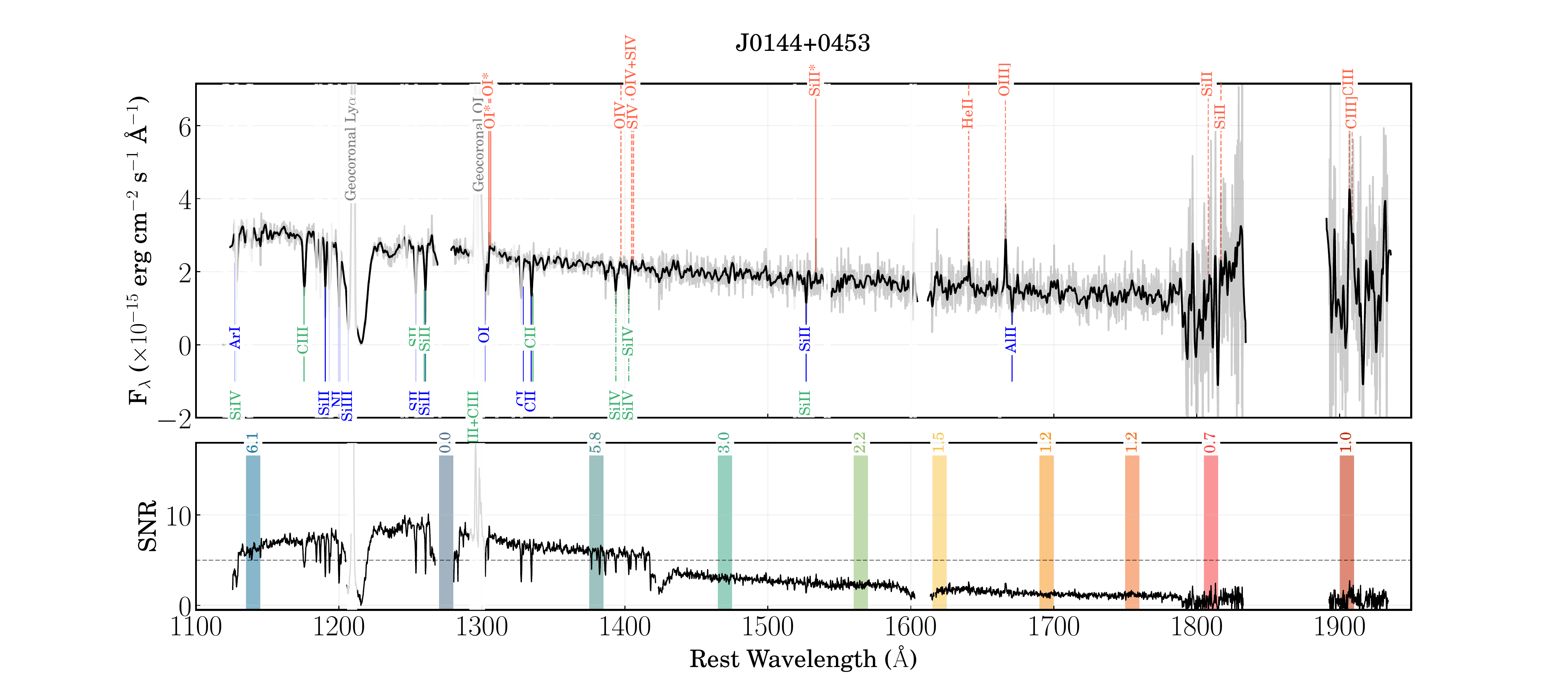} \\
    \includegraphics[width=1.0\textwidth,trim=10mm 15mm 10mm 1mm,clip]{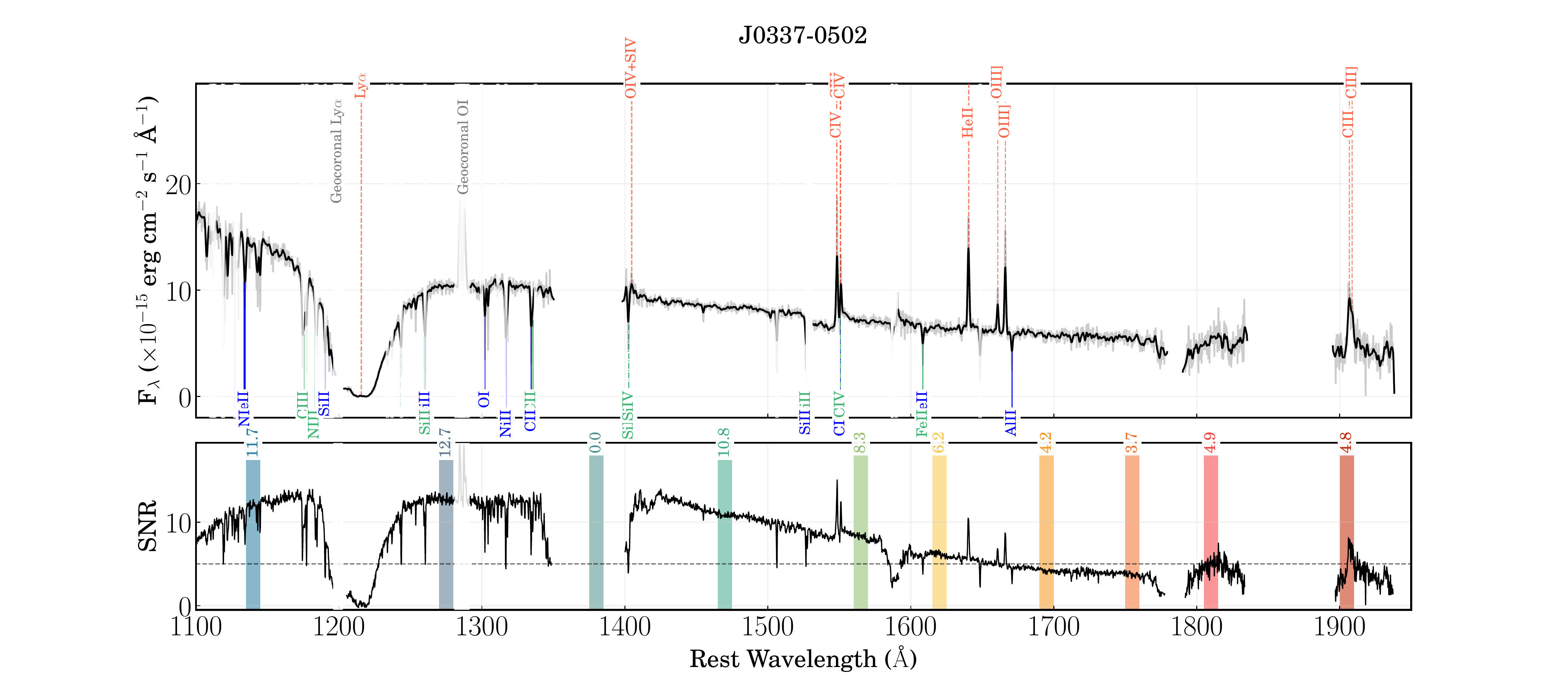} \\
    \includegraphics[width=1.0\textwidth,trim=10mm 0mm 10mm 1mm,clip]{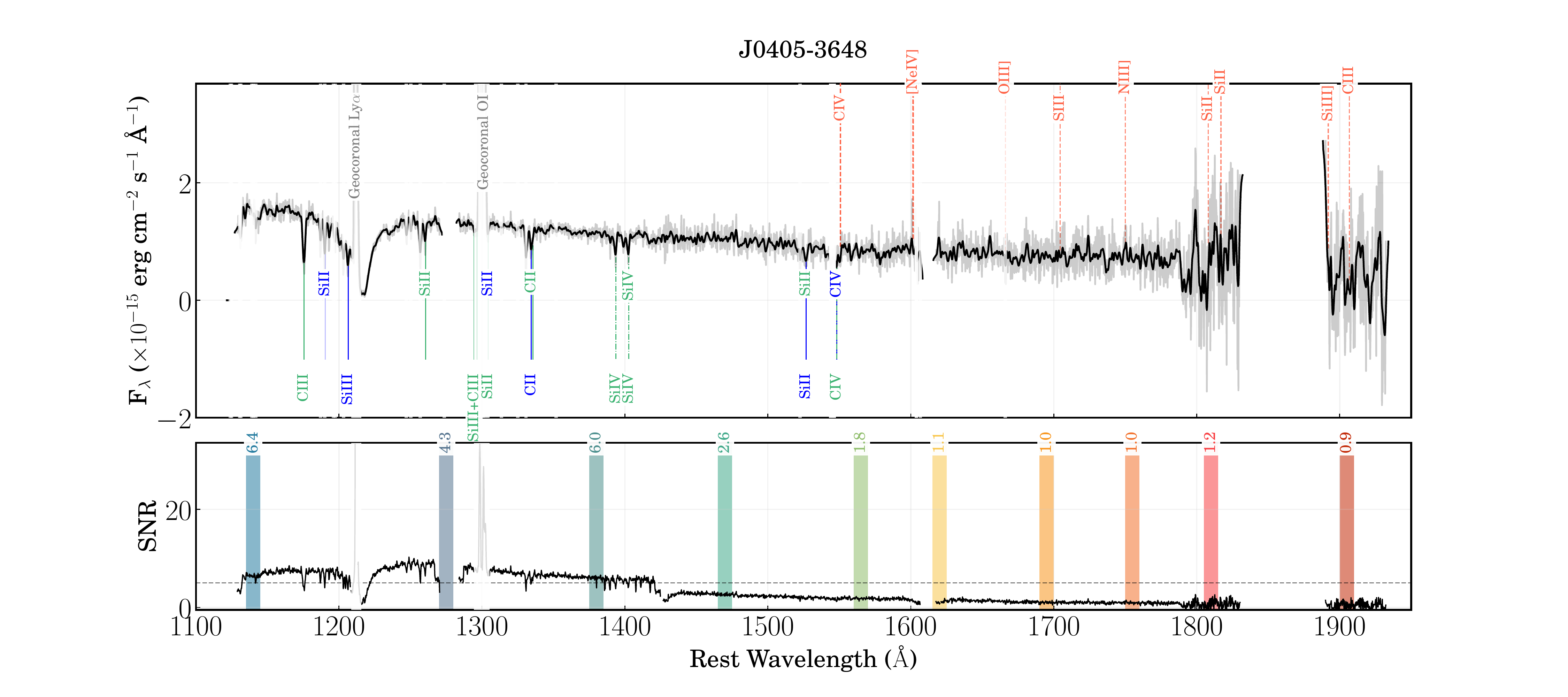} 
\end{center}
\vspace{-4ex}
\caption{(\textit{continued})}
\end{figure*}

\begin{figure*}
\begin{center}
    \includegraphics[width=1.0\textwidth,trim=10mm 15mm 10mm 1mm,clip]{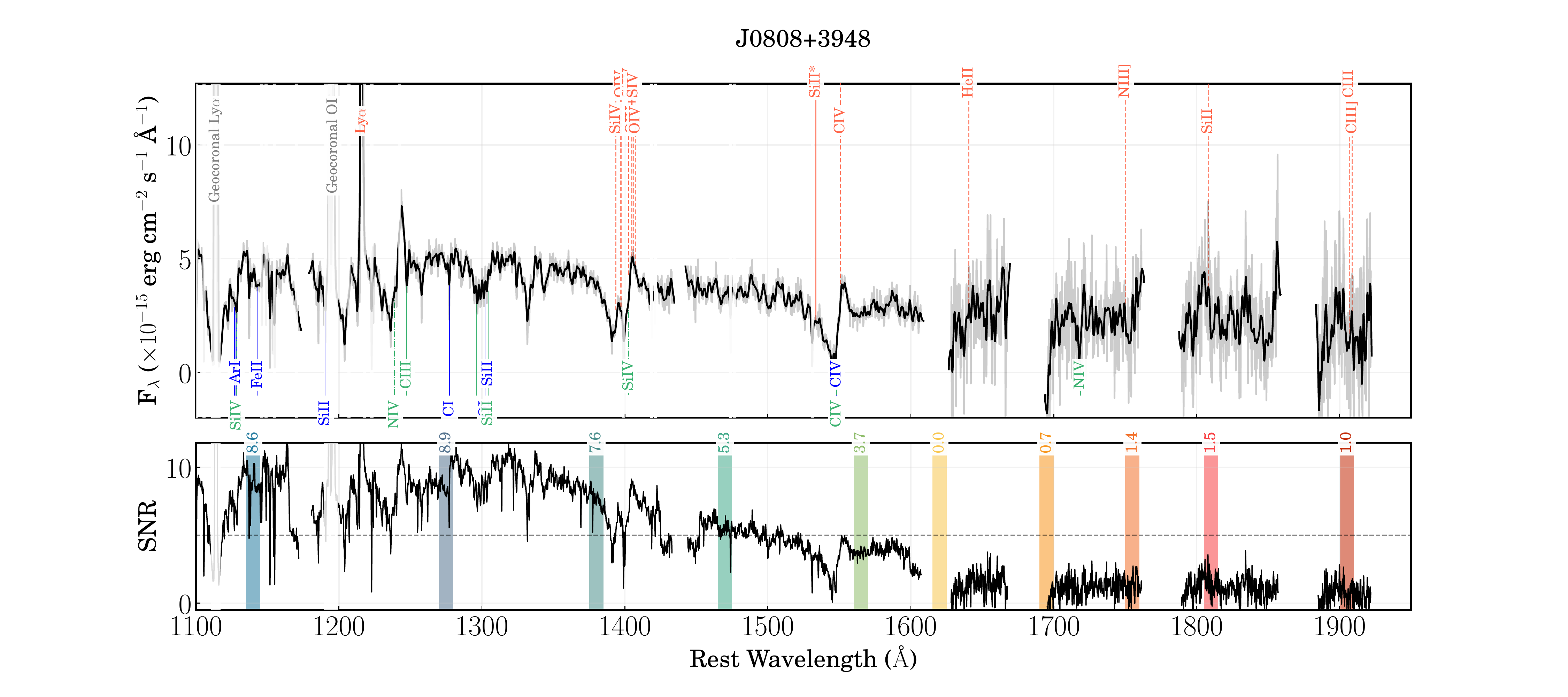} \\
    \includegraphics[width=1.0\textwidth,trim=10mm 15mm 10mm 1mm,clip]{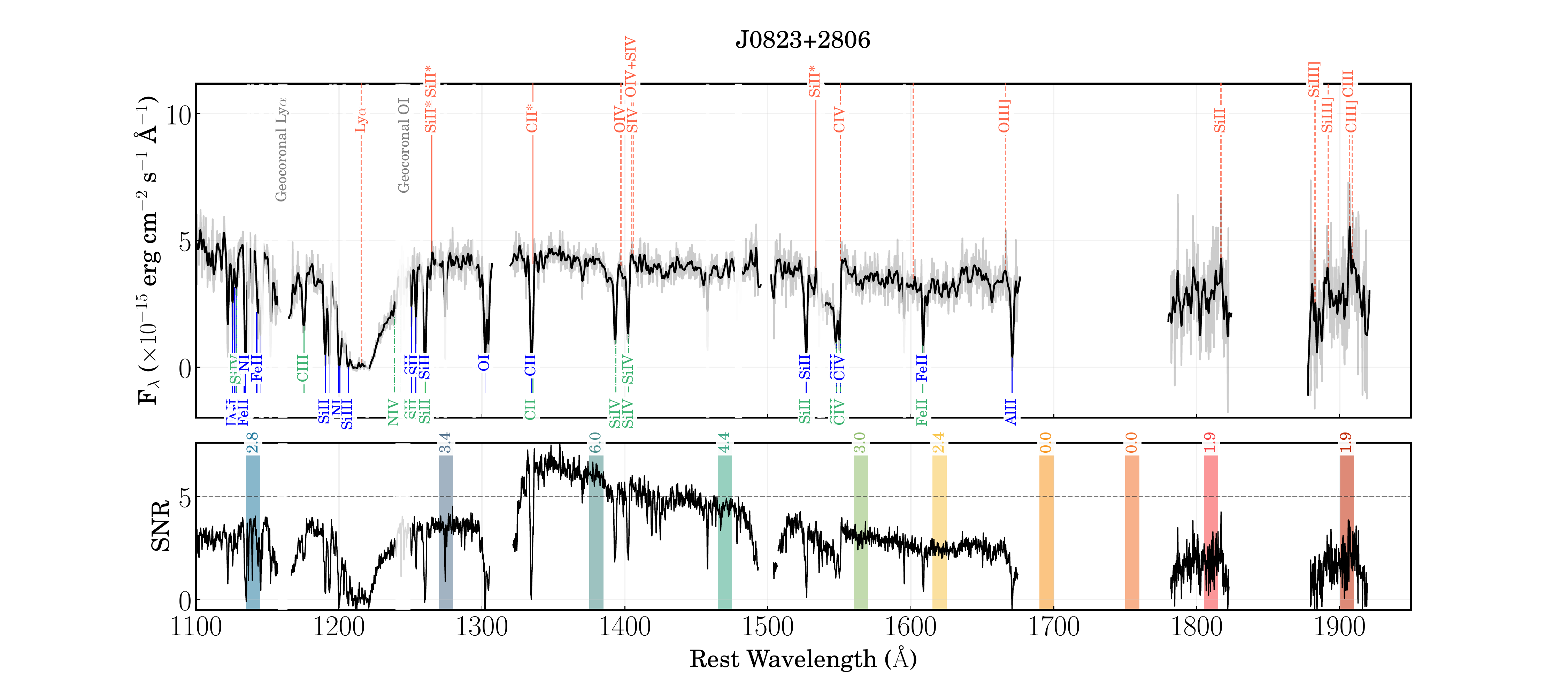} \\
    \includegraphics[width=1.0\textwidth,trim=10mm 0mm 10mm 1mm,clip]{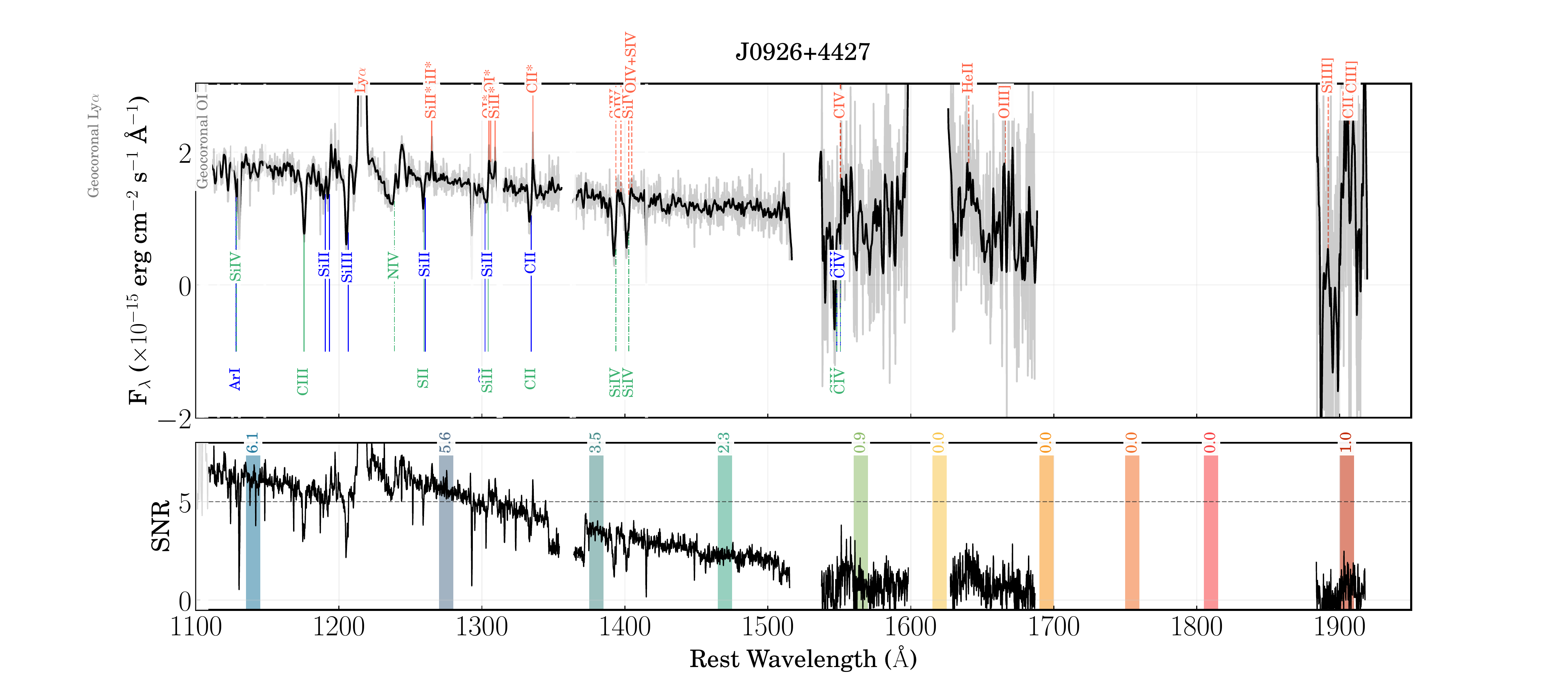} 
\end{center}
\vspace{-4ex}
\caption{(\textit{continued})}
\end{figure*}

\begin{figure*}
\begin{center}
    \includegraphics[width=1.0\textwidth,trim=10mm 15mm 10mm 1mm,clip]{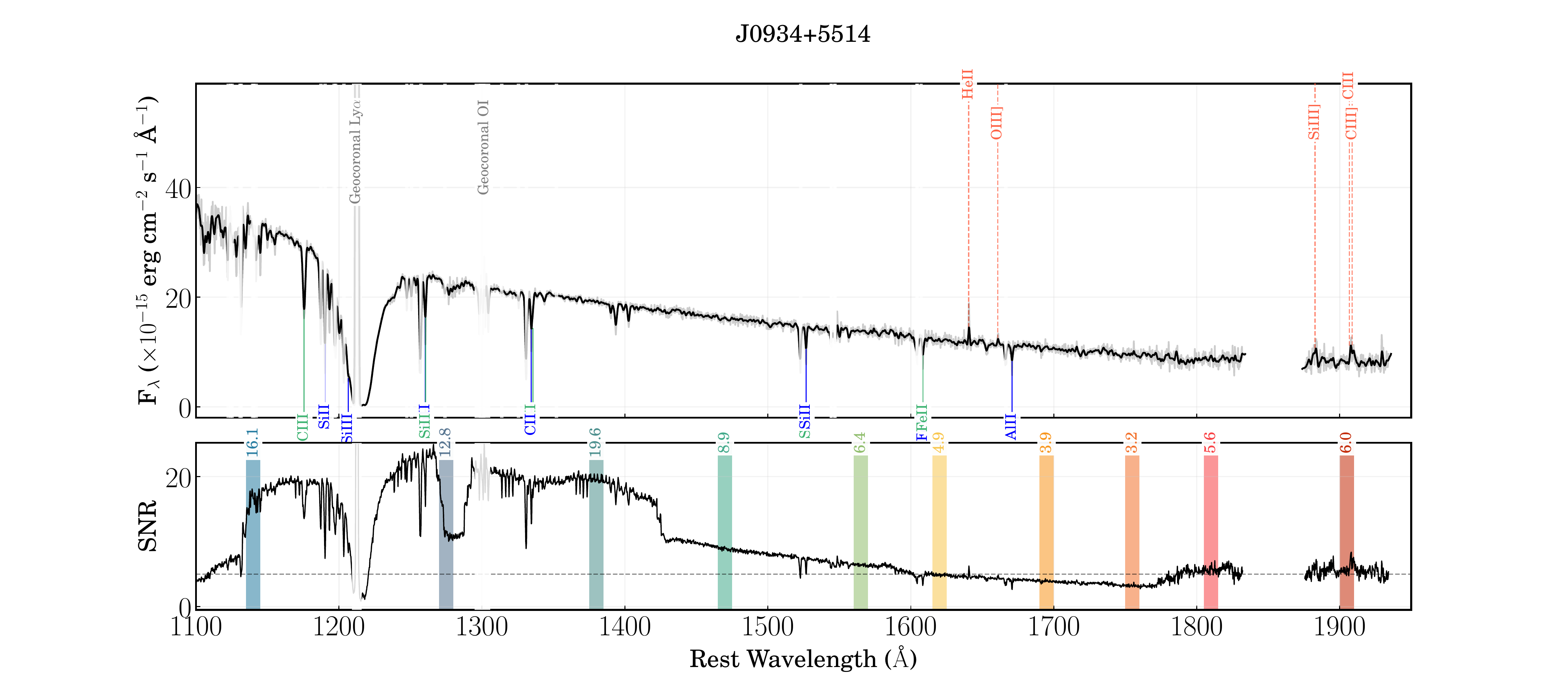} \\
    \includegraphics[width=1.0\textwidth,trim=10mm 15mm 10mm 1mm,clip]{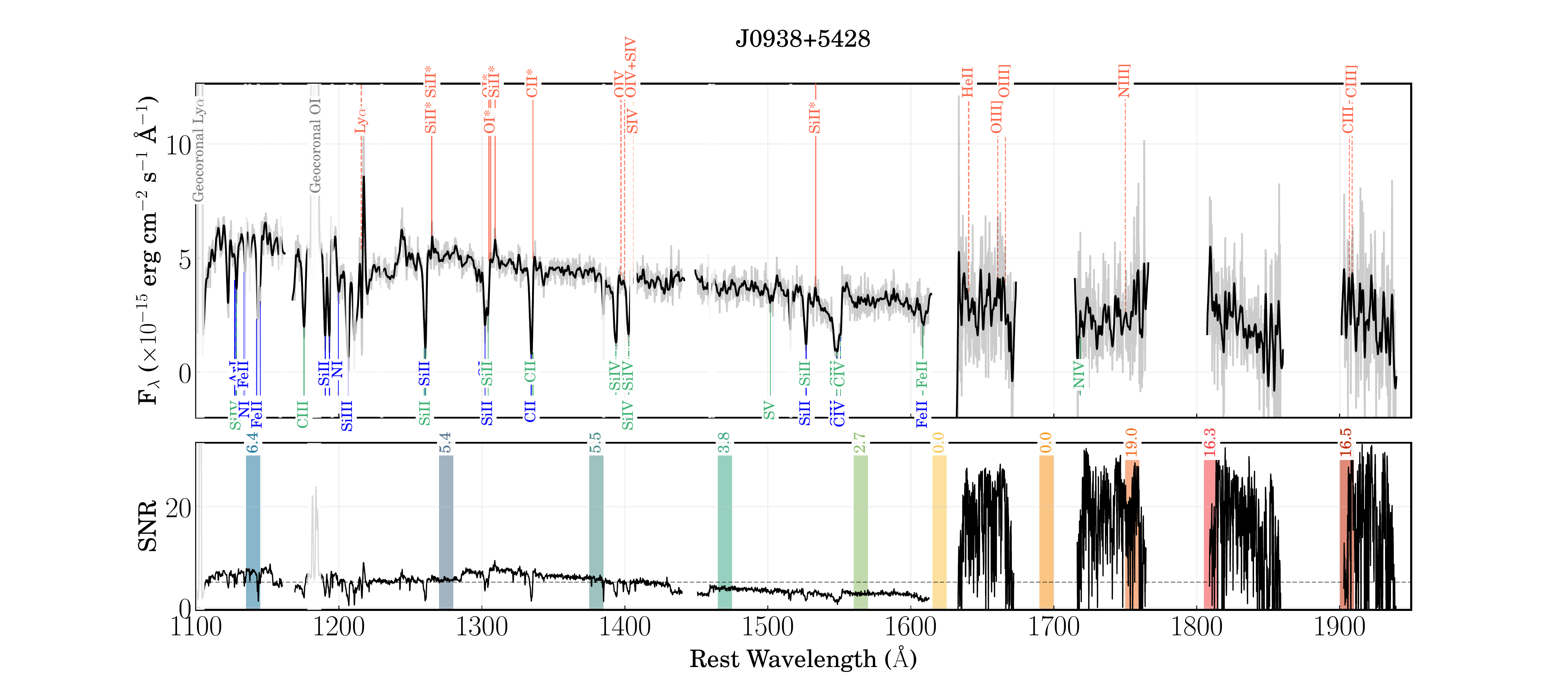} \\
    \includegraphics[width=1.0\textwidth,trim=10mm 0mm 10mm 1mm,clip]{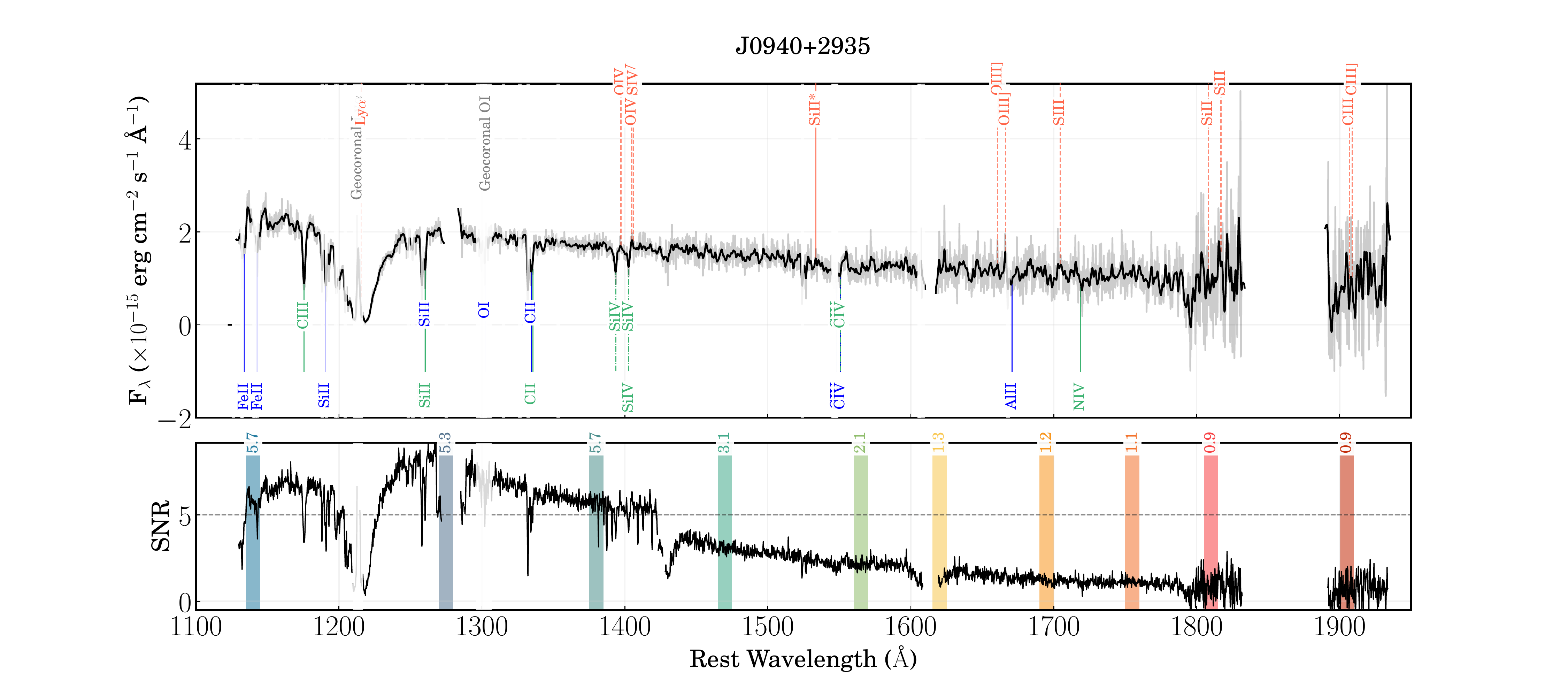} 
\end{center}
\vspace{-4ex}
\caption{(\textit{continued})}
\end{figure*}

\begin{figure*}
\begin{center}
    \includegraphics[width=1.0\textwidth,trim=10mm 15mm 10mm 1mm,clip]{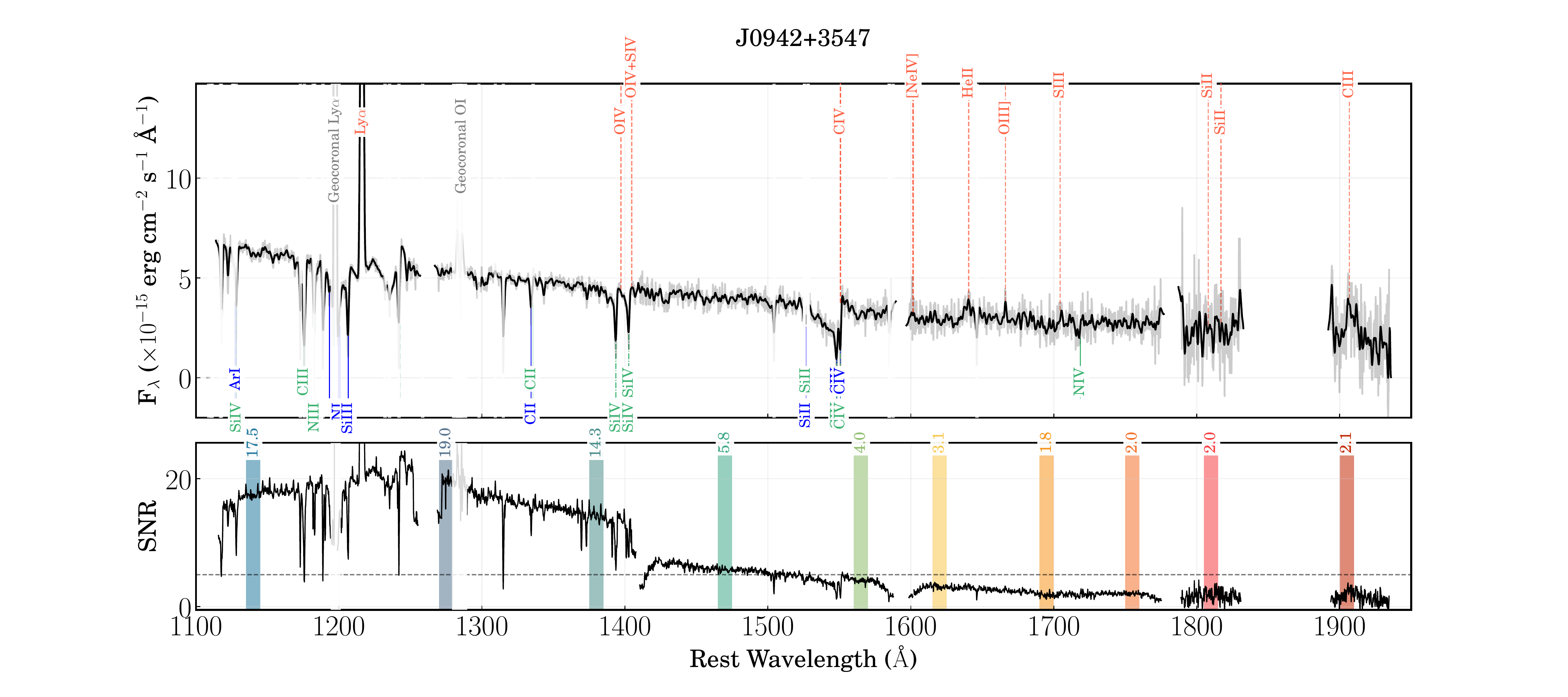} \\
    \includegraphics[width=1.0\textwidth,trim=10mm 15mm 10mm 1mm,clip]{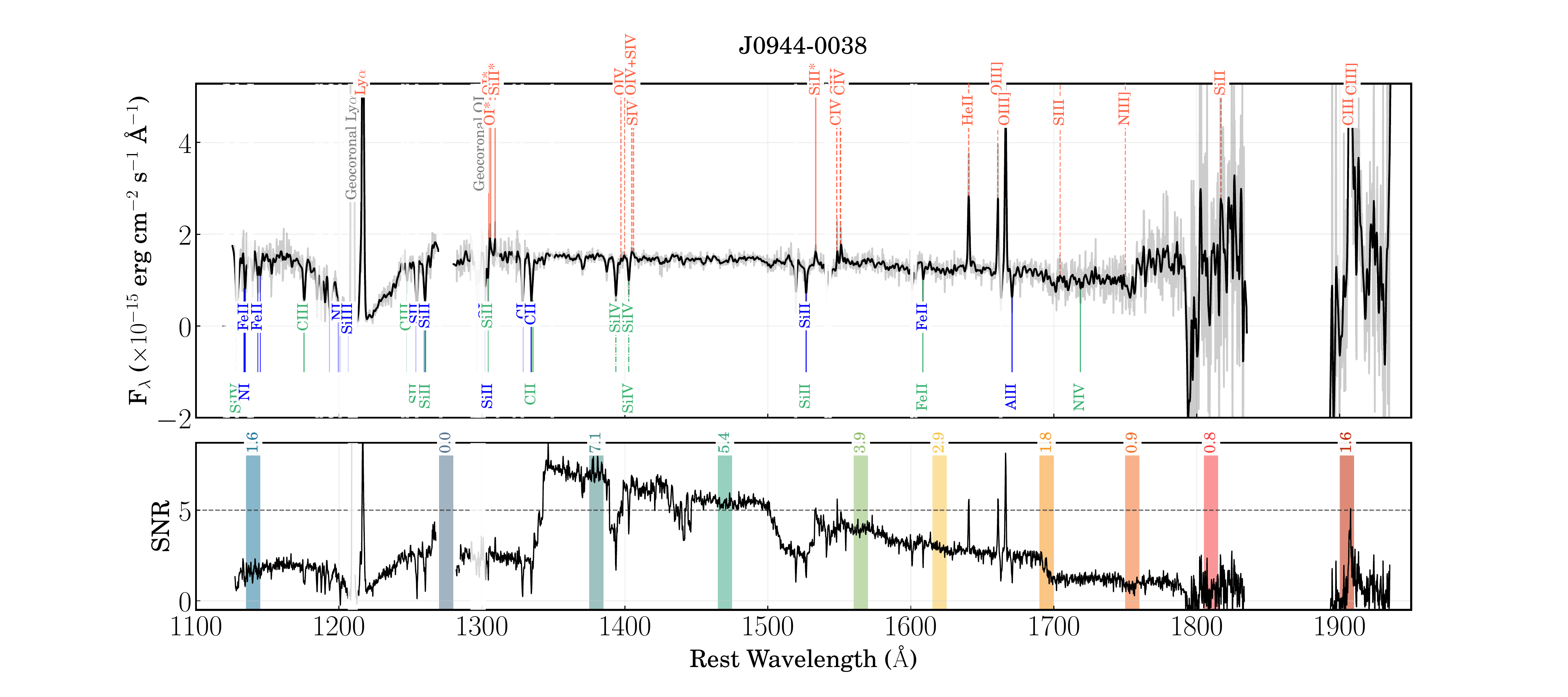} \\
    \includegraphics[width=1.0\textwidth,trim=10mm 0mm 10mm 1mm,clip]{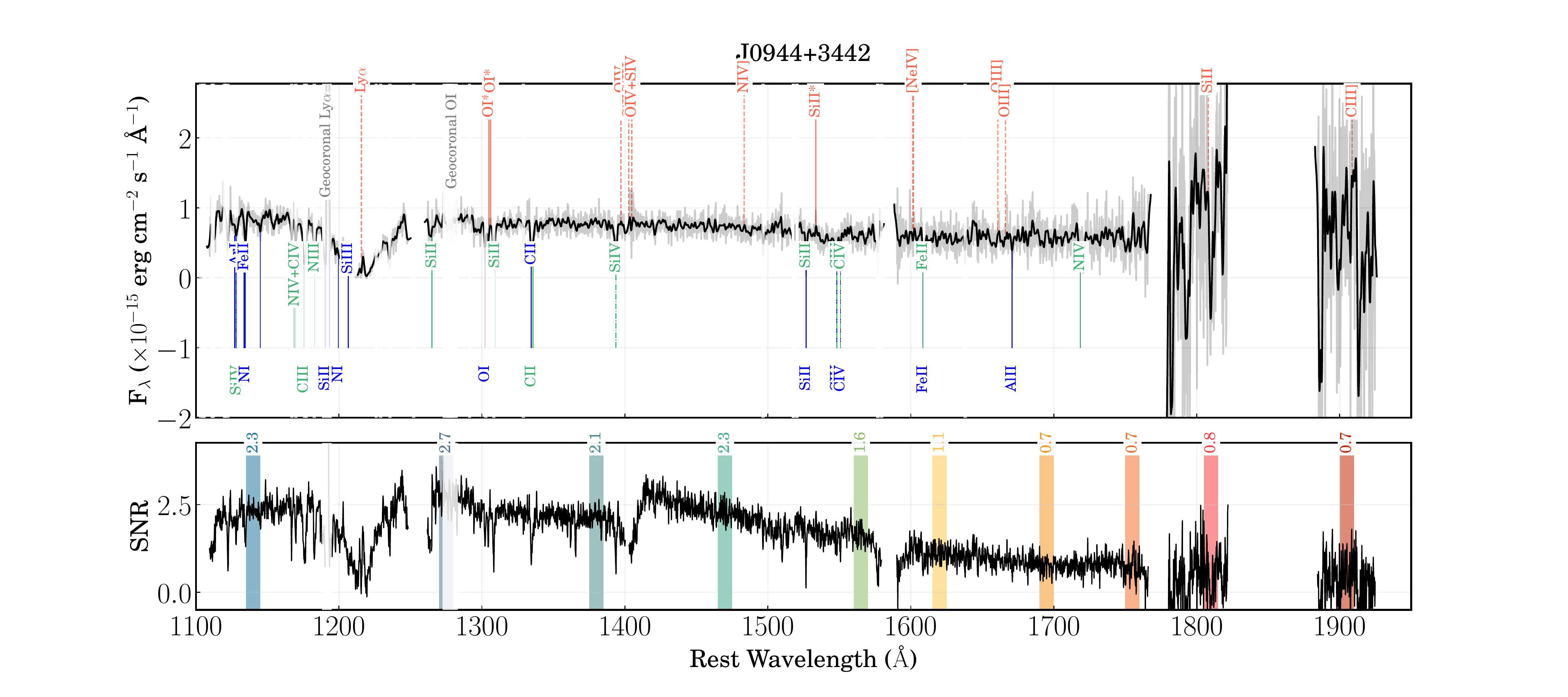} 
\end{center}
\vspace{-4ex}
\caption{(\textit{continued})}
\end{figure*}

\begin{figure*}
\begin{center}
    \includegraphics[width=1.0\textwidth,trim=10mm 15mm 10mm 1mm,clip]{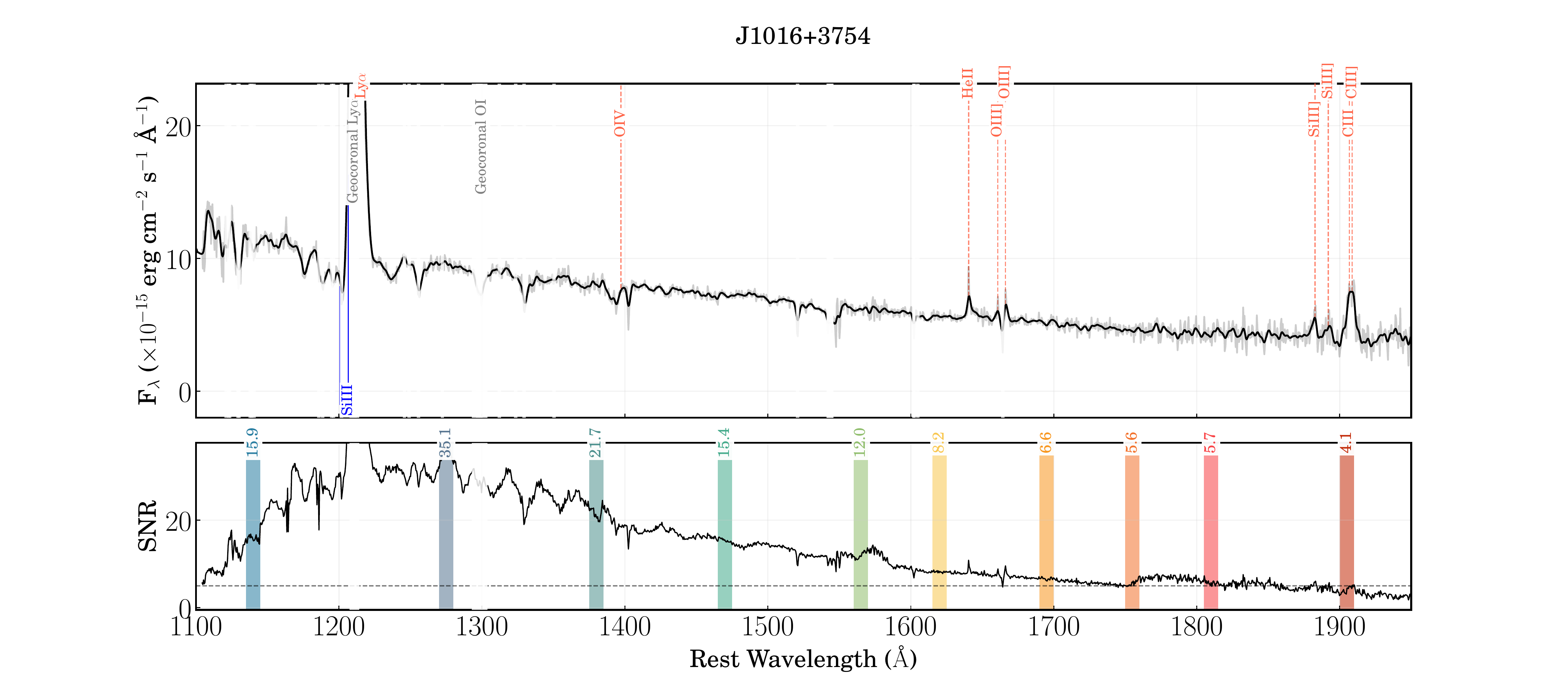} \\
    \includegraphics[width=1.0\textwidth,trim=10mm 15mm 10mm 1mm,clip]{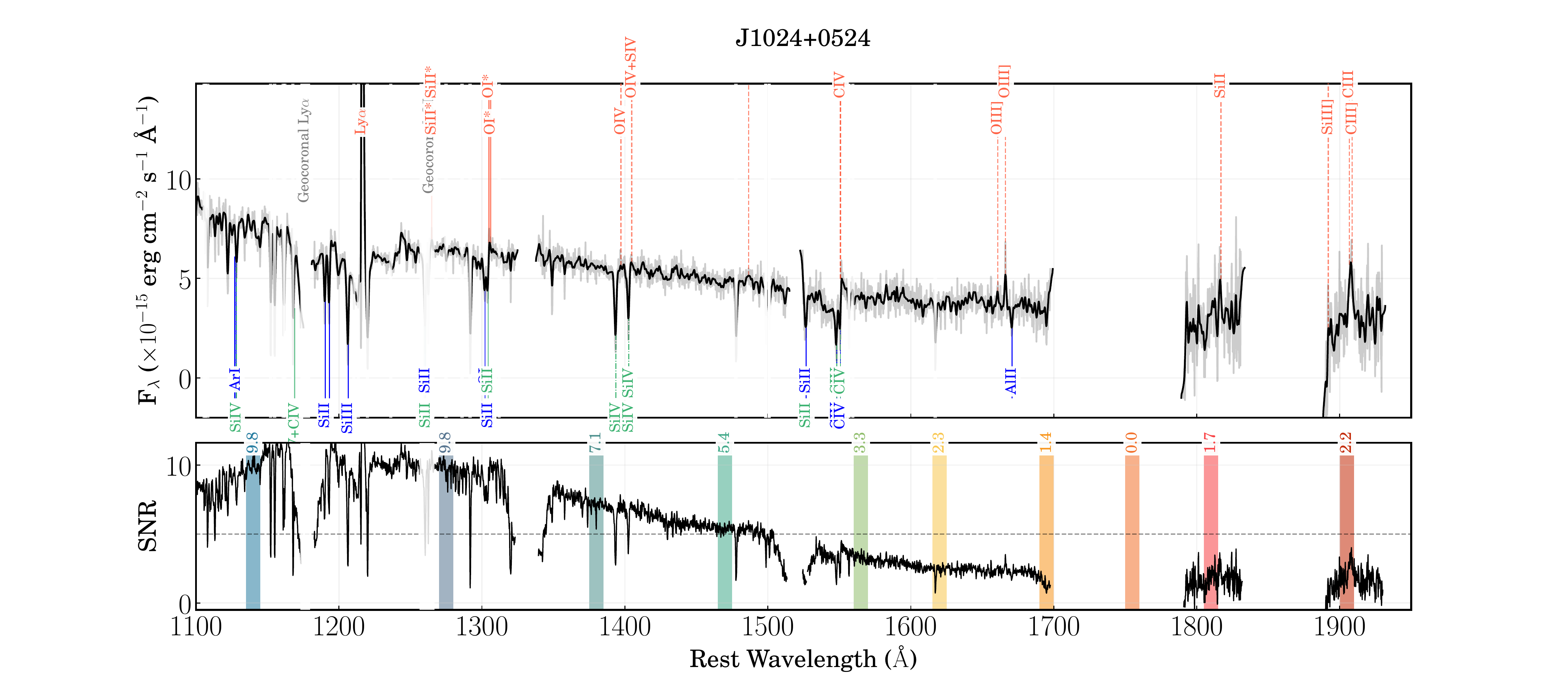} \\
    \includegraphics[width=1.0\textwidth,trim=10mm 0mm 10mm 1mm,clip]{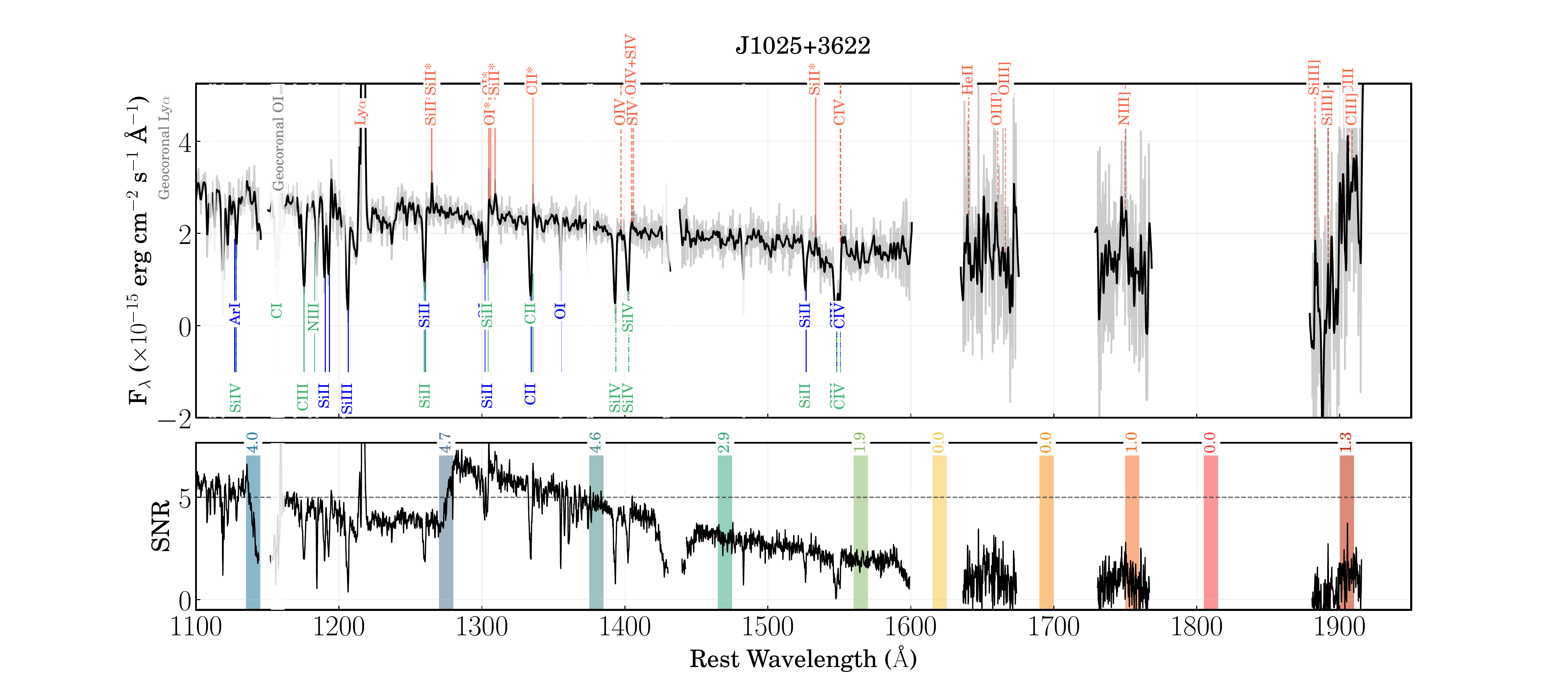} 
\end{center}
\vspace{-4ex}
\caption{(\textit{continued})}
\end{figure*}

\begin{figure*}
\begin{center}
    \includegraphics[width=1.0\textwidth,trim=10mm 15mm 10mm 1mm,clip]{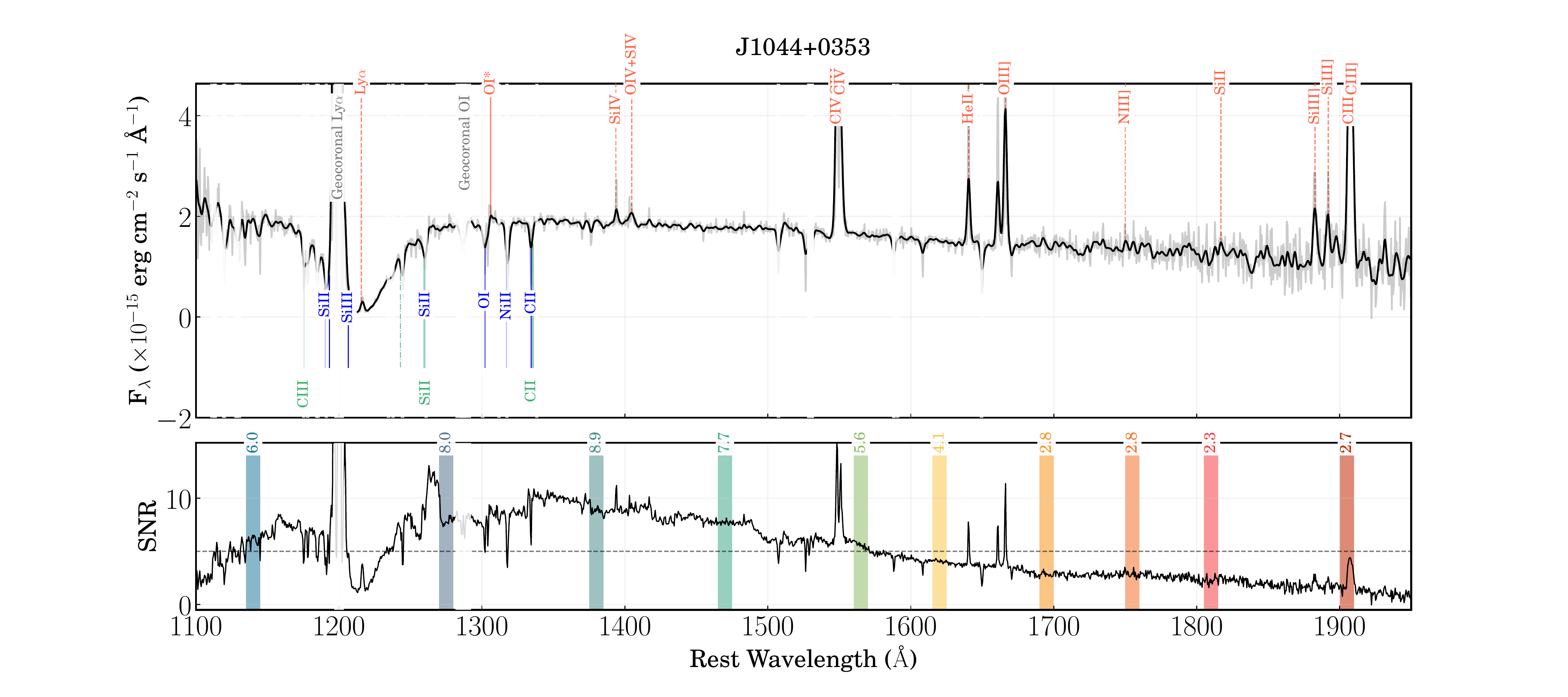} \\
    \includegraphics[width=1.0\textwidth,trim=10mm 15mm 10mm 1mm,clip]{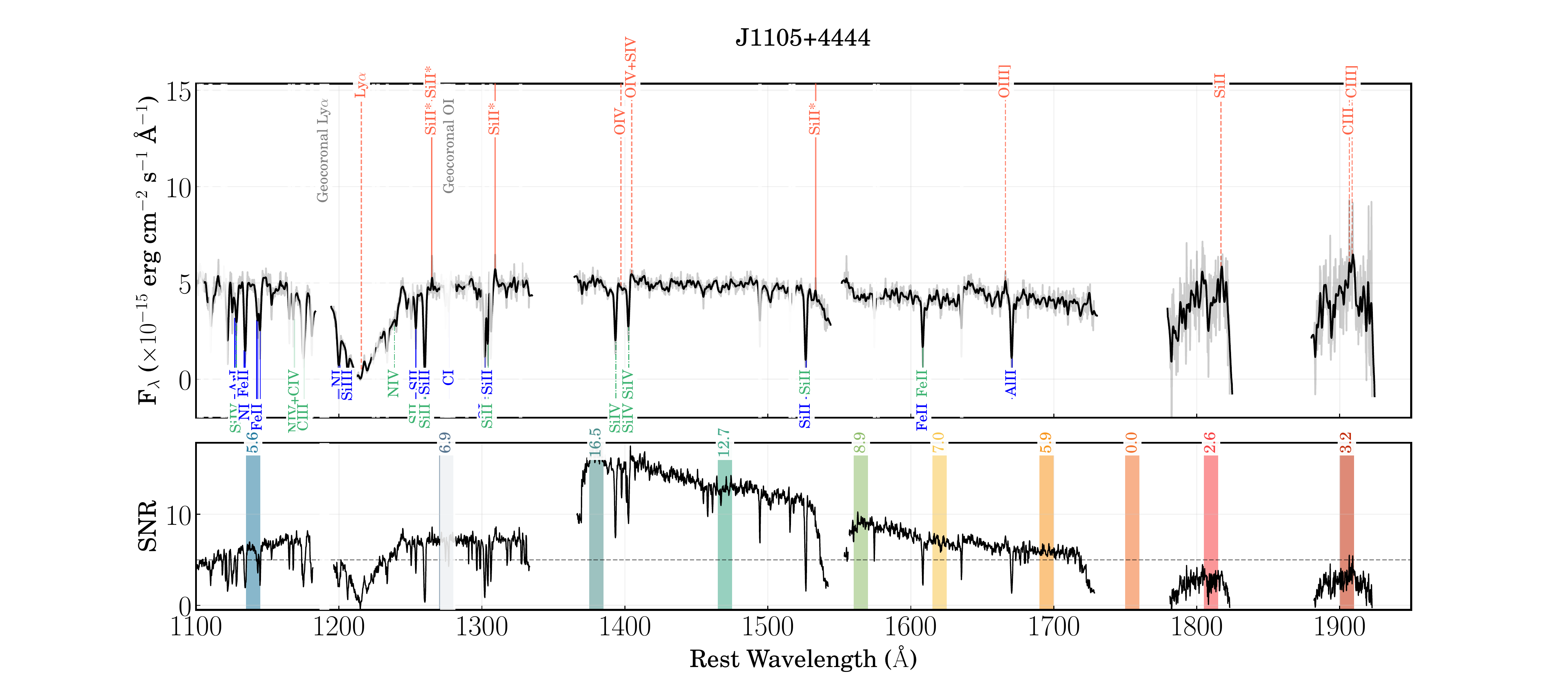} \\
    \includegraphics[width=1.0\textwidth,trim=10mm 0mm 10mm 1mm,clip]{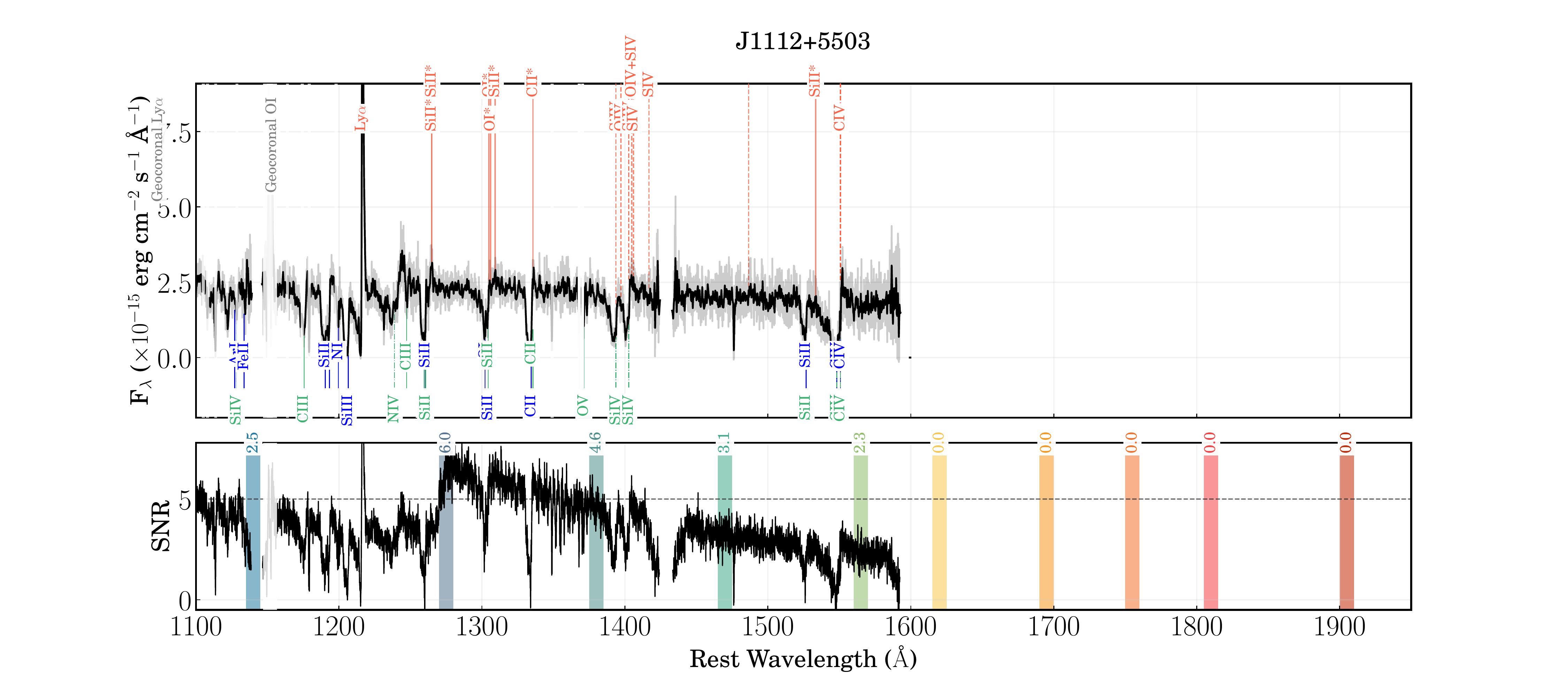} 
\end{center}
\vspace{-4ex}
\caption{(\textit{continued})}
\end{figure*}

\begin{figure*}
\begin{center}
    \includegraphics[width=1.0\textwidth,trim=10mm 15mm 10mm 1mm,clip]{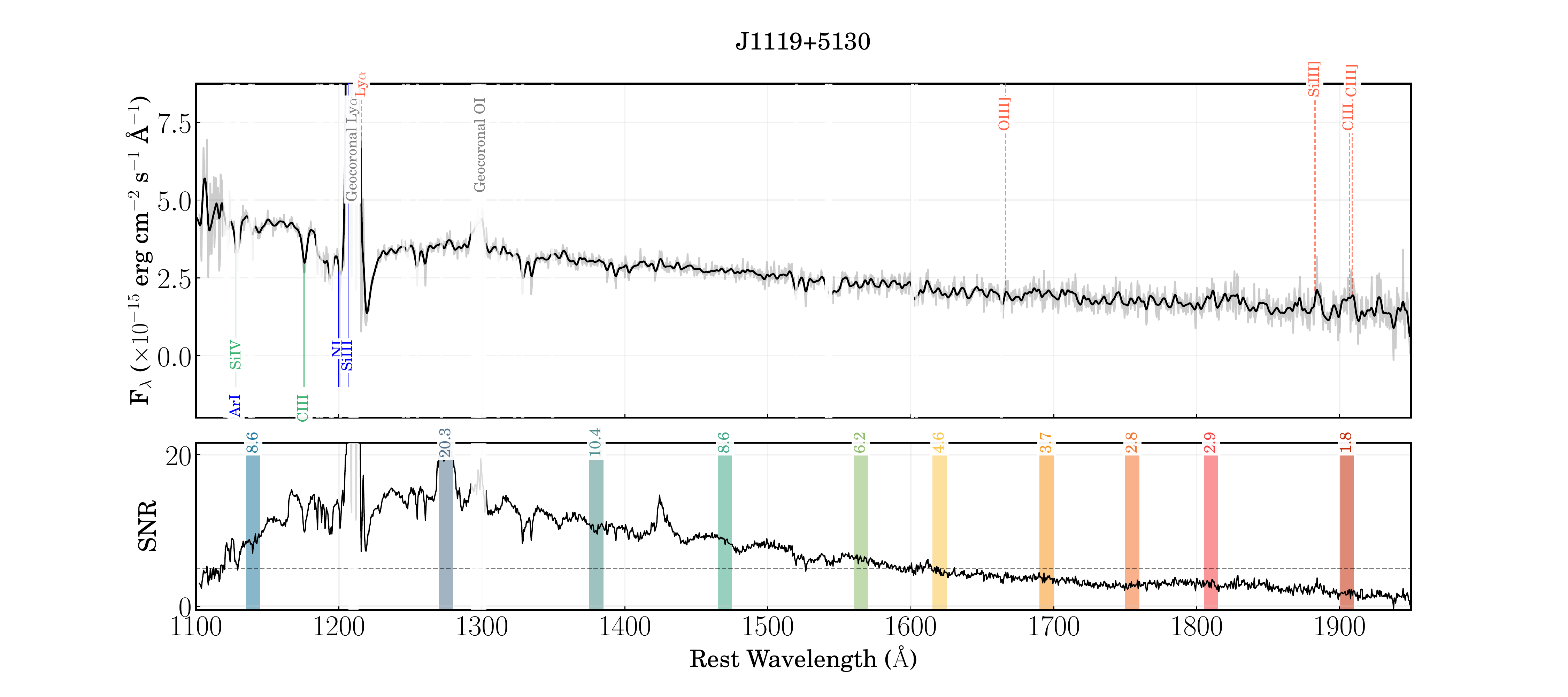} \\
    \includegraphics[width=1.0\textwidth,trim=10mm 15mm 10mm 1mm,clip]{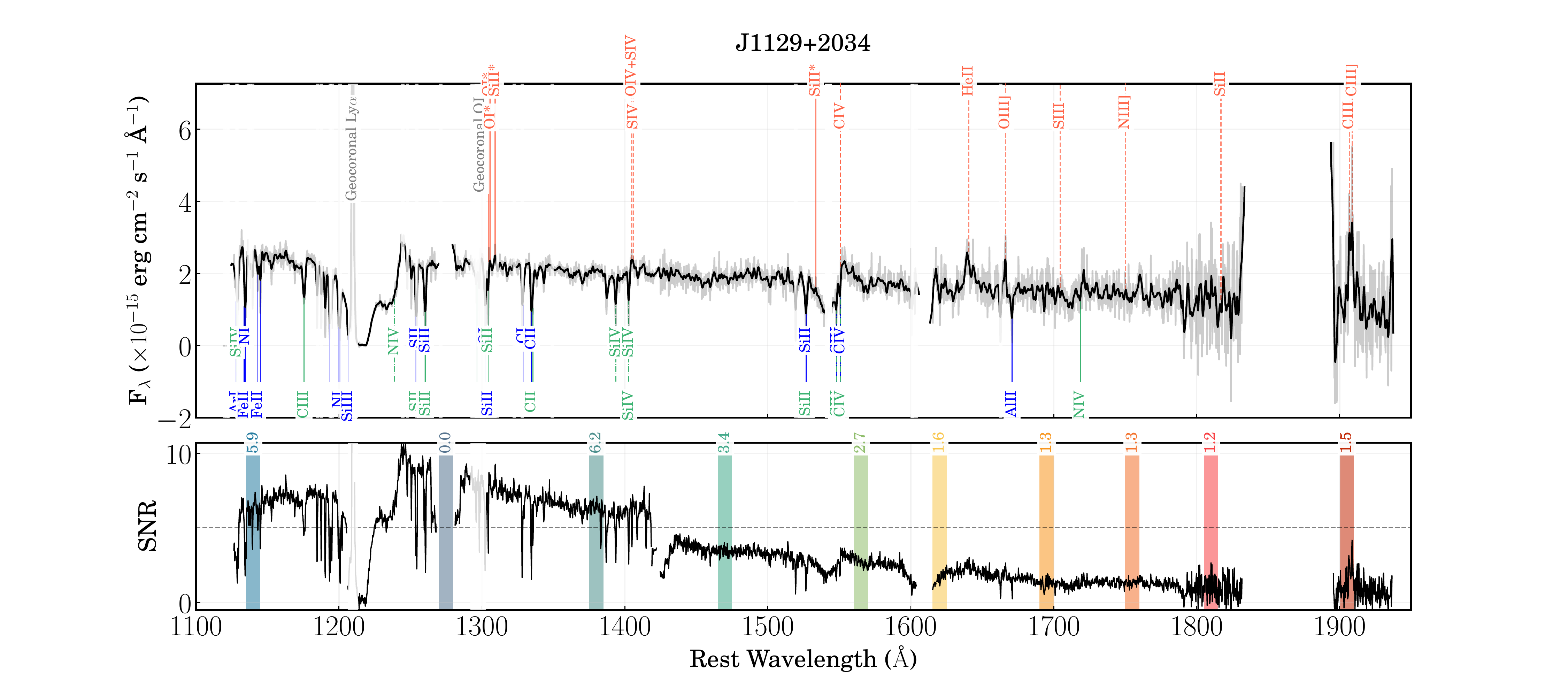} \\
    \includegraphics[width=1.0\textwidth,trim=10mm 0mm 10mm 1mm,clip]{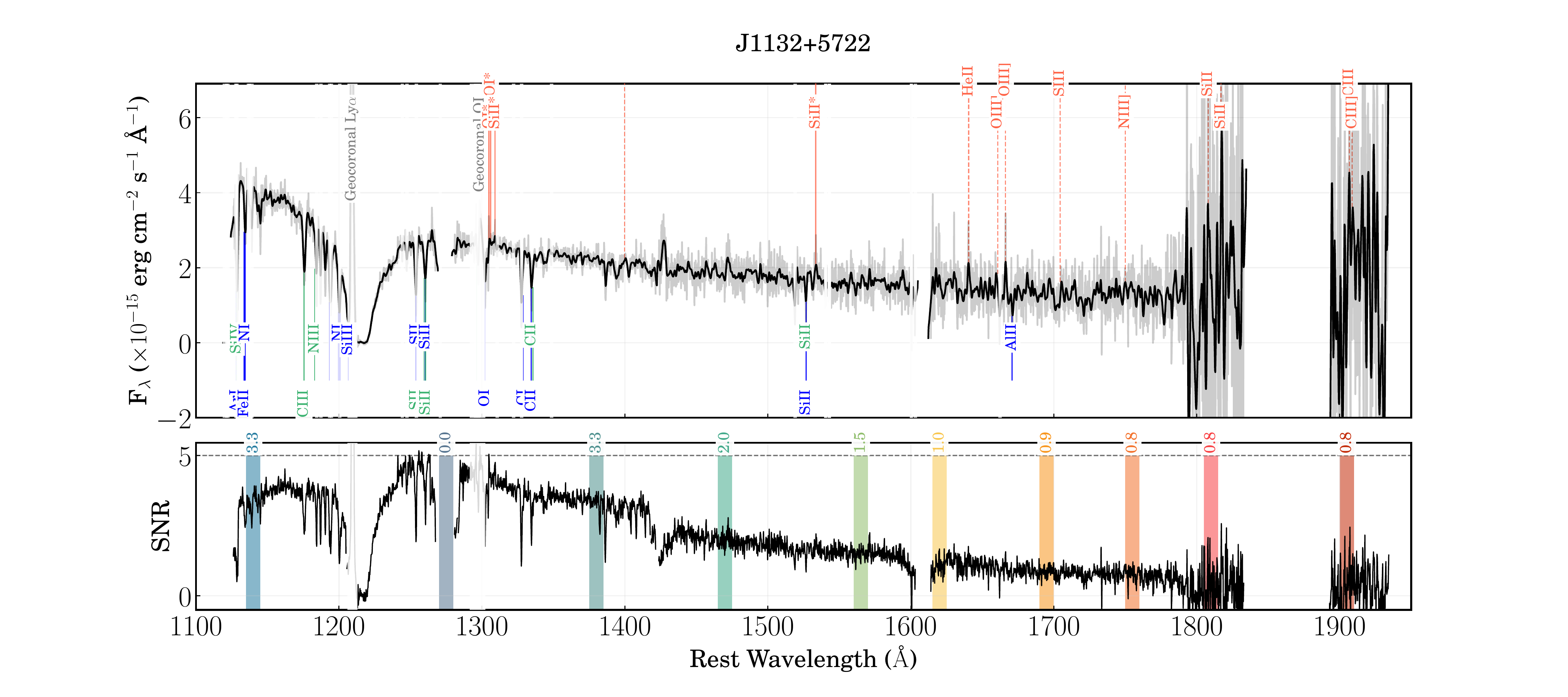} 
\end{center}
\vspace{-4ex}
\caption{(\textit{continued})}
\end{figure*}

\begin{figure*}
\begin{center}
    \includegraphics[width=1.0\textwidth,trim=10mm 15mm 10mm 1mm,clip]{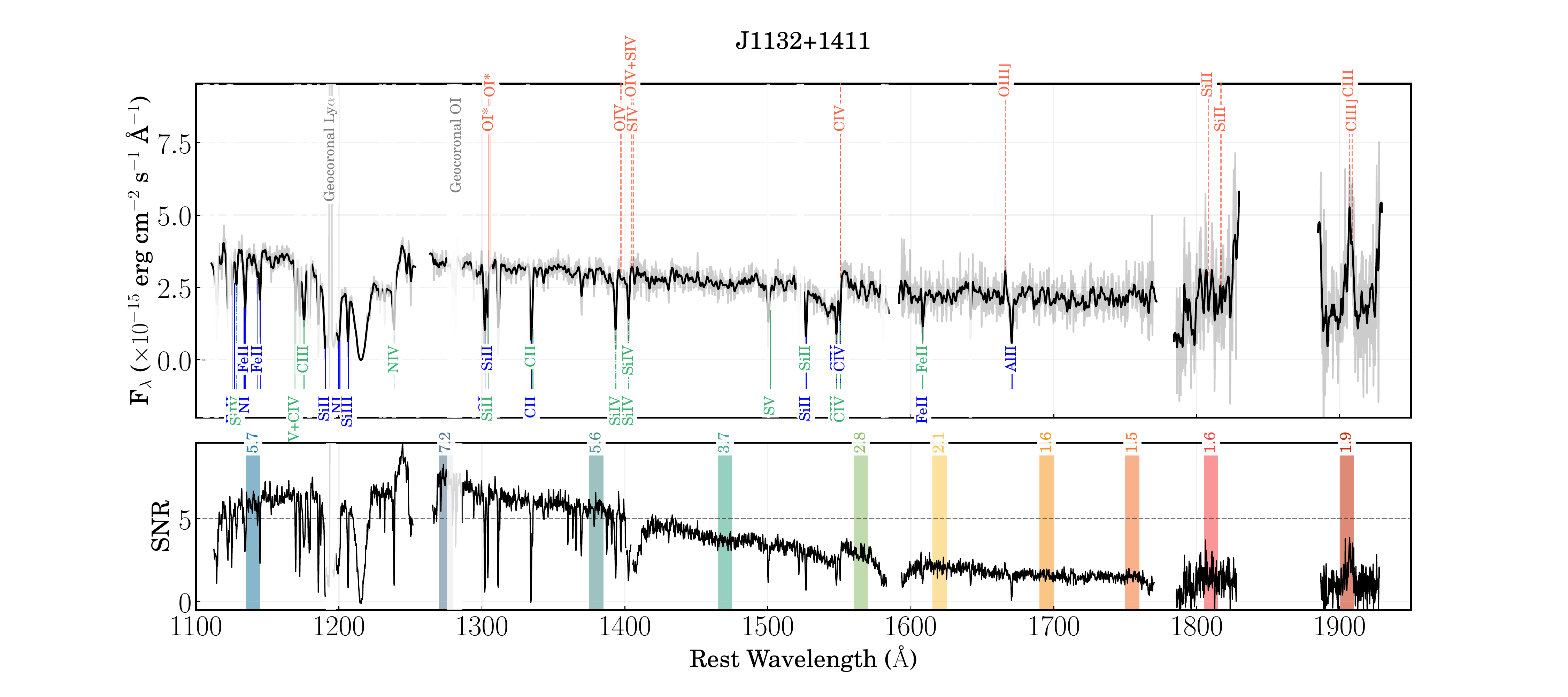} \\
    \includegraphics[width=1.0\textwidth,trim=10mm 15mm 10mm 1mm,clip]{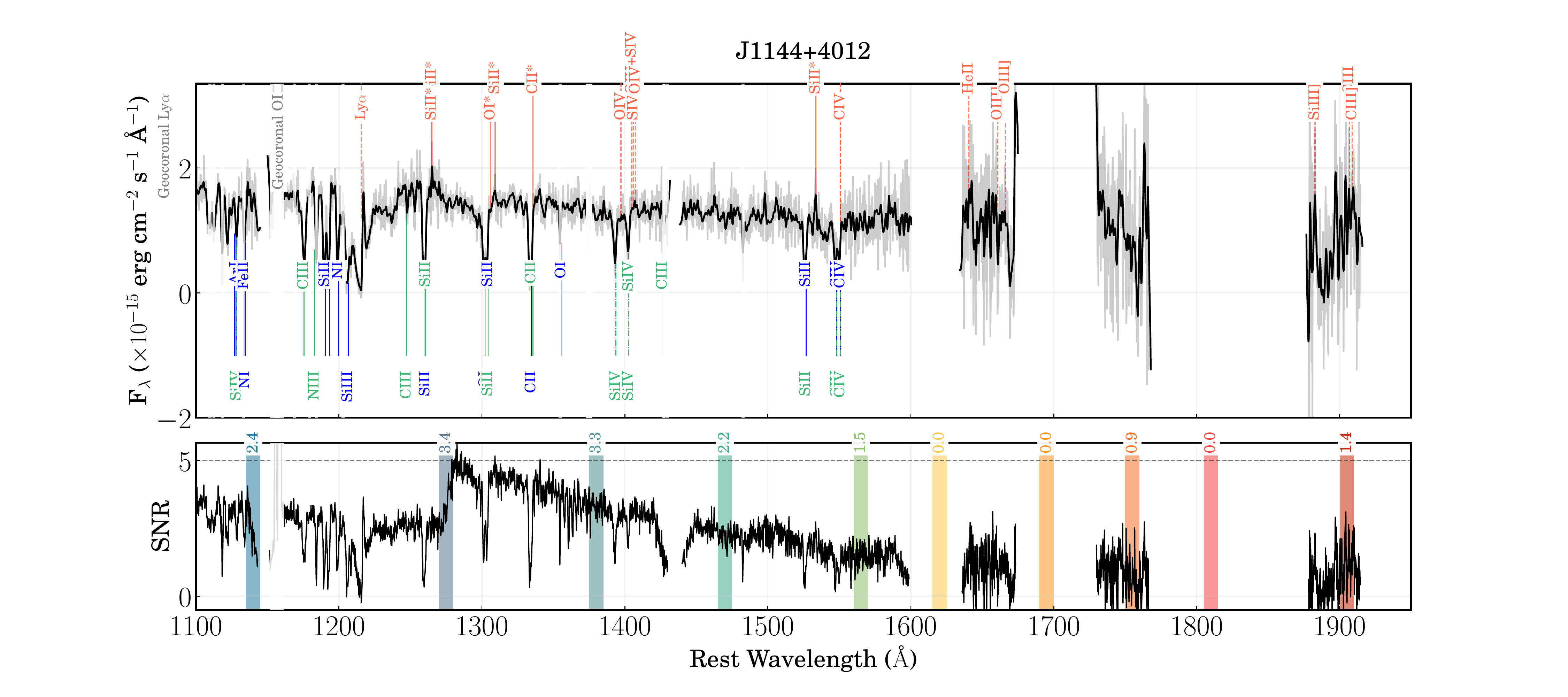} \\
    \includegraphics[width=1.0\textwidth,trim=10mm 0mm 10mm 1mm,clip]{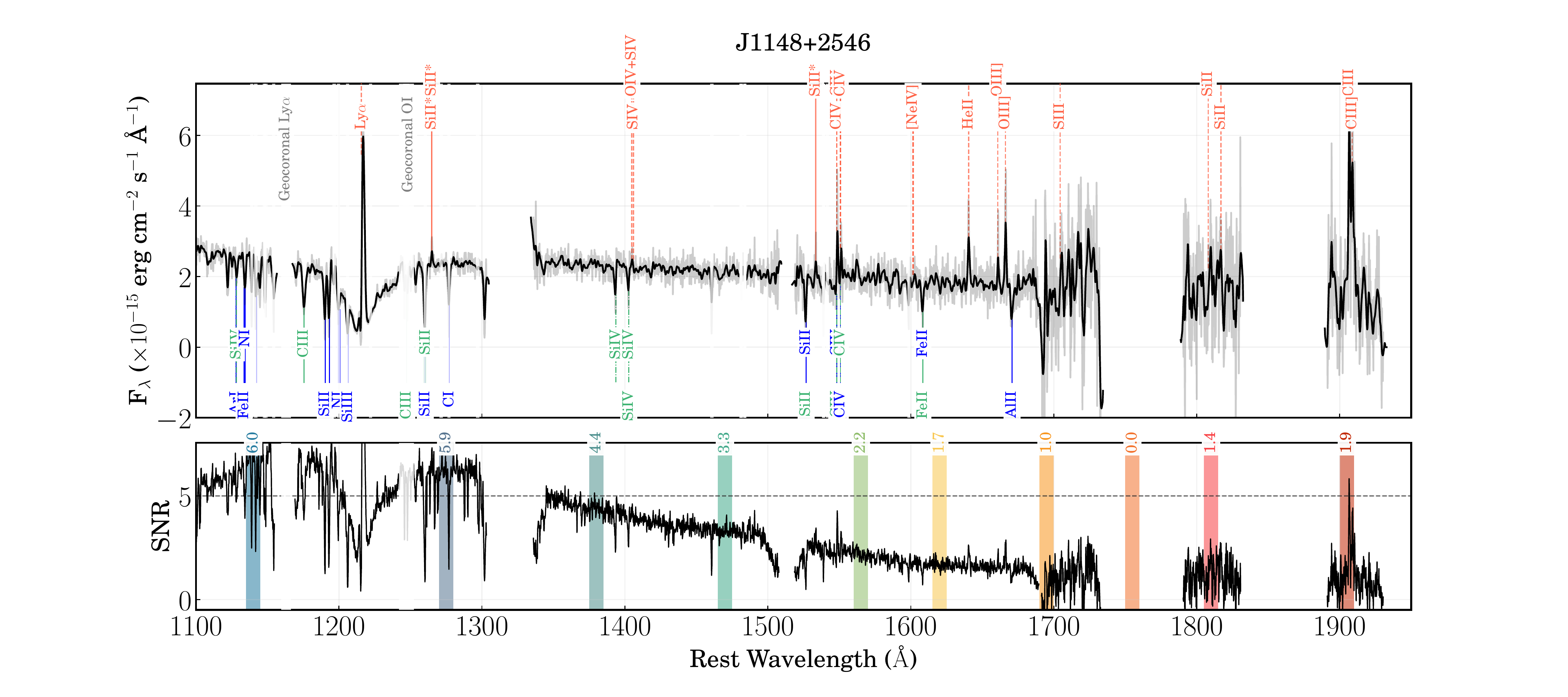} 
\end{center}
\vspace{-4ex}
\caption{(\textit{continued})}
\end{figure*}

\begin{figure*}
\begin{center}
    \includegraphics[width=1.0\textwidth,trim=10mm 15mm 10mm 1mm,clip]{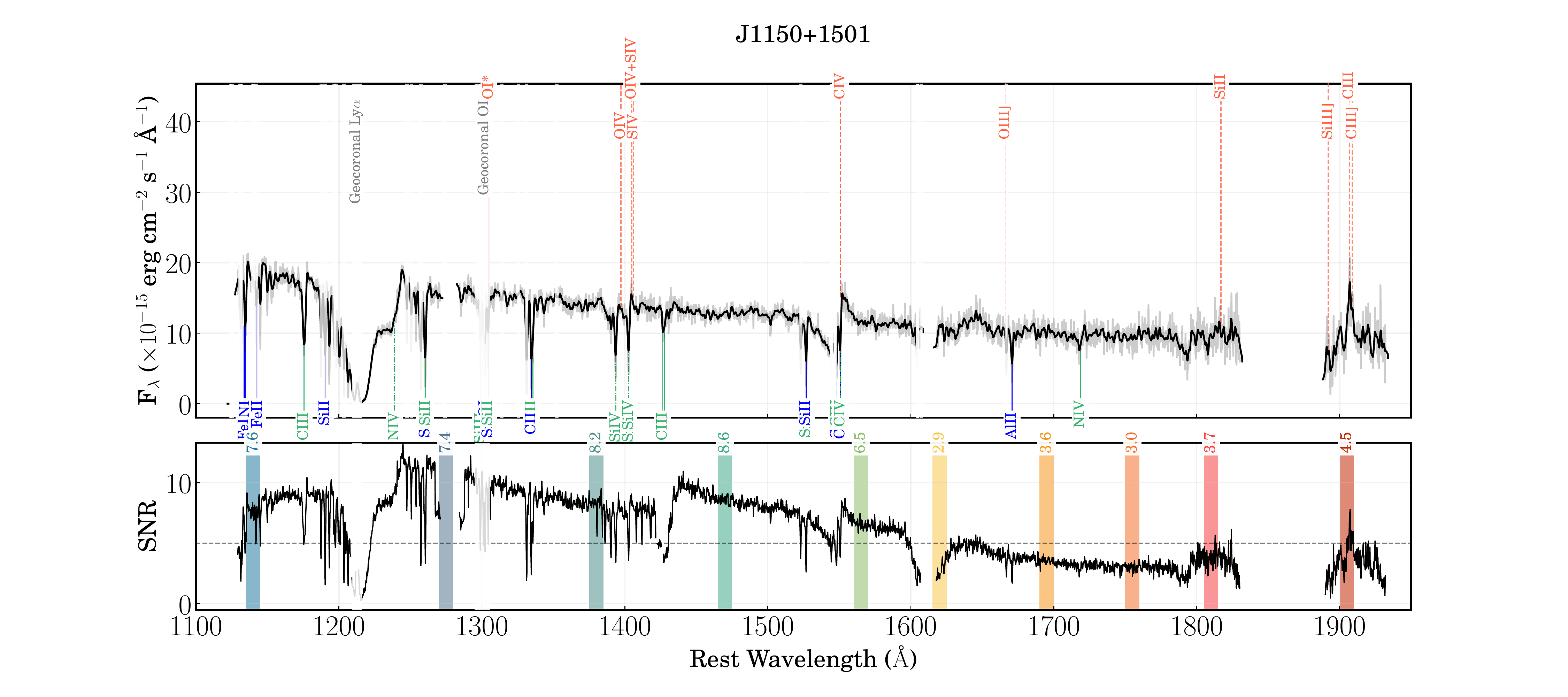} \\
    \includegraphics[width=1.0\textwidth,trim=10mm 15mm 10mm 1mm,clip]{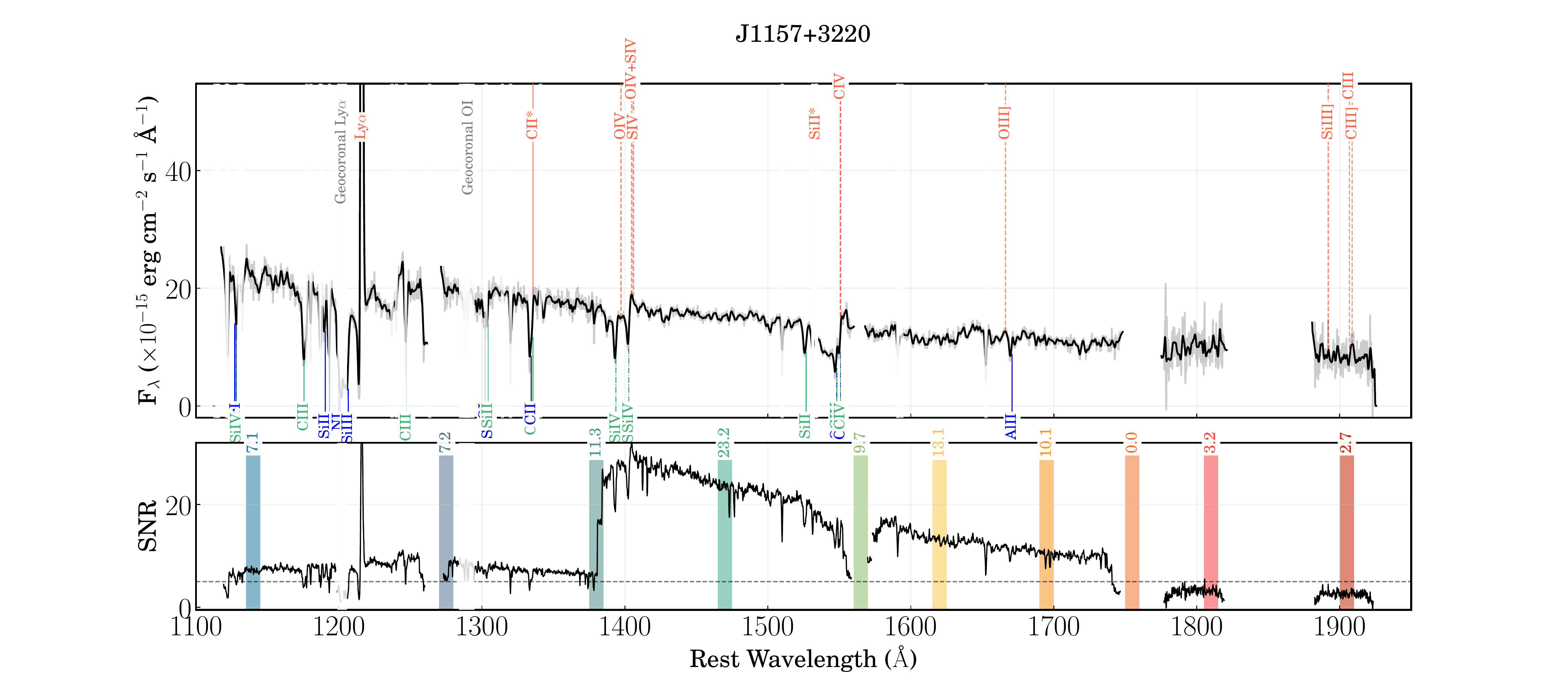} \\
    \includegraphics[width=1.0\textwidth,trim=10mm 0mm 10mm 1mm,clip]{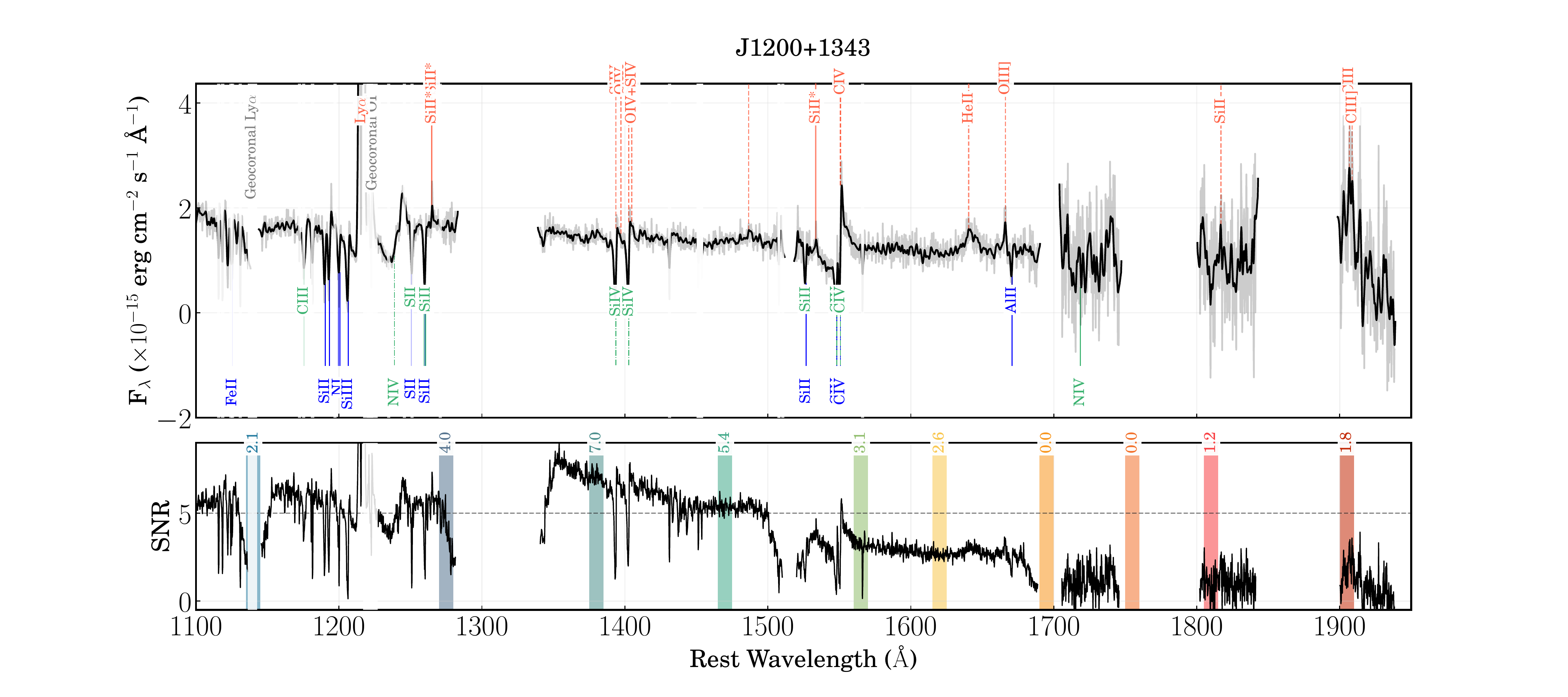} 
\end{center}
\vspace{-4ex}
\caption{(\textit{continued})}
\end{figure*}

\begin{figure*}
\begin{center}
    \includegraphics[width=1.0\textwidth,trim=10mm 15mm 10mm 1mm,clip]{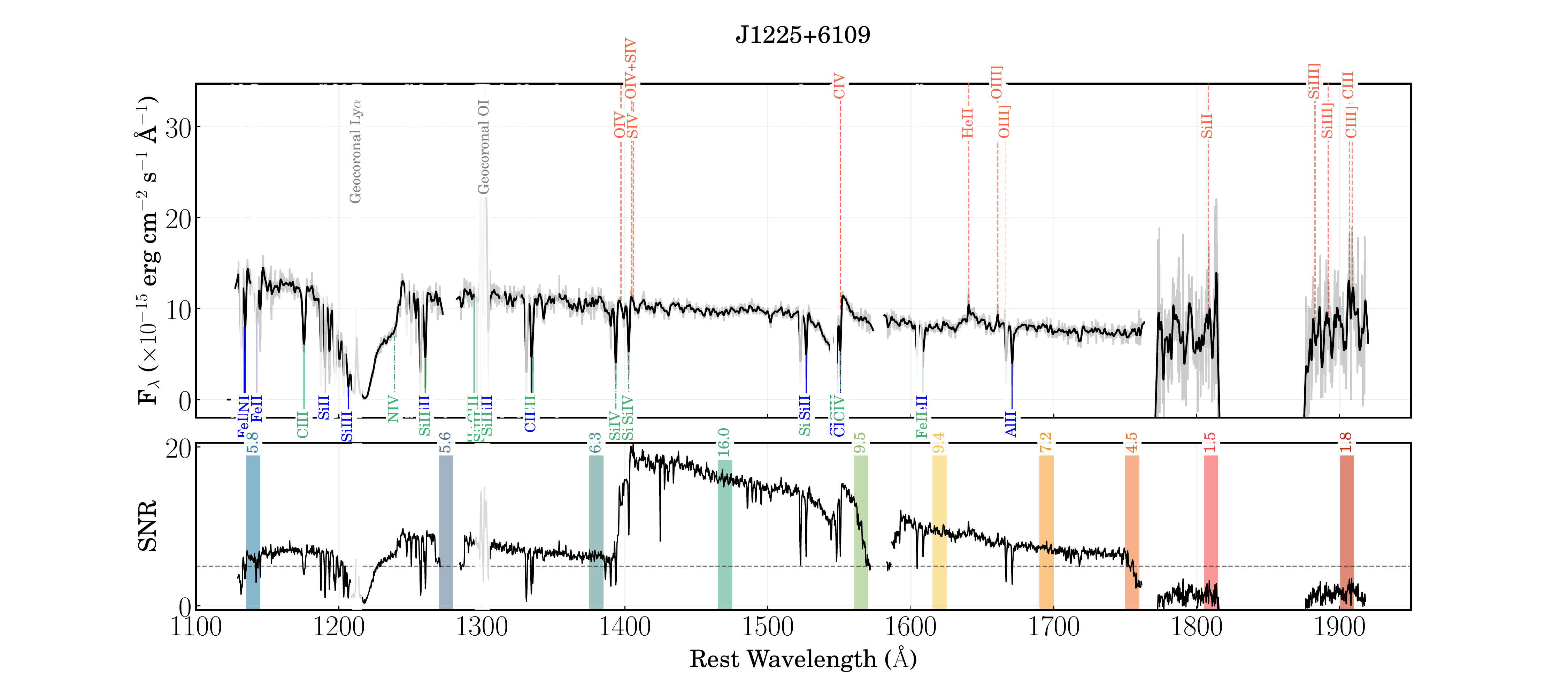} \\
    \includegraphics[width=1.0\textwidth,trim=10mm 15mm 10mm 1mm,clip]{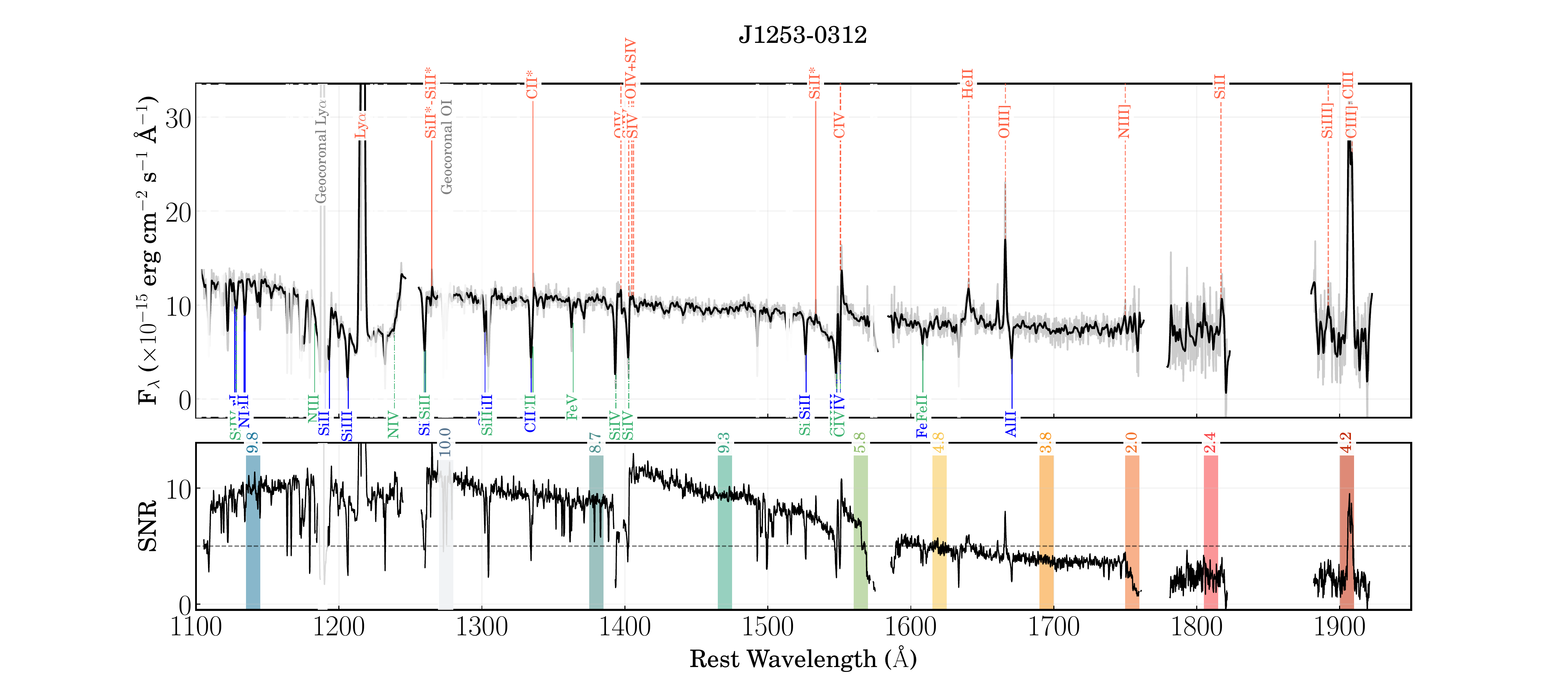} \\
    \includegraphics[width=1.0\textwidth,trim=10mm 0mm 10mm 1mm,clip]{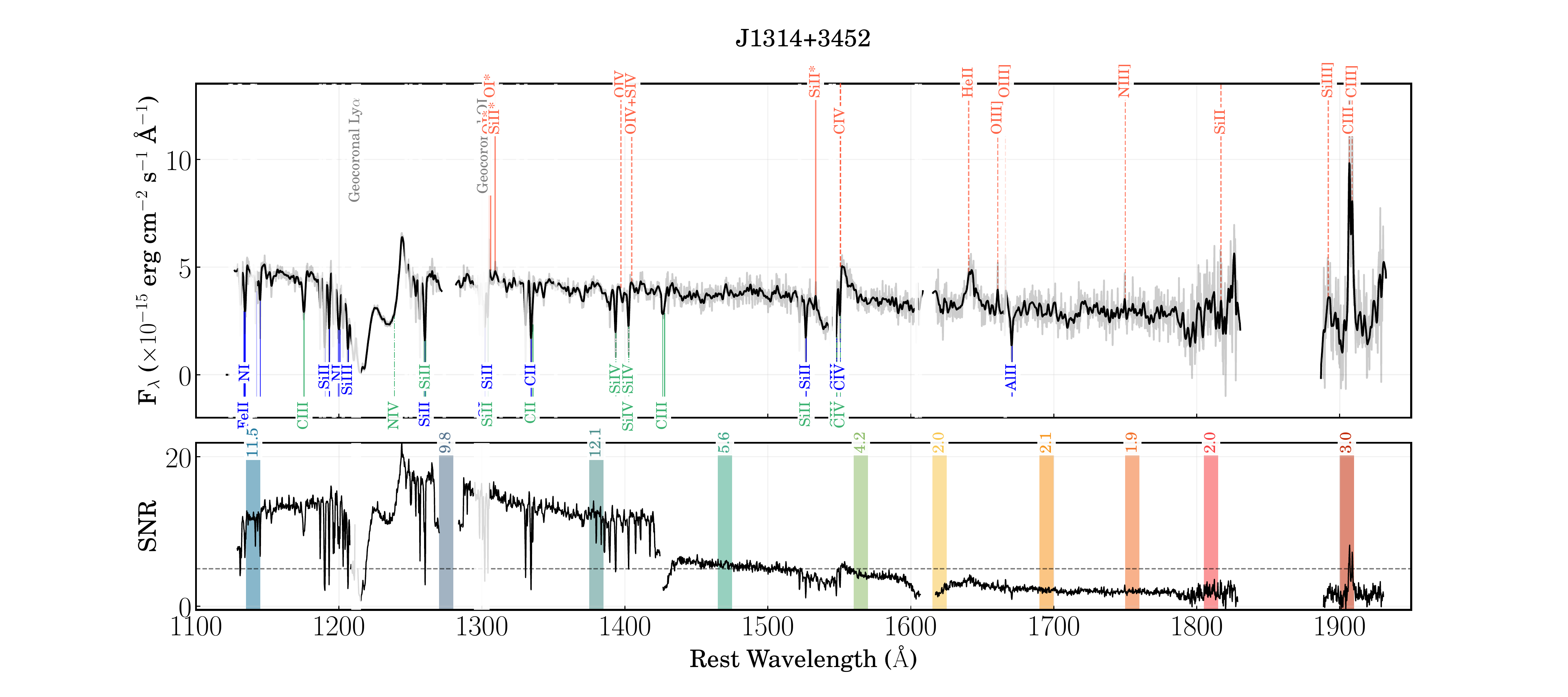} 
\end{center}
\vspace{-4ex}
\caption{(\textit{continued})}
\end{figure*}

\begin{figure*}
\begin{center}
    \includegraphics[width=1.0\textwidth,trim=10mm 15mm 10mm 1mm,clip]{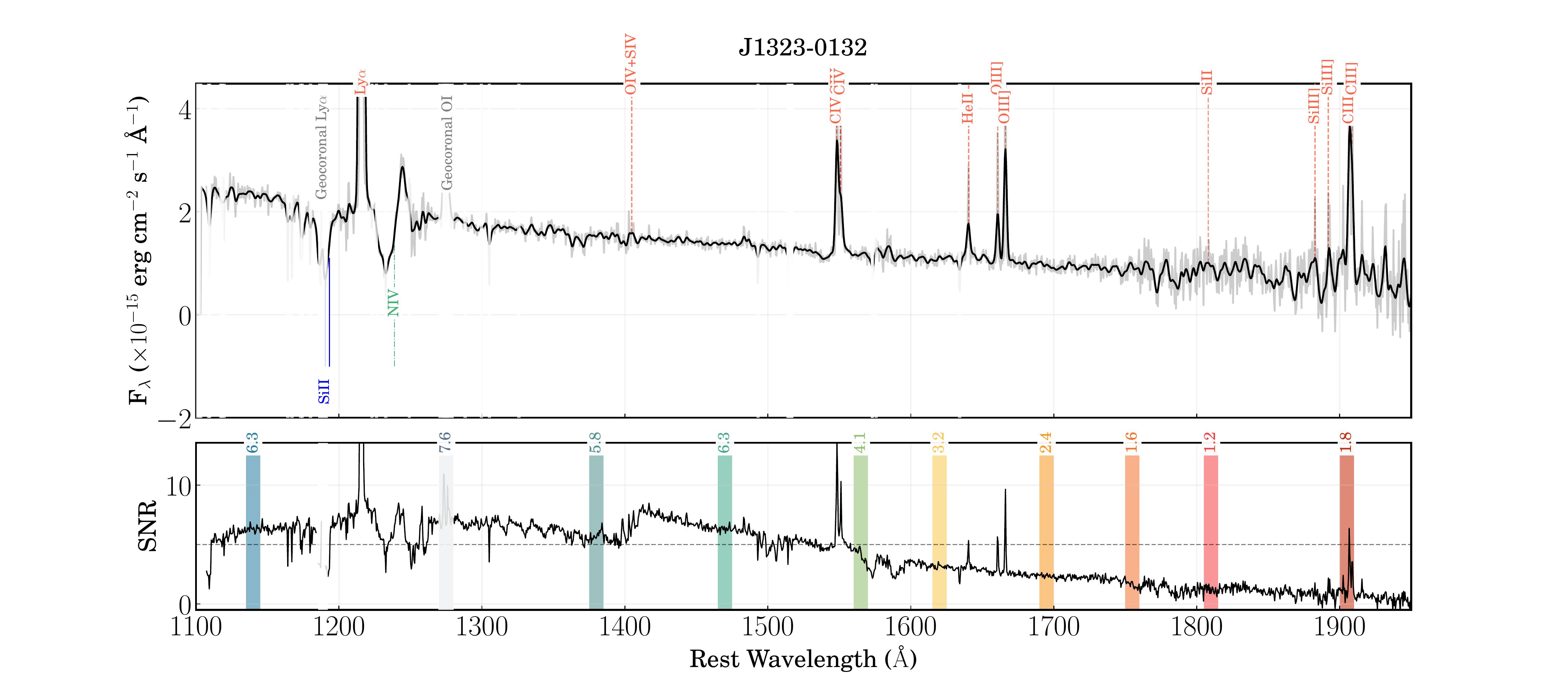} \\
    \includegraphics[width=1.0\textwidth,trim=10mm 15mm 10mm 1mm,clip]{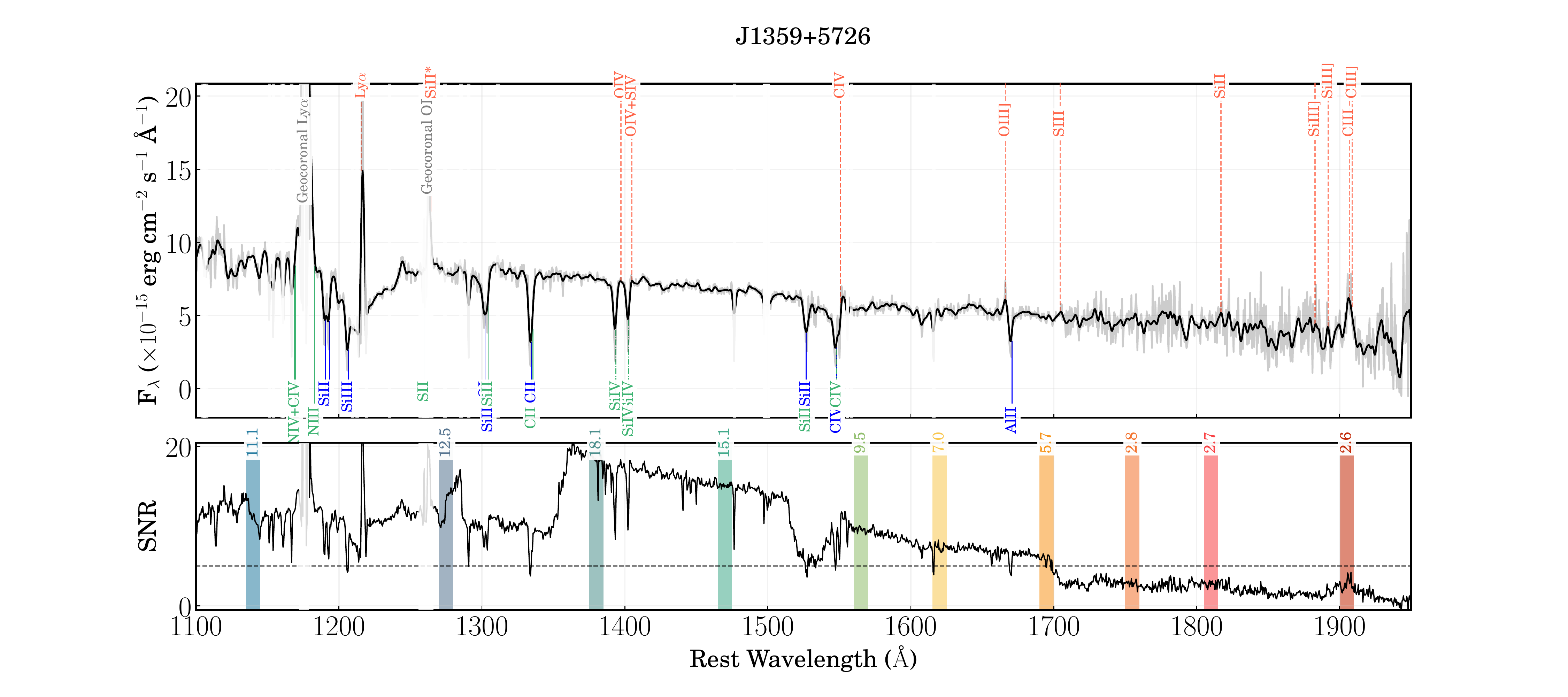} \\
    \includegraphics[width=1.0\textwidth,trim=10mm 0mm 10mm 1mm,clip]{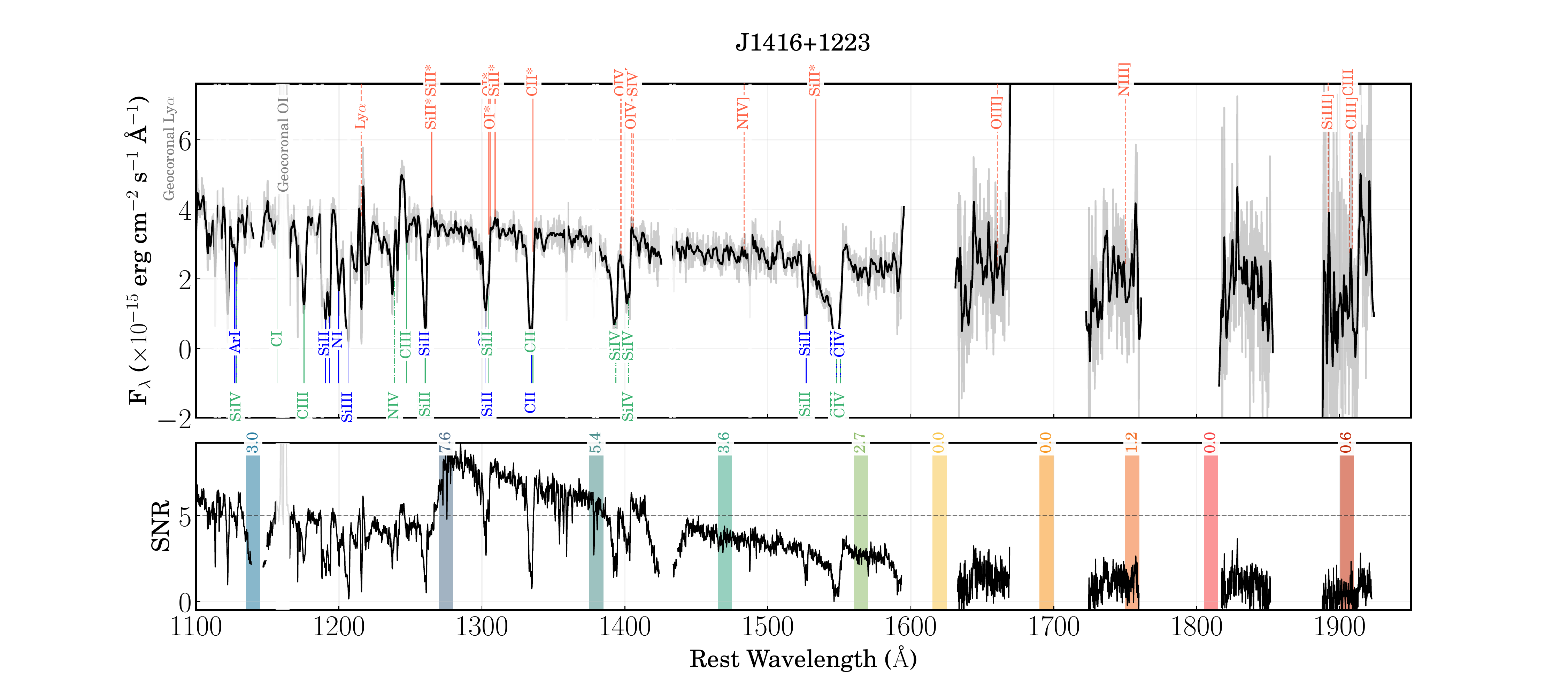} 
\end{center}
\vspace{-4ex}
\caption{(\textit{continued})}
\end{figure*}

\begin{figure*}
\begin{center}
    \includegraphics[width=1.0\textwidth,trim=10mm 15mm 10mm 1mm,clip]{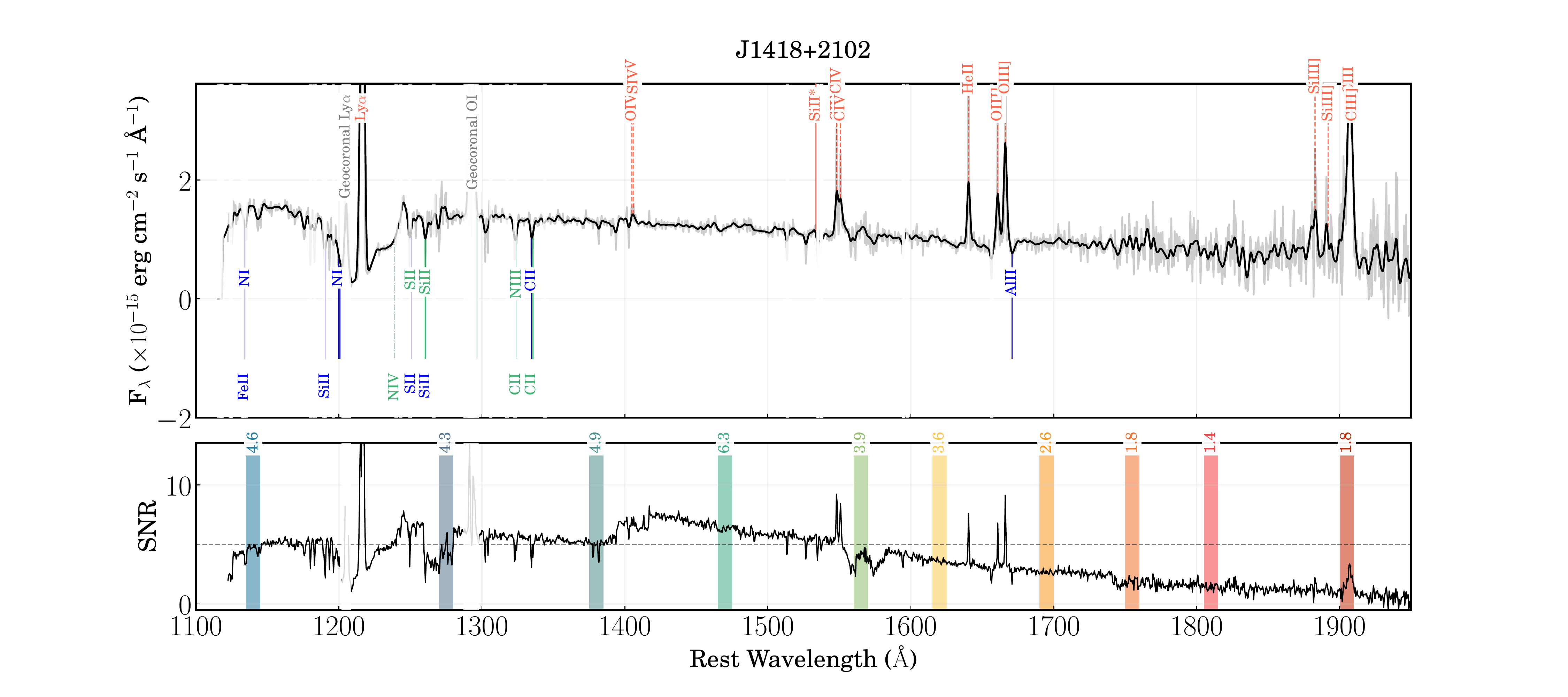} \\
    \includegraphics[width=1.0\textwidth,trim=10mm 15mm 10mm 1mm,clip]{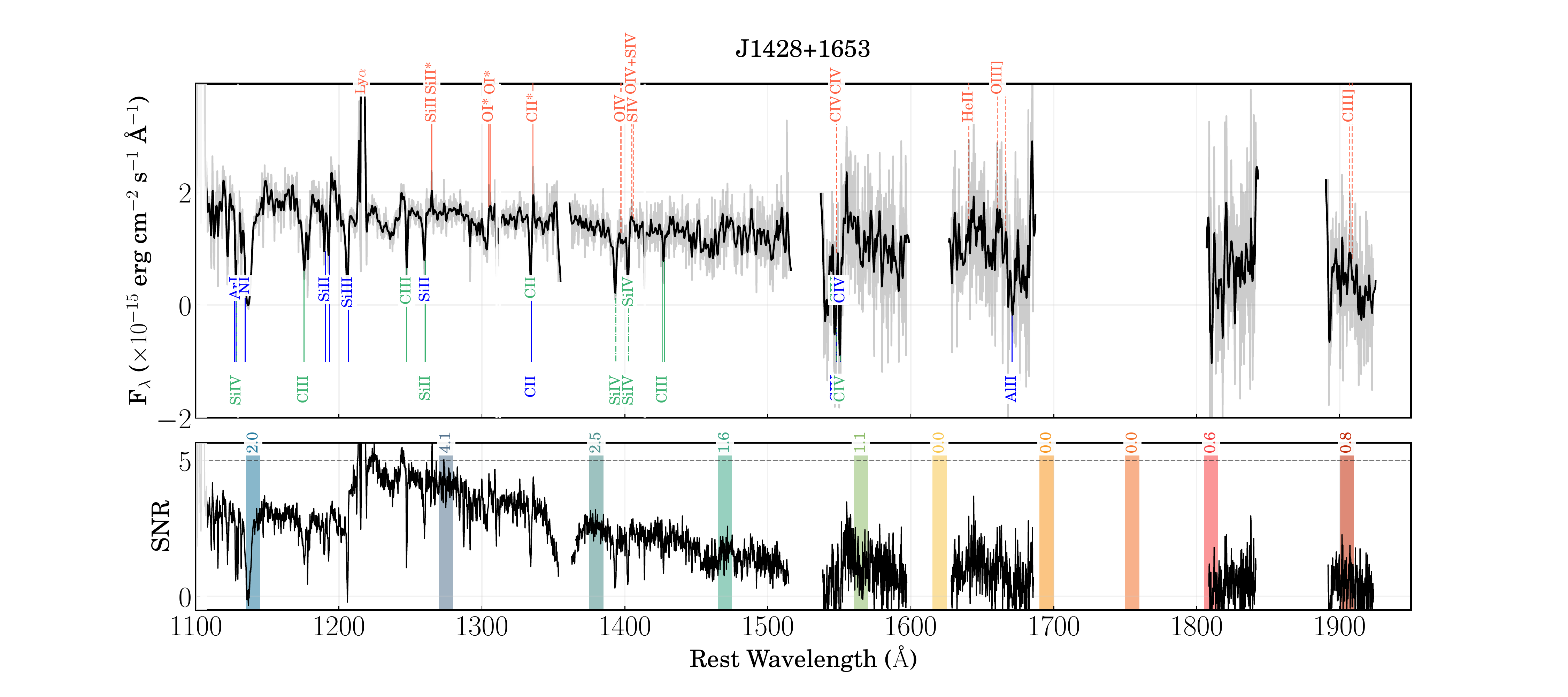} \\
    \includegraphics[width=1.0\textwidth,trim=10mm 0mm 10mm 1mm,clip]{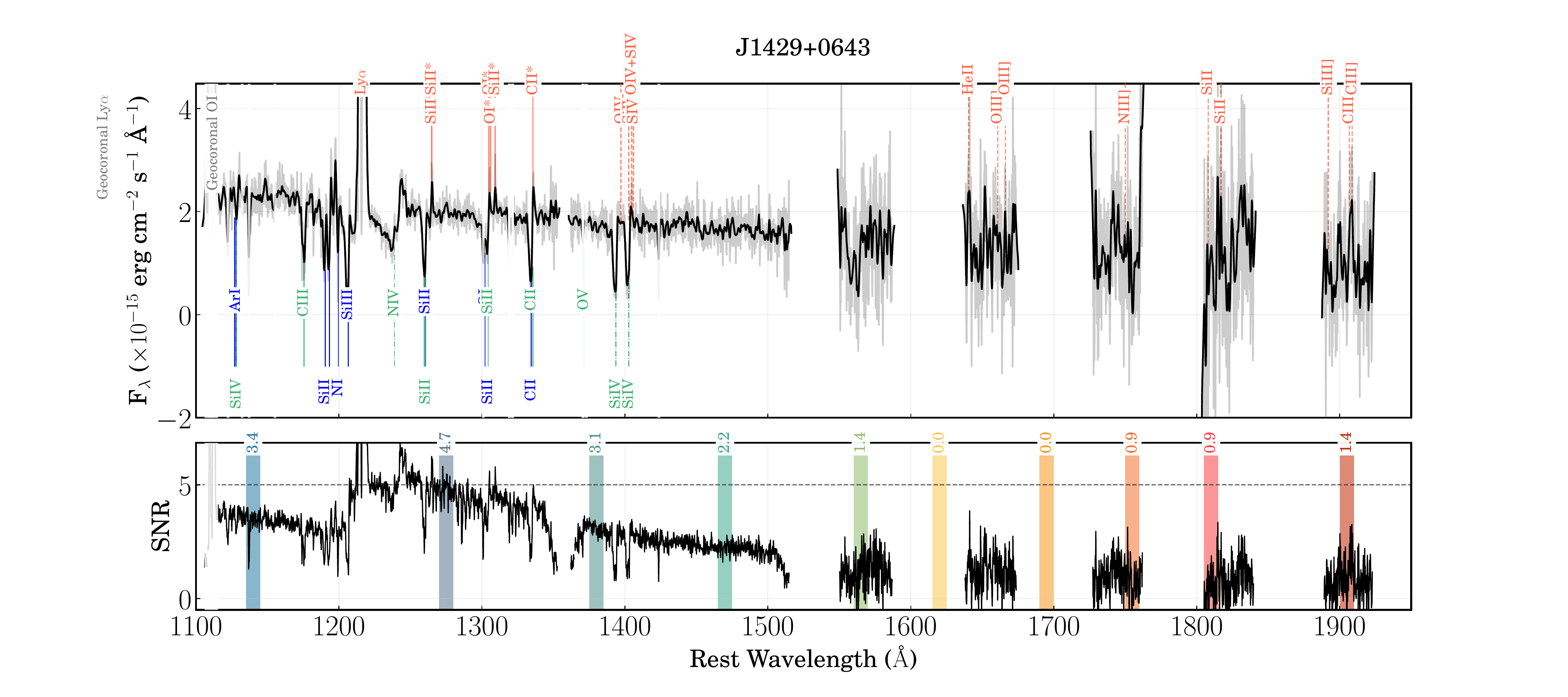} 
\end{center}
\vspace{-4ex}
\caption{(\textit{continued})}
\end{figure*}

\begin{figure*}
\begin{center}
    \includegraphics[width=1.0\textwidth,trim=10mm 15mm 10mm 1mm,clip]{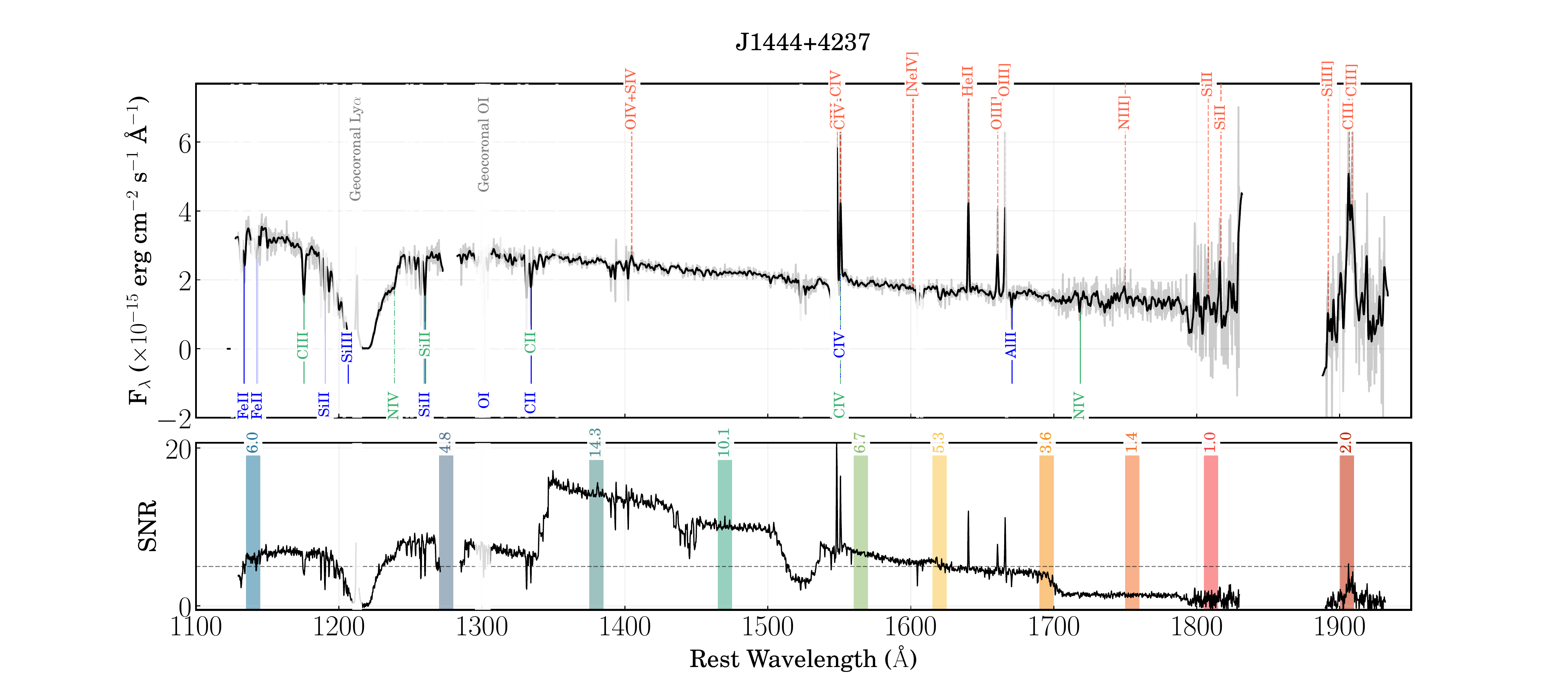} \\
    \includegraphics[width=1.0\textwidth,trim=10mm 15mm 10mm 1mm,clip]{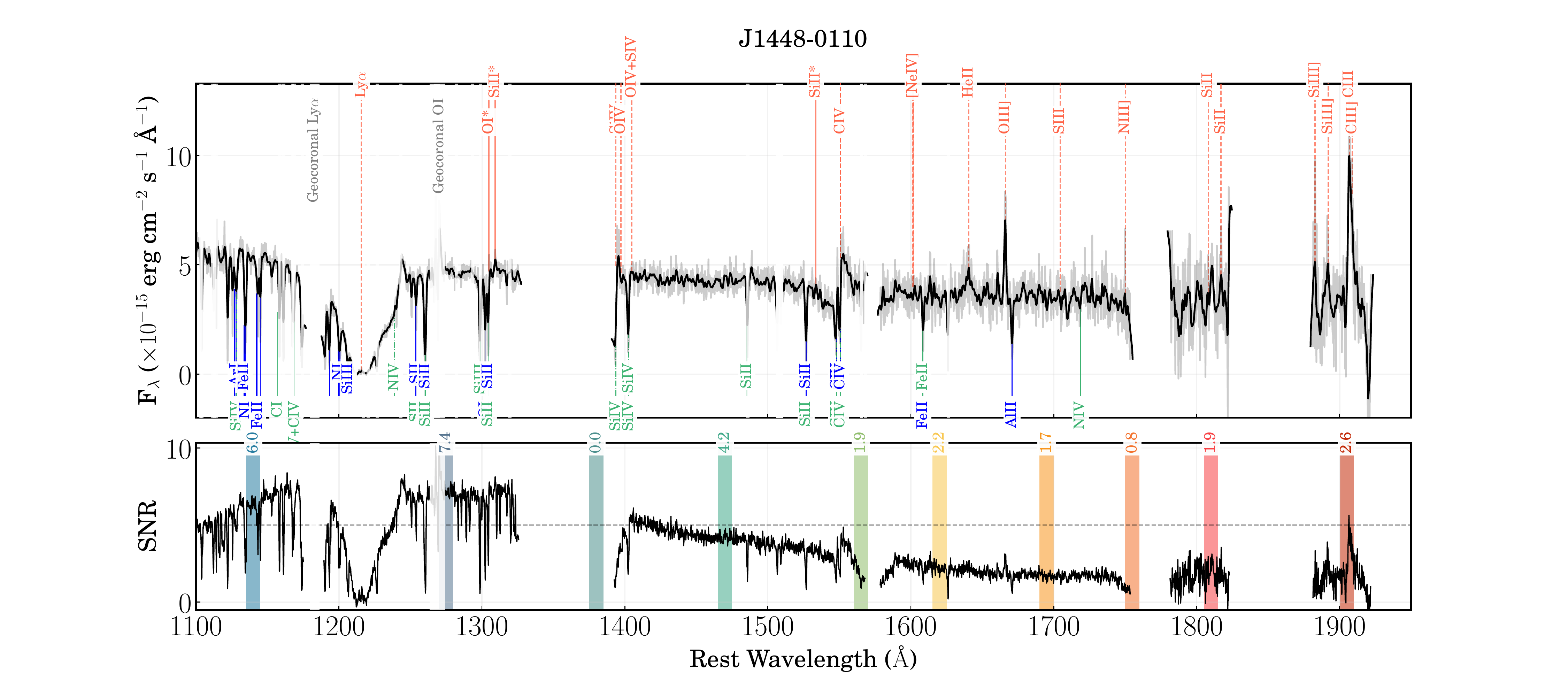} \\
    \includegraphics[width=1.0\textwidth,trim=10mm 0mm 10mm 1mm,clip]{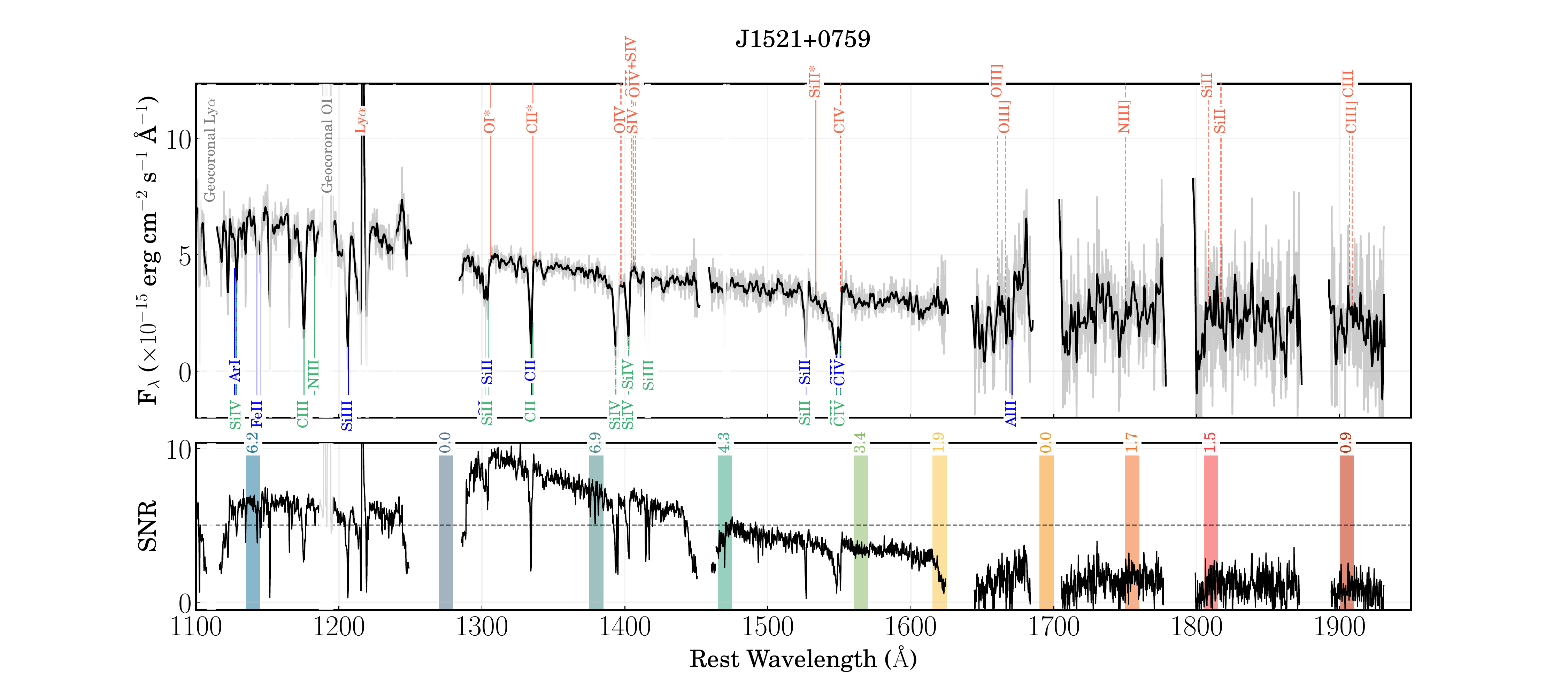} 
\end{center}
\vspace{-4ex}
\caption{(\textit{continued})}
\end{figure*}

\begin{figure*}
\begin{center}
    \includegraphics[width=1.0\textwidth,trim=10mm 15mm 10mm 1mm,clip]{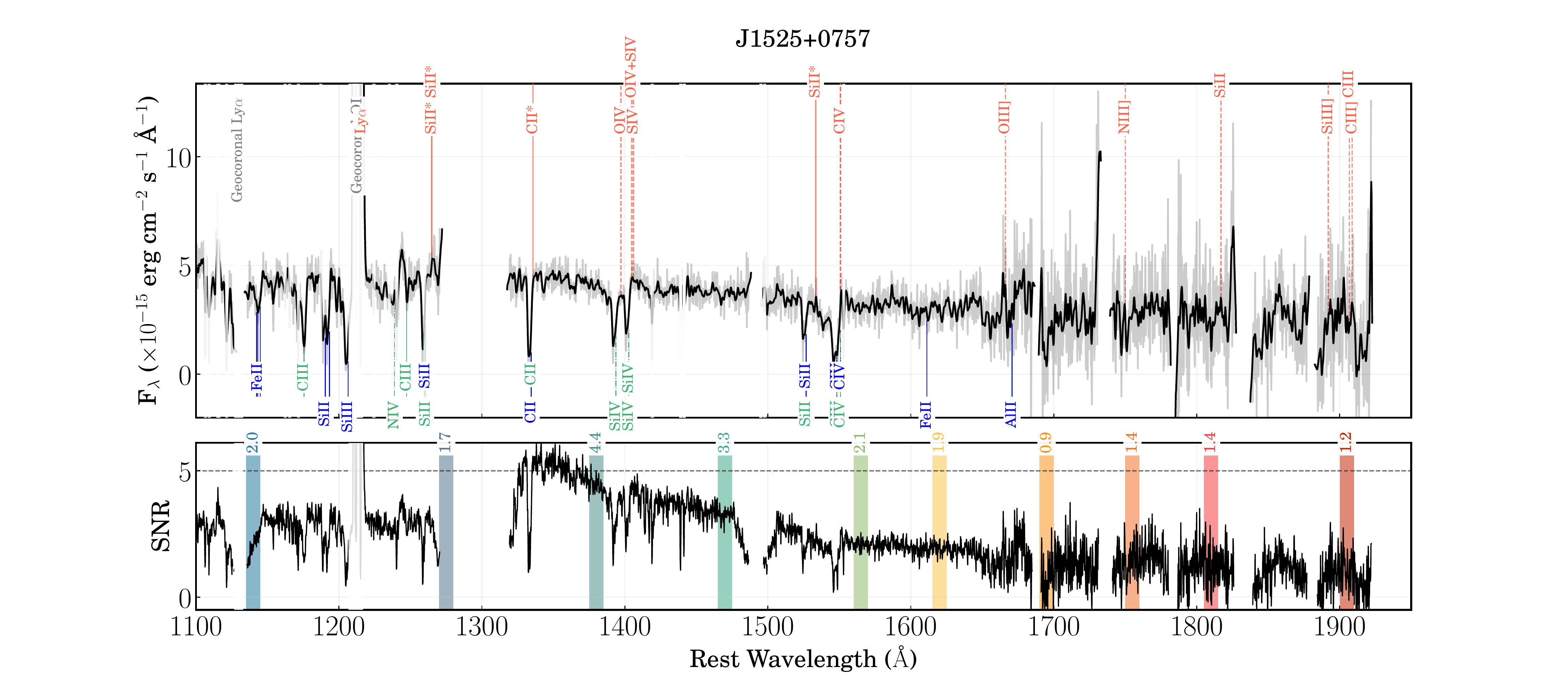} \\
    \includegraphics[width=1.0\textwidth,trim=10mm 15mm 10mm 1mm,clip]{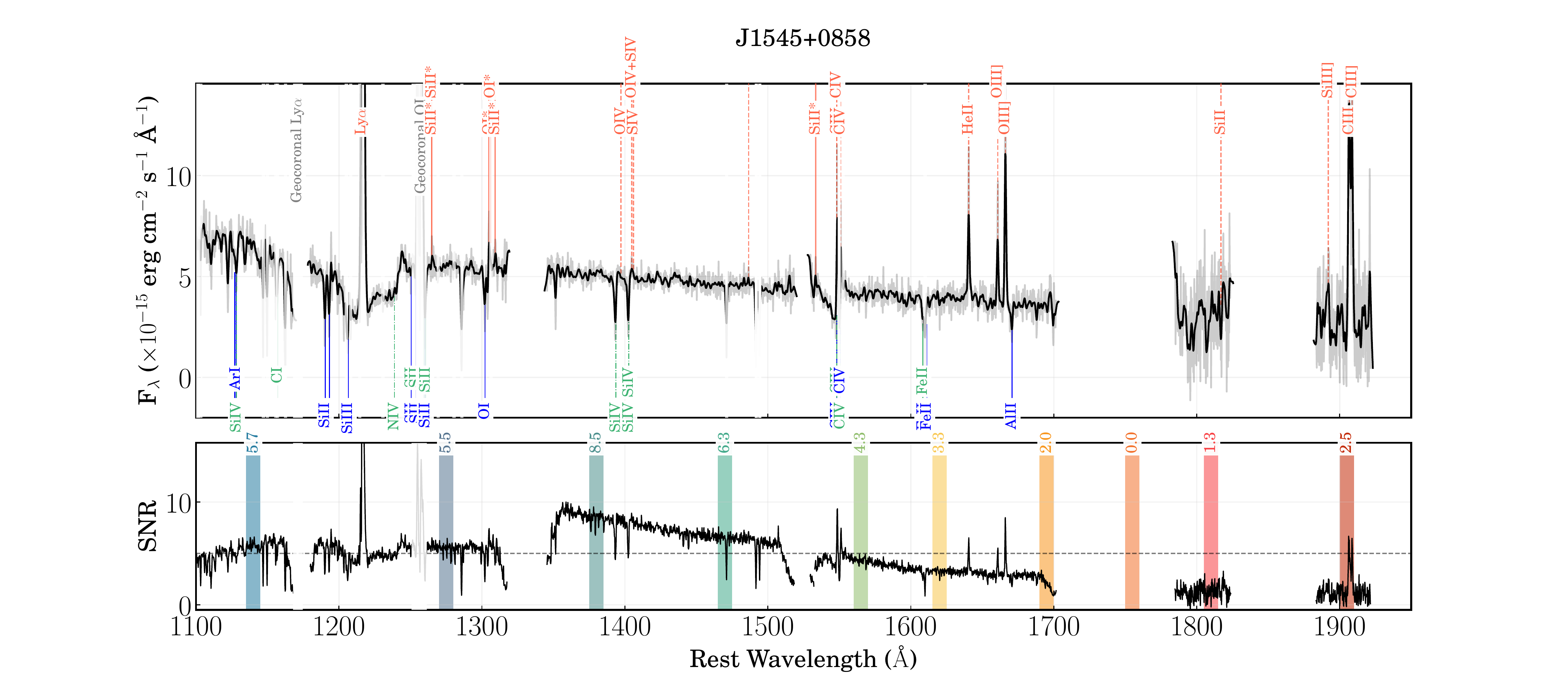} \\
    \includegraphics[width=1.0\textwidth,trim=10mm 0mm 10mm 1mm,clip]{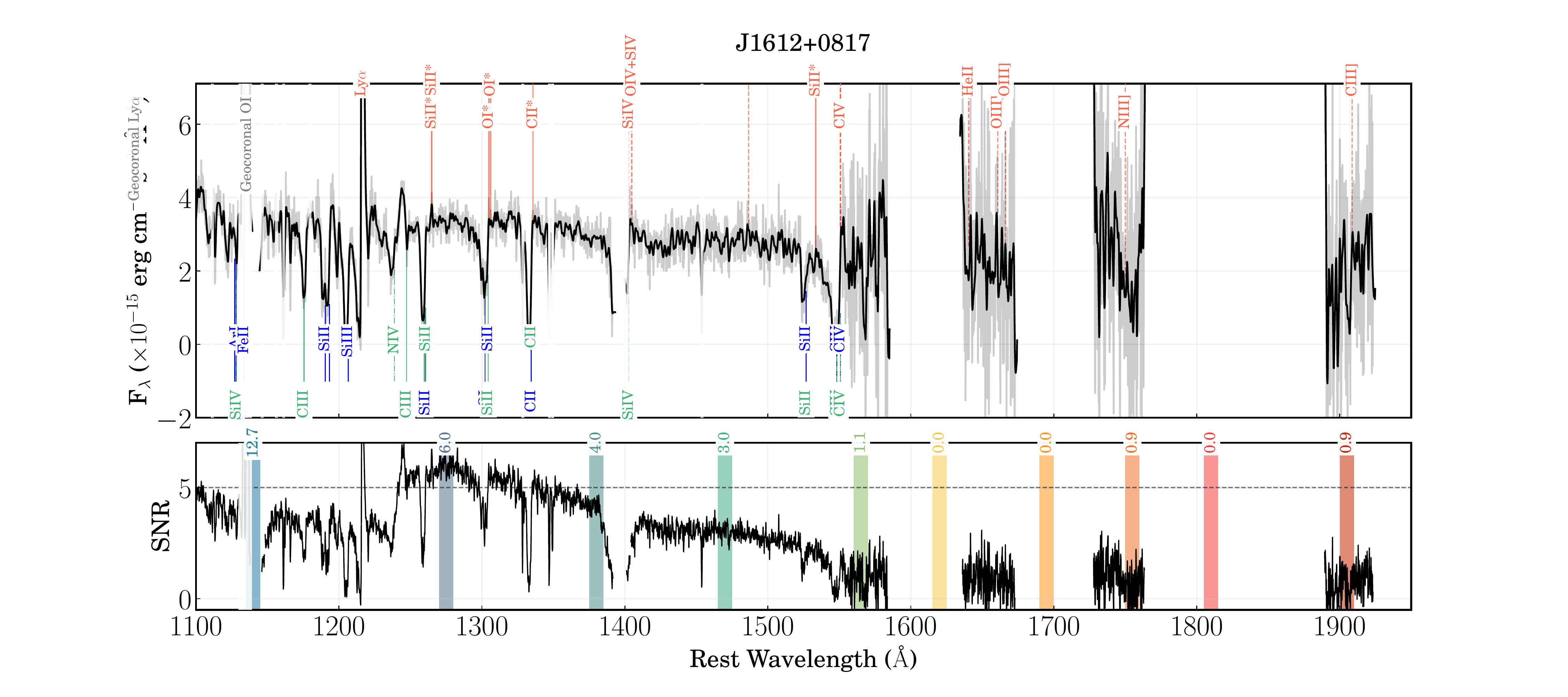} 
\end{center}
\vspace{-4ex}
\caption{(\textit{continued})}
\end{figure*}


\typeout{} 
\bibliography{CLASSY}

\clearpage

\end{document}